\documentclass[a4paper,10pt]{article}
\usepackage[utf8]{inputenc}
\usepackage{graphicx}
\usepackage[hmarginratio=1:1,top=32mm,columnsep=20pt]{geometry} 
\usepackage[hang, small,labelfont=bf,up,textfont=it,up]{caption}

\usepackage{amsmath}
\usepackage{blkarray}
\usepackage{abstract} 

\usepackage[sc]{mathpazo} 
\usepackage[T1]{fontenc} 
\usepackage{url}
\usepackage{lettrine}
\usepackage{xcolor}

\usepackage{abstract}
\usepackage{microtype}
\usepackage[english]{babel}
\usepackage[switch]{lineno}
\usepackage{dblfloatfix}
\usepackage{hyperref} 
\usepackage{algpseudocode}
\usepackage{algorithm}
\usepackage{longtable}
\usepackage{comment}
\usepackage[resetlabels,labeled]{multibib}
\newcites{New}{Supplementary - References}
\makeatletter
\def\BState{\State\hskip-\ALG@thistlm}
\makeatother

\title{\Large MULTI-MODAL NETWORKS REVEAL PATTERNS OF \\ OPERATIONAL SIMILARITY OF  TERRORIST ORGANIZATIONS}
\author{Gian Maria Campedelli$^{1, \dagger}$\\
Iain Cruickshank$^{2}$\\
Kathleen M. Carley$^{2}$
\\
\\

\normalsize $^{1,}$ \textit{Department of Sociology and Social Research, University of Trento, Italy}\\
\normalsize $^{2}$ \textit{School of Computer Science, Carnegie Mellon University, Pittsburgh, PA, United States of America}\\
\normalsize $\dagger$ Corresponding Author: gianmaria.campedelli@unitn.it}
\date{\color{red}Published in \textit{Terrorism \& Political Violence}\\
DOI: \href{https://doi.org/10.1080/09546553.2021.2003785}{https://doi.org/10.1080/09546553.2021.2003785}}

\begin{document}

    \maketitle
    \begin{abstract}
      \noindent Capturing dynamics of operational similarity among terrorist groups is critical to provide actionable insights for counter-terrorism and intelligence monitoring. Yet, in spite of its theoretical and practical relevance, research addressing this problem is currently lacking. We tackle this problem proposing a novel computational framework for detecting clusters of terrorist groups sharing similar behaviors, focusing on groups’ yearly repertoire of deployed tactics, attacked targets, and utilized weapons. Specifically considering those organizations that have plotted at least 50 attacks from 1997 to 2018, accounting for a total of 105 groups responsible for more than 42,000 events worldwide, we offer three sets of results. First, we show that over the years global terrorism has been characterized by increasing operational cohesiveness. Second, we highlight that year-to-year stability in co-clustering among groups has been particularly high from 2009 to 2018, indicating temporal consistency of similarity patterns in the last decade. Third, we demonstrate that operational similarity between two organizations is driven by three factors: (a) their overall activity; (b) the difference in the diversity of their operational repertoires; (c) the difference in a combined measure of diversity and activity. Groups’ operational preferences, geographical homophily and ideological affinity have no consistent role in determining operational similarity.
      \\
      \\
    \end{abstract}




\section*{Introduction}
\lettrine[lraise=0.1, nindent=0em, slope=-.5em]{R}{esearch} on terrorism has examined the decision-making processes of terrorist organizations in terms of tactical behaviors and operations from various perspectives \cite{McCormickTerroristDecisionMaking2003, ShapiroTerroristDecisionMakingInsights2012a}. The empirical relevance of the topic is both theoretical and practical, transcending the boundaries of pure scientific inquiry. In fact, unraveling behaviors of terrorist organizations in terms of the epitome of their decision-making processes—i.e., their attacks—can significantly aid the design of counter-terrorism strategies.

Of particular interest for researchers has been the choice of targets, tactics, and timing of terrorist organizations, which is defined by McCormick as a group's operating profile \cite{McCormickTerroristDecisionMaking2003}. A group’s operating profile lies in between the necessity to optimize influence and maximize security: on the one hand, influence determines the group’s ability to grow in terms of support and resources. On the other hand, security is fundamental to avoid external interventions and disruption. Beyond the competitive influence-security dichotomy, a group’s operating profile is known to be constrained by a set of interrelated factors. Among these, the literature has identified ideology, resources, group size, and goals.

Terrorist groups may decide to attack specific targets because of their symbolic or strategic values in conveying ideological messages during terrorism campaigns \cite{DrakeIdeology1998,AsalSoftestTargetsStudy2009c, AhmedTerroristIdeologiesTarget2018b}. Furthermore, a group’s operating profile can be determined by the availability of financial or material resources, as more resources tend to guarantee easier technology adoption and, consequently, higher effectiveness in terrorist operations \cite{JacksonTechnologyAcquisitionTerrorist2001a, DolnikUnderstandingTerroristInnovation2007b}. Concerning goals, previous works for instance outlined how different aims are associated with different choices in target selection \cite{PoloTwistingarmssending2016e}. Adding a layer of complexity to the spectrum of variables that determine groups’ profiles, research on terrorists’ life cycles further reveals that organizations are subject to variations in their goals, strategies, support, and resources, thus in turn also reflecting temporal variability in their decision-making processes \cite{ClausetDevelopmentalDynamicsTerrorist2012b, SantifortTerroristattacktarget2013c, YangQuantifyingfuturelethality2019b}. A group's operating profile is hence a dynamic concept and is also inherently connected to the study of terrorists' tactical innovations.


In opposition to a traditional strand of research that claimed terrorists' limited ability to innovate \cite{ClutterbuckTrendsterroristweaponry1993, MerariTerrorismstrategystruggle1999a, HoffmanTerrorismSignalingSuicide2004}, based on a restrictive definition of innovation, more recent scholarship hints that innovation not only pertains to the invention of novel weapons or the development of unprecedented strategies \cite{Koehler-DerrickChooseYourWeapon2019a,LubranoNavigatingTerroristInnovation2021}. Instead, it refers to a broad set of changes in activity that may be related to tactical shifts, such as campaigns against alternative types of targets, or modifications of a group’s usual modus operandi, in line with the definition provided by Crenshaw, who argued that terrorist innovation broadly regards the adoption of new patterns of behaviors \cite{CrenshawInnovationDecisionPoints2001}. 

Much has been written on terrorists’ decision-making, innovation, and the consequent evolution of terrorist operations. Both in quantitative and qualitative terms, scholars mostly concentrated on individual groups to map changes in these aspects. However, research overlooked comprehensive comparative accounts of terrorist behaviors, and hence the variegate literature on these topics currently lacks knowledge on operational similarity patterns among different organizations. Addressing this gap may shed light on global characteristics of terrorist violence and can help both research and practice in unfolding the complexity of terrorist behaviors by highlighting behavioral tendencies that cannot be captured when analyzing organizations independently.

In this work we therefore tackle this research problem, focusing on operational patterns of similarity among terrorist groups from 1997 to 2018, by analyzing their dynamic mechanisms and studying the factors explaining operational affinity, building on the representational power of multi-modal networks.

In the last two decades, network science has gained a prominent role in the study of terrorism \cite{PerligerSocialNetworkAnalysis2011b, BouchardSocialNetworksTerrorism2017}. Most applications revolve around the study of connections among affiliates within terrorist organizations \cite{KrebsUncloakingTerroristNetworks2002a, MedinaSocialNetworkAnalysis2014a} or alliance relations among groups \cite{AsalResearchingTerroristNetworks2006, AsalFriendsTheseWhy2016, PhillipsTerroristGroupRivalries2019a}. Additionally, the literature has lately explored the flexibility of networks to map terrorist behaviors going beyond mere physical or communication networks, experimenting with computational techniques that exploit operational or strategical relational features of terrorist groups’ behaviors \cite{DesmaraisForecastinglocationaldynamics2013, Campedellicomplexnetworksapproach2019, CampedelliLearningfutureterrorist2021a}. Our work fits into this latter evolving strand of transdisciplinary research. 

By exploiting data retrieved from the Global Terrorist Database \cite{LaFreeIntroducingGlobalTerrorism2007}, we examine all the groups that have plotted at least 50 attacks from 1997 to 2018, accounting for a total of 105 organizations and more than 42,000 events. We specifically investigate patterns of operational similarity deriving clusters of organizations sharing similar behaviors at the yearly level using event-level information on three distinct attack dimensions—i.e., tactics, targets, and weapons. Clusters are detected using Multi-view Modularity Clustering (MVMC), an ensemble multi-view clustering technique developed to specifically deal with networks measuring or describing phenomena that can be expressed through multi-modal representation. We use the number of detected clusters as a measure of overall heterogeneity in terrorist operations and co-clustering, i.e., being assigned to the same cluster, as evidence of operational similarity between groups. How clustering, and co-clustering, change in time help us better capture the dynamic range of macro-behavioral profiles that characterize terrorism complexity and what factors explain to explain pairwise patterns of similarity.

We first detect an overall reduction in operational heterogeneity over time as documented by the dynamics displayed by the modal tactic, target, and weapon networks and, foremost, by the ratio between the yearly number of the detected clusters and the yearly number of active terrorist organizations. Second, we observe increased stability of co-clustering over time. Prior to 2002 we note high year-to-year co-clustering variability, attesting that groups that were clustered together at  $t$ were then mostly separated in the following years. This signals overall frequent changes of modi operandi in the first years of our analyses. Over time more stable co-clustering emerges, especially during the 2009–2018 period, showcasing consistency in operational similarity patterns. Third, and finally, we demonstrate that operational similarity between two organizations, as defined by co-clustering, is explained by their overall amount of activity, the diversity of their repertoires and a measure of the two combined. On the contrary, convergence of preferences in tactics, targets and weapons, as well as geographical and ideological homophily, do not have a role in determining operational similarity.

\section*{Materials and Methods}
\subsection*{Data}
The primary data source of this work is the Global Terrorism Database (GTD), which is the most comprehensive open-access database on terrorist events available worldwide \cite{LaFreeIntroducingGlobalTerrorism2007}. We here consider the world's most prominent and active terrorist organizations that we operationalize as the groups that have plotted at least 50 attacks from January 1, 1997, to December 31, 2018. \color{black} The time-frame is restricted to the 1997-2018 window because, as illustrated by the GTD Codebook \cite{STARTGTDCodebookInclusion2017},  events that occurred before 1997 often have issues related to incomplete information. One of the variables affected by this problem is the “doubtterr” one, mapping whether there is a degree of uncertainty regarding the actual terrorist nature of a certain event. Since we relied on this variable in our data processing phase to exclude all uncertain events avoiding results driven by problematic records, we chose to focus on a time-frame that guarantees systematic information in this regard.

The sample accounts for a total of 105 organizations that have plotted more than 42,000 attacks worldwide \color{black} (the complete list of groups is available in the Supplementary Information  Table \ref{groups}). \color{black} The GTD gathers more than one hundred variables associated with each event, including the perpetrator, the location (at various resolutions), the time, and operational information on the attack such as the deployed tactics, selected targets, and employed weapons.

In the present work, we specifically use these three latter sources of information to study the operational behaviors of terrorist groups at the yearly level. All events in the GTD can be featured by up to three different tactics (originally labeled in the dataset as “attack type”), three different targets, and four different weapons. To exemplify, an attack may be plotted using a mix of tactics, using multiple weapons against several targets simultaneously. This allows gaining a comprehensive picture of the operational yearly repertoire of each group, starting from rich event-level information.

To analyze the composition of the detected clusters, we moreover exploit other variables included in the GTD, and particularly the regions in which organizations operate, and information on group ideologies retrieved from three ancillary sources: the Big Allied and Dangerous (BAAD) Database \cite{AsalBigAlliedDangerous2011}, the Extended Data on Terrorist Groups (EDTG) dataset \cite{HouIntroducingExtendedData2020a} and the TRAC platform maintained by the Terrorism Research \& Analysis Consortium \cite{TheBeachamGroupTRACTerrorismResearch2021}. 

\paragraph*{Tactics}
IIn the period under consideration, the 105 analyzed groups have deployed a total of nine different tactics, as labeled by the GTD. Information on the hierarchical structure of the information collected in the GTD, as well as on the other variables that we do not consider in this work are available at  \cite{STARTGTDCodebookInclusion2017}. These are: \textit{Armed Assault,	Assassination,	Bombing/Explosion,	Facility/Infrastructure Attack,	Hijacking,	Hostage Taking (Barricade Incident),	Hostage Taking (Kidnapping),	Unarmed Assault,	Unknown}.

\paragraph*{Weapons}
Concerning weapons, organizations in our sample utilized a total of eleven distinct weapon types: \textit{Biological,	Chemical,	Explosives,	Fake Weapons,	Firearms,	Incendiary,	Melee,	Other,	Sabotage Equipment,	Unknown,	Vehicle (not to include vehicle-borne explosives, i.e., car or truck bombs).
}

\paragraph*{Targets}
The universe of total attacked targets consists instead of twenty-two different targets. These are: \textit{Abortion Related,	Airports \& Aircraft,	Business,	Educational Institution,	Food or Water Supply,	Government (Diplomatic),	Government (General),	Journalists \& Media,	Maritime,	Military,	NGO,	Other,	Police,	Private Citizens \& Property,	Religious Figures/Institutions,	Telecommunication,	Terrorists/Non-State Militia,	Tourists,	Transportation,	Unknown,	Utilities,	Violent Political Party.}

\begin{figure}[!hbt]
    \centering
    \includegraphics[scale=0.7]{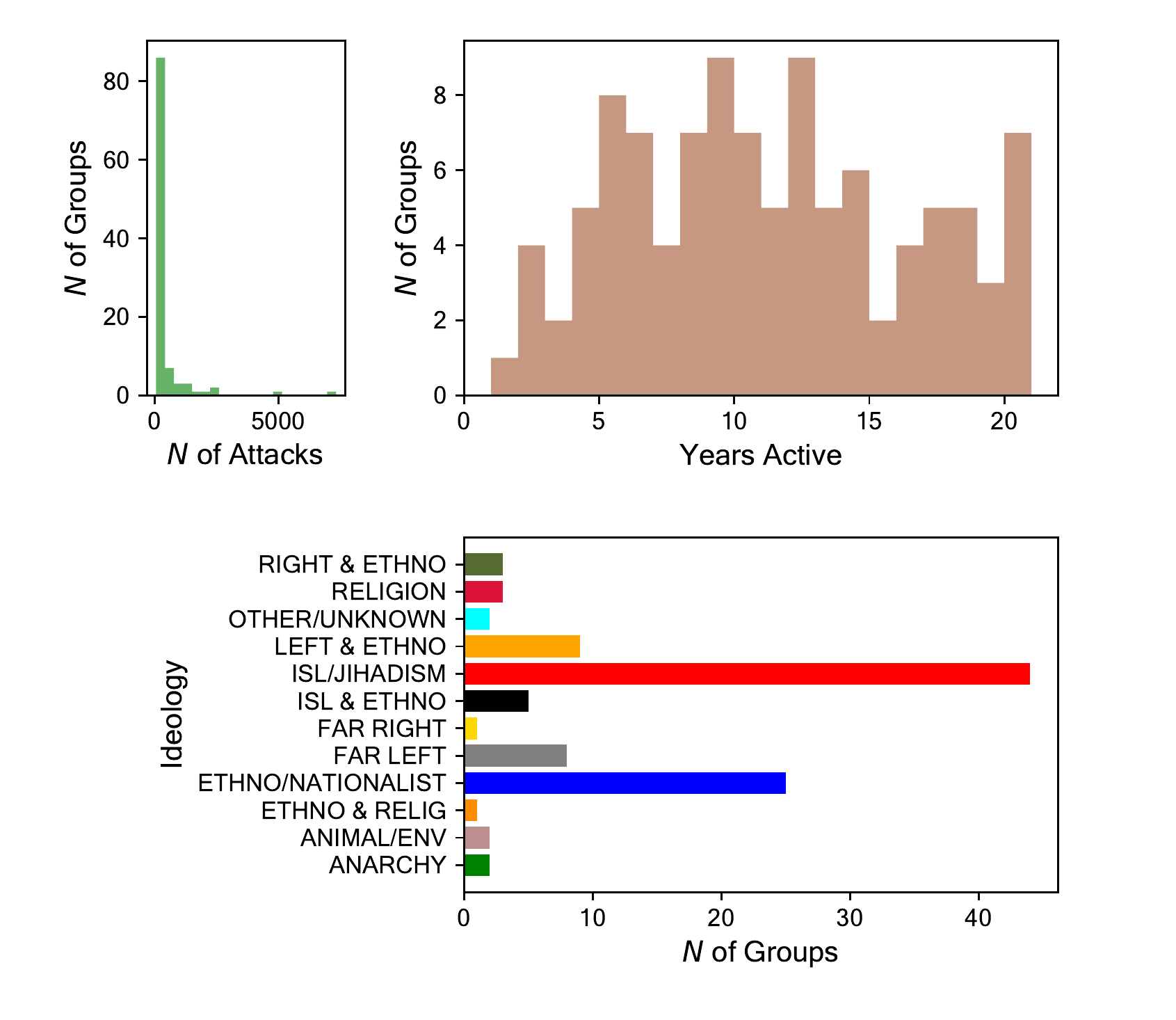}
    \caption{Top-left: histogram displaying the distribution of groups per number of attacks in our sample. Top-right: histogram displaying the distribution of groups per number of years active in our sample. Bottom: Bar chart showing the distribution of groups in terms of ideologies in our sample.}
    \label{fig:desc}
\end{figure}

\paragraph*{Descriptive Characteristics of Terrorist Organizations}
Descriptive statistics of organizations are visually provided in Figure \ref{fig:desc}. The top-left subplot shows that most groups plotted a relatively low number of attacks (minimum 50), while a tiny minority accounts instead for the majority of events (the ten most active groups have plotted a sum of 25,589 attacks, more than 60 percent of the total). The top right subplot instead shows the distribution of terrorist organizations by the number of years they have been active. The subplot denotes that a considerable share of groups has plotted attacks for at least 15 years out of 21. Finally, the bottom subplot shows the distribution of ideologies in our sample. The two most recurring ideologies are Isl/Jihadism, mapping Islamist or Jihadist groups, and Ethno/Nationalist, mapping groups acting for nationalist or separatist purposes, such as the “Corsican National Liberation Front (FLNC)” group. The third and fourth most common ideological categories are “Left \& Ethno”, capturing groups that have both far-left and ethno-nationalist ideological textures, and “Far left”, which trivially considers engaging in political violence in name of far-left ideologies, including Communism.

\subsection*{Multi-View Modularity Clustering}

\subsubsection*{Overview}

Our data are naturally structured in a multi-modal form, with tactics, targets, and weapons being the three modes defining terrorists’ attacks and, in turn, terrorists’ operations. Given this characteristic, and given the independent nature of the three  (deploying tactic $x$ does not strictly imply attacking target $y$ or using weapon $z$), we approach this problem as a multi-view clustering problem. From the GTD data, for each organization in each year $t$, and for each view\footnote{The term ``view'' is used as synonymous of ``mode'' throughout the paper, and refers to the particular structure of the data utilized in our computational framework.} of the data, we form yearly multimodal bipartite graphs, $\mathcal{G}_{t}^{M}=\left \{ G_{t}^{\mathrm{group}\times \mathrm{tactic}}, G_{t}^{\mathrm{group}\times \mathrm{target}}, G_{t}^{\mathrm{group}\times \mathrm{weapon}} \right \}
$, in which weights are integer entries denoting whether and how many times each terrorist organization used a certain kind of tactic, employed a certain weapon, or hit a given target. 

Having established these bipartite graphs for each view and year of the data, we use the multi-view clustering technique of Multi-view Modularity Clustering (MVMC). MVMC is a novel technique designed to work with multiple modes or views, of any data type, of the same underlying social phenomena to produce one set of clusters \cite{CruickshankMultiviewClusteringSocialbased2020}. The technique works in two main steps. First, we take as input each graph in $\mathcal{G}_{t}^{M}$ and learn as many unipartite weighted networks using a Radius Ball Graph (RBG henceforth) procedure, with each new graph connecting groups that are behaviorally similar in each respective mode. Second, an iterative process clusters all of the modal unipartite graphs by optimizing a view-weighted, resolution-adjusted modularity function.

\begin{figure}[!htb]
\centering
\includegraphics[scale=0.57]{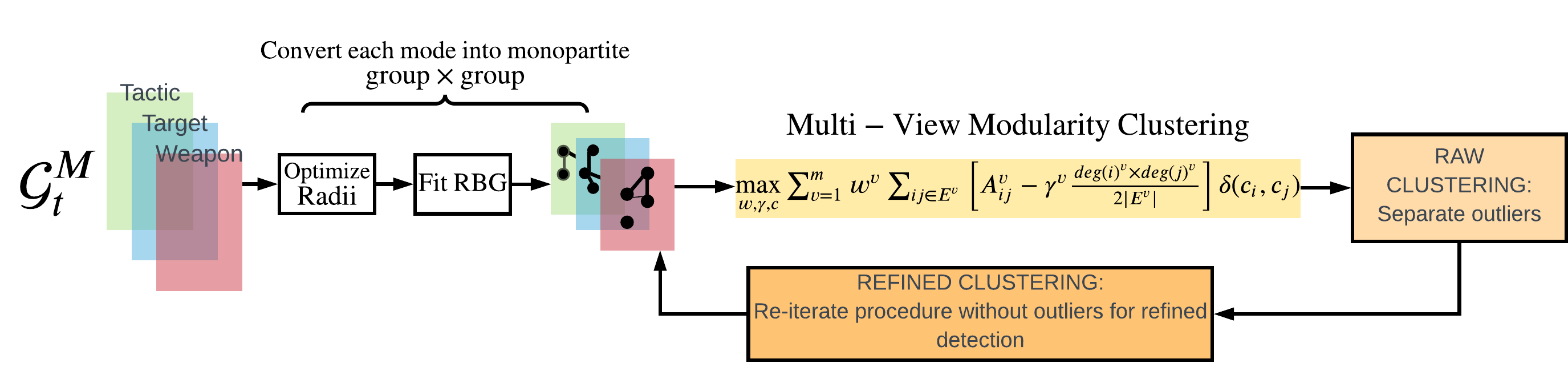}
\caption{\textit{Diagram of the MVMC method as used in this study. We first find networks using a Radius Ball Graph (RBG) procedure for each of the modes of the terrorist attacks, transforming the original bipartite networks in each $\mathcal{G}_{t}^{M}$ into weighted group $\times$ group monopartite ones, and then cluster the organizations to obtain the first raw cluster aimed at separating outliers. We then go through the MVMC process one more time with just the organizations that end up in the single large clusters to produce the refined clusters.}}
    \label{fig:method_overview}
\end{figure}

Since the yearly samples of terrorist organizations frequently contain a small number of outlier groups with significantly more attacks than most of the other terrorist organizations, we iteratively perform the MVMC procedure twice to get a higher resolution clustering of the organizations. The first, or ‘raw’ clustering separates the outlier organizations. The second, or ‘refined’ clustering is directed to the remaining organizations. In this way, we can get more nuanced clusters by working on more than one resolution of the data, instead of just separating the anomalous organizations. Finally, we perform this procedure multiple times for the same data using cluster ensembling. \cite{StrehlClusterensemblesknowledge2003a}. Since the MVMC algorithm is stochastic, we performed an ensembling process to guarantee robustness and make sure the discovered clusters were consistent and not a result of chance, also exploiting the computational efficiency of the algorithm. Figure \ref{fig:method_overview}, provides a graphical overview of the multi-view clustering technique we deploy in the current study. Below we provide technical details for each step of the whole clustering technique.

\subsubsection*{Graph Learning Step}

For the first step of MVMC, we used an RBG procedure to create graphs from each view of the data. An RBG places an edge connecting each group to every other actor that is within a certain distance to the group to create the graph \cite{QiaoDatadrivengraphconstruction2018}. So, for each view of the data $v$ we created a distance-weighted graph, $D^v$, by the equation:

\begin{equation}
    D_{ij}^{v} = \begin{cases}
    dist(x_i^{v} , x_j^{v}), & \text{if $dist(x_i^{v} , x_j^{v}) \leq r^{v}$}.\\
    0, & \text{otherwise}.
    \end{cases}
\end{equation}

where $dist(x_i^{v} , x_j^{v})$ is the distance between two organizations for a particular view of the data and $r^v$ is the radius inside of which we create an edge in the distance-weighted view graph $D^v$. For this study, the distance function of $dist(\cdot, \cdot)$ was defined as the Euclidean distance. We also experimented with Cosine distance, but only noticed a modest difference between the two metrics, especially in the refined clustering step. To determine the radius for each view $r^v$ we adopted the technique for finding the appropriate epsilon parameter in DBSCAN \cite{RahmahDeterminationOptimalEpsilon2016}. In essence, we select a small number of neighbors for each actor, compute the distances to all of these neighbors, and then look for the ‘knee’ point at which these distances increase significantly, indicating measurements outside of a dense region of organizations.

For the small number of neighbors, we followed guidance from \cite{MaierOptimalconstructionknearestneighbor2009} and choose $\sqrt{n}$ neighbors for each actor. In this way, we could heuristically set an appropriate radius for each RBG which would maintain connectivity in dense areas of the data but not place edges across sparse areas of the data. Finally, having found the RBGs for each view, we converted the distance to a similarity with a Gaussian kernel, which is a common means of doing this transformation,\cite{QiaoDatadrivengraphconstruction2018}, and is done by:

\begin{equation}
    a_{ij}^{v} = \begin{cases}
    exp(\frac{-{d_{ij}^{v}}^2}{2 \sigma^2} ), & \text{if $d_{ij}^{v}  >0$}.\\
    0, & \text{otherwise}.
    \end{cases}
\end{equation}

where $A^{v}$ is now a similarity weighted graph and sigma is that standard deviation of all of the edge weights of $D^v$ (i.e. standard deviation of the nonzero elements of $D^v$). By this process, we obtain the monopartite graphs for every mode of the data, and for every year.

\subsubsection*{Clustering Step}

Having transformed each of the modes into representative graphs, we then collectively cluster those mode graphs. To find the clusters from all of the mode graphs simultaneously, we optimize a view-weighted, resolution corrected from of modularity $Q$ given by:

\begin{equation}
    Q=\sum_{v=1}^m w^{v} \sum_{ij \in E^v} [ A^v_{ij} - \gamma^v \frac{deg(i)^v \times deg(j)^v}{2 |E^v|} ]\delta( {c_i, c_j})
\end{equation}

where $w^v$ is the weight assigned to each mode, $v$, $A^v$ is the adjacency matrix for each mode, $\gamma^v$ is the resolution for each mode, $c_i$ is the cluster assignment for actor $i$ and $\delta(\cdot, \cdot)$ is the delta function which returns 1 if both items are the same or 0 otherwise. Since this function is optimizing over both cluster assignments and view weights and modularities, an iterative optimization procedure is opted whereby the the view modularities ($\gamma^v$) and view weights ($w^{v} $) are fixed and we solve for provisional cluster assignments ($c$), using a modularity maximization technique (i.e. Louvain or Leiden). Then the provisional cluster assignments are fixed and we update the view modularities and view weights. The view modularities are updated by the following function:

\begin{equation}
    \gamma = \frac{\theta_{in} - \theta_{out}}{\text{log} \theta_{in} - \text{log} \theta_{out}}
\end{equation}

where $\gamma$ is the resolution parameter, and $\theta_{in}$ and $\theta_{out}$ are the propensities of having edges internal to clusters or external to clusters respectively. The view weights are updated by the following function:

\begin{equation}
    w_v = \frac{\text{log} \theta^v_{in} - \text{log} \theta^v_{out}}{<\text{log} \theta^v_{in} - \text{log} \theta^v_{out}>_v}
\end{equation}

where $w_v$ is the weight given to a view, $v$, and $\theta^v_{in}$ $\theta^v_{out}$ are the propensities for edges to form internal to a cluster or external for the $v$th view, respectively. $<.>_v$ is the average across all of the views. 

This process is repeated until the resolution and weight parameters no longer change. If the resolution and weight parameters do not converge (which can happen in practice), the clustering with the highest modularity value is chosen as the final clustering.  The pseudocode in \ref{alg:Multi-view_Modularity_Clustering} describe the procedure in detail.

The algorithm begins by initializing all the resolution parameters, $\gamma^v_1$, and weight parameters, $w^v_1$ to one (or whatever the user may specify). The algorithm then goes on to cluster the view graphs, $A^v$, by a modularity maximization technique (i.e. Louvain, Leiden), $cluster()$, with the current resolution and weight settings. This step of the algorithm can be done with a fast approximate, modularity optimization routine, which typically has a computational time complexity of $O(NlogN)$ for each graph. So this step of the algorithm will have a computational time complexity of $O(m \times NlogN)$. This step of the output of this is then used to determine the propensities for internal edge formation $\theta_{in}^v$, and external edge formation, $\theta_{out}^v$ for each view. These values are then used to update the resolution, $\gamma^v$, and weight parameters, $w^v$, for each of the views. If the new weight and resolution parameters are the same as the previous ones (within tolerance), the algorithm then exists and returns the final clustering. If the algorithm fails to converge to stable resolution and weight parameters, within the maximum number of iterations allowed, then the algorithm returns whichever clustering produced the highest modularity. Note that modularity for this algorithm is the view-weighted, Reichardt and Bornholdt modularity which incorporates the view resolutions \cite{Reichardt2006resolutionmodularity}.

\begin{algorithm}[!hbt]
\caption{Multi-view Modularity Clustering (MVMC)}
\label{alg:Multi-view_Modularity_Clustering}
\begin{algorithmic}
\BState \textbf{input}:
\begin{itemize}
    \item Graph for each view: $A^{v}$
    \item Max number of iterations: $max\_iter=20$
    \item Starting resolutions: $\gamma_1^{v}=1$, $\forall v \in m$
    \item Starting weights: $w_1^{v}=1$, $\forall v \in m$ 
    \item Convergence tolerance: $tol=0.01$
\end{itemize}
\BState \textbf{output}: Cluster assignments

\State $clustering^* \gets None$
\State $modularity^* \gets - \infty$
\For{$i=1:max\_iter$}
    \State $clustering_i \gets cluster(A, w_i, \gamma_i)$
    \State $modularity_i \gets RBmodularity(A, clustering_i, w_i, \gamma_i)$
    \State $\theta_{in}, \theta_{out} \gets calculate\_thetas(A, clustering_i)$
    \State $\gamma_{i+1}^{v} \gets \frac{\theta_{in}^{v} - \theta_{out}^{v}}{\text{log} \theta_{in}^{v} - \text{log} \theta_{out}^{v}}$, $\forall v \in m$ 
    \State $w_{i+1}^{v} \gets \frac{\text{log} \theta^v_{in} - \text{log} \theta^v_{out}}{<\text{log} \theta^v_{in} - \text{log} \theta^v_{out}>_v}$, $\forall v \in m$ 
    \If{$abs(\gamma_{i+1}-\gamma_i)<tol$ AND $abs(weights_{i+1}-weights_{i})<tol$}
        \State $clustering^* \gets clustering_i$
        \State $modularity^* \gets modularity_i$
        \State BREAK
    \EndIf
    \If{$iter>=max\_iter$}
        \State $best\_iteration \gets argmax(modularity)$
        \State $clustering^* \gets clustering[best\_iteration]$
        \State $modularity^* \gets modularity[best\_iteration]$
    \EndIf
\EndFor
\State \Return $clustering^*$
\end{algorithmic}
\end{algorithm}

\subsubsection*{Two-Step Clustering and Ensembling for Robust Results}

\begin{figure}[!hbt]
    \centering
    \includegraphics[scale=0.7]{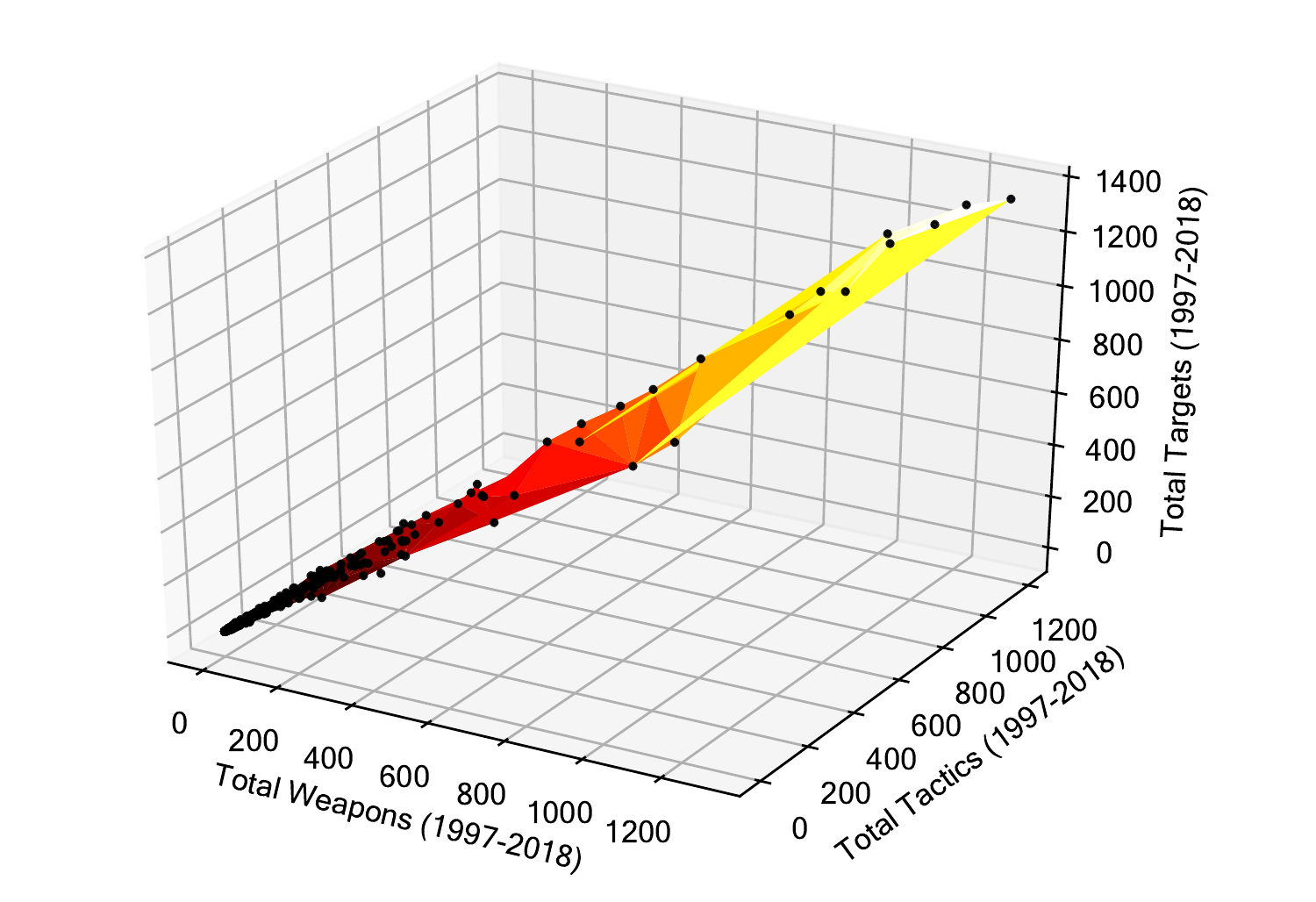}
    \caption{Distribution of terrorist organizations in a three-dimensional space with each dimension mapping the sum of weights of tactics, weapons and targets featured in the period 1997-2018 for each group}
    \label{fig:3d_dist}
\end{figure}

During the preliminary analysis of the groups’ profiles, we noted that the yearly sets of terrorist organizations frequently contain a small number of outlier groups with significantly more attacks than most of the other terrorist organizations  (see Figure \ref{fig:3d_dist}). As such, we run the MVMC procedure twice to get a more nuanced clustering of the organizations. The first, or ‘rough’ clustering separates the outlier terrorist groups exhibiting anomalously high activity. Following this clustering, we extract those organizations that are placed in the largest cluster by the number of organizations. Such cluster represents the bulk of groups that, at first glance, appear similar because of their low or average activity, but possess instead peculiar weights distributions across the three modes, and then require further computational inspection. Hence, we then filter the view data to just these organizations and repeat the MVMC procedure with just these organizations to produce the ‘refined’ clustering. Following the two-step clustering, the clusters from both steps are combined into the final mutually exclusive clustering assignments.

Furthermore, since the MVMC algorithm, like many clustering algorithms, is a stochastic algorithm we also took one final step to ensure the robustness of the results. To ensure the clusterings of the organizations that we found were not a result of chance, we used cluster ensembling to get robust clusters of the data \cite{Strehl2003clusterensembling}. More specifically, we ran the MVMC clustering procedure 10 separate times for each year of the data to create the ensemble of cluster labels. Then, we used a modified version of the BGPA algorithm to create a single, robust clustering from the ensemble clusterings. The BGPA algorithm works by treating the organizations and cluster labels, across all of the ensemble clusterings, as a bipartite graph (actor by cluster labels) and then clusters that bipartite graph \cite{Fern2004BGPA}. We use biLouvain as the final bipartite graph clustering procedure \cite{cruickshank2020multiviewclusteringhashtags}. This procedure then results in one, robust set of cluster labels for the data.

\subsection*{Temporal stability in clustering assignments}
To quantify the degree of co-clustering stability among terrorist organizations over the years we employ two alternative metrics: the Adjusted Rand Index (ARI) and the Fowkles Mallows score (FMS). For each pair of years, both metrics are calculated on the intersection sub-sample of active groups, given the level of variability in organizations active over time. This means that given two years $t_i$ and $t_j$, in which two sets of groups $O_{t=i}=\left \{ o_{1_{t=i}},\cdots,o_{n_{t=i}} \right \}$ and $O_{t=j}=\left \{ o_{1_{t=j}},\cdots,o_{n_{t=j}} \right \}$ have plotted terrorist attacks, the metrics are computed on $S = O_{t=i}\cap O_{t=j}$.

The ARI is a version of the Rand Index which is adjusted for randomness. In an unsupervised learning setting with ground-truth labeling, it is generally calculated as a similarity score between two clusterings by counting pairs of organizations assigned to the same clusters comparing predicted and ground-truth clusterings. In our work, we do not have ground-truth clusters labels, therefore the calculation is made contemplating pairs of clusterings of groups active in a given year $C(O_t=j)$, for all possible pairs. The equation is:
\begin{equation}
    \mathrm{ARI} = \frac{RI - \mathrm{expected}(RI)}{\mathrm{max}(RI) - \mathrm{expected}(RI)}
\end{equation}
where $RI$ stands for Rand Index and is the ratio between the number of agreeing pairs of groups clustered together and the total number of pairs possible. The score is symmetric as $\mathrm{ARI}(C(O_t=i);C(O_t=j)) = \mathrm{ARI}(C(O_t=j);C(O_t=i))$ and takes values in the range [-1;1], with higher values certifying higher stability.

The FMS score is instead defined as the geometric mean of precision and recall through the equation: 
\begin{equation}
    \mathrm{FMS} = \frac{TP}{\sqrt{(TP + FP) \times (TP + FN)}}
\end{equation}
where $TP$ is the number of true positives cases, i.e the pairs of organizations co-clustered together at $t=i$ and $t=j$, $FP$ is the count of pairs of organizations belonging to the same cluster at $t=i$ but not in $t=j$ and $FN$ is the number of pairs of organizations clustered together in $t=j$ but not in $t=i$. The FMS is bounded in the range [0;1], with higher values indicating higher similarity between clusterings in different years $C(O_t=i)$ and $C(O_t=j)$.

\subsection*{Inference on co-clustering}
To understand what drives terrorist organizations’ co-clustering we have applied Exponential Random Graph Modeling (ERGM) per each year, using as the network of interest the two-mode graph connecting a terrorist organization to its cluster of reference. The ERGM is a well-known statistical approach to investigate the factors that contribute to explain the structure of a particular network, allowing inferential reasoning about drivers of outcomes that are not independent of one another (i.e., nodes in a graph).

ERGM has been first specified and presented by Wasserman and Pattison \cite{WassermanLogitmodelslogistic1996}, building on previous statistical breakthroughs \cite{HollandExponentialFamilyProbability1981, FienbergCategoricalDataAnalysis1981a, StraussPseudolikelihoodEstimationSocial1990a}, and gained wide success in a number of fields, including political science \cite{CranmerInferentialNetworkAnalysis2011b, CranmerComplexDependenciesAlliance2012}, criminology \cite{DuxburyNetworkStructureOpioid2018a} and terrorism research \cite{DesmaraisForecastinglocationaldynamics2013, AsalFriendsTheseWhy2016}. Exponential Random Graph models can be generally written as:
\begin{equation}
    P_{\theta ,\mathcal{Y}}(\mathbf{Y}=\mathbf{y}|\mathbf{X})=\frac{\mathrm{exp}\left \{ \theta^{\mathrm{T}}g(y, X) \right \}}{\kappa (\theta, \mathcal{Y})}
    \label{ergm1}
\end{equation}
with $Y$ representing a bipartite network with realization $\mathbf{y}$ where $y_{o,c}=1$, meaning that a connection exists, if organization $o$ belongs to cluster $c$, $g(y,X)$ being a vector of model statistics for network realization $y$, $\theta$ being the vector of coefficient of statistics $g(y)$ and $\kappa (\theta, \mathcal{Y})$ representing a normalizing constant, namely the numerator summed across all possible network realizations. 

Although developments in network modeling led to the extension of the traditional ERGM approach in order to take into account temporal dependence and sequentiality in network realizations, for instance through Temporal ERGM (TERGM) \cite{HannekeDiscretetemporalmodels2010a, DesmaraisStatisticalmechanicsnetworks2012} and separable temporal ERGM \cite{KrivitskySeparableModelDynamic2014a}, we estimated yearly separate cross-sectional models. This enables us to observe year-to-year variations in coefficients and study how the influence of certain factors evolved over time. Moreover, the use of TERGM is not justified as our yearly clusterings are not temporally dependent (that is, clusters in year $t$ do not influence clusters in year $t+1$), and therefore previous realizations of group-to-cluster networks cannot be used to generate successive ones.

The estimated models include eight covariates each. First, “Sum of Features Weights” measures the sum of all feature weights in the original bipartite multi-modal yearly networks $\mathcal{G}_{t}^{M}$ as a proxy of groups’ yearly level of activity and resources.

Second, “Difference in Number of non-zero features” assesses the role of organizations’ pairwise difference in the number of features characterizing attacks in each year, aiming at understanding how diversity in operational heterogeneity impacts the probability of co-clustering. Each organization’s number of non-zero features is simply the count in the number of (binary) links in each year’s $\mathcal{G}_{t}^{M}$. 

Third, “Difference in Non-zero Features to Weights Ratio” is a more comprehensive indicator of operational diversity. For each group, we calculate a ratio between the number of non-zero feature weights and features weights. The ratio is bounded in the range $(0,1]$, with 1 indicating that a terrorist group always differentiates its operations. In the ERGM, the covariate maps the absolute difference between this ratio for each pair of groups to understand if operational diversity in relation to resources impacts co-clustering.

Fourth, fifth, and sixth, categorical variables “Shared most common target,” “Shared most common tactic,” and “Shared most common weapon” capture the role that homophily in operational preferences has in co-clustering, investigating whether two organizations sharing the same most common target, tactic and weapon—which are obtained by group’s highest weight in the respective mode of $\mathcal{G}_{t}^{M}$ --- have increased likelihood of being in the same cluster.

Seventh, the categorical variable “Shared Region” addresses geographical homophily, mapping whether two terrorist groups that have carried out the majority of their operations in the same geographic region are also operationally more similar. The rationale is to examine if macro-geographical affinity patterns characterizing terrorist operations exist.

Eighth, the categorical covariate “Shared Ideology” analyzes whether two groups have a higher or lower likelihood of being co-clustered together when sharing the same ideological motives.

\section*{Results}

\subsection*{Global trends in operational patterns} \label{section: trends}
\begin{figure}[!hbt]
    \centering
    \includegraphics[scale=0.55]{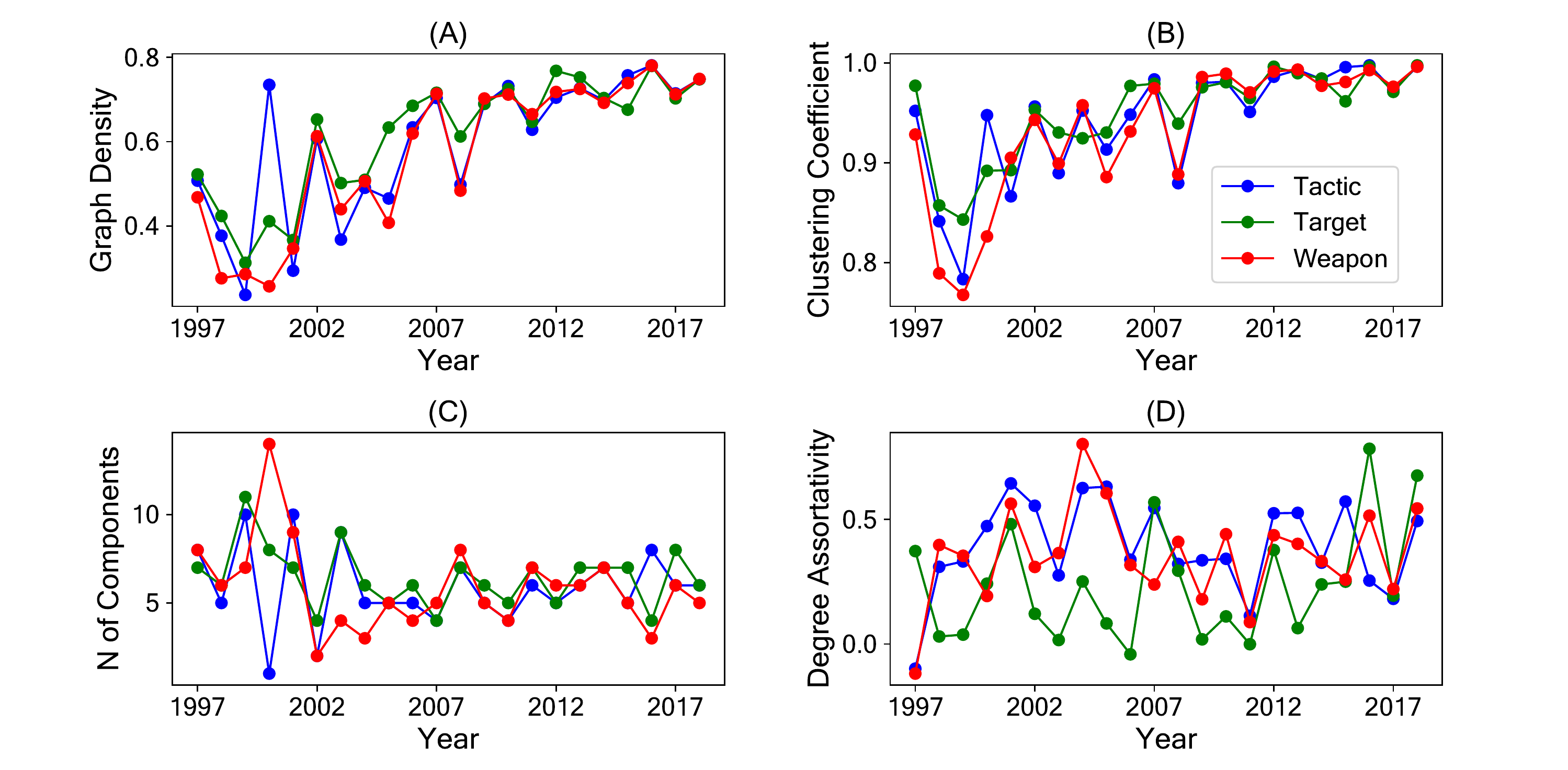}
    \caption{\textit{Temporal trends of RBG networks for each mode. Subplot (A) maps graph density, subplot (B) displays clustering coefficient, subplot (C) shows the number of components, and subplot (D) degree assortativity. Graph density increases over the years for all modes, testifying to increasing homogeneity at the modal level. Similarly, the trends in clustering coefficient and the number of components suggest a reduction in heterogeneity. Degree assortativity instead reveals misaligned patterns across modes, with network structures differing from mode to mode.}}
    \label{fig:rbgnetsstats}
\end{figure}

Network-level statistics for the RBG modal networks from which clusters are computed already anticipate clear trends in global terrorism dynamics over time  (Figure \ref{fig:rbgnetsstats}). First, as shown in the top-left subplot, all modes are characterized by increasing network density. This means that, over time, organizations in each of the modal graphs have on average more links, thus suggesting an increase in similarity among actors. This trend in modal graphs starts consistently after 2000.

The clustering coefficient results are in line with this intuition. After the first years marked by a downward trend, RBG networks have continued to become more and more clustered, testifying to a reduction in the overall heterogeneity of terrorists’ operational patterns.

The number of components in the RBG, which substantially remains stable over time, confirms the tendency towards homogeneity of behaviors and increasing cohesiveness of RBG modal networks.

Finally, degree assortativity values unravel additional patterns on the dynamics of the RBG networks. Degree assortativity assesses the extent to which nodes with high degree are generally more likely to be connected to other nodes with high degree. The coefficient is bounded in the range [-1;1], with -1 meaning that all the connections join nodes of different “status,” 0 being the asymptotic value for a randomly connected graph, and 1 indicating that all connections connect similar nodes. When compared to the other three measurements, degree assortativity does not behave synchronously across modes. Weapons and tactics started with a random level of assortativity, which grew consistently up until 2005 circa, delineating increasingly core-periphery structures. For weapons, assortativity then decreased, and the same happened (with slightly less magnitude) for tactics. Contrarily, the target mode oscillated from year to year until 2011. After 2011, however, the mode witnessed an upward trend, reaching its peak in 2016. The asynchronous mode-level patterns of assortative mixing highlight the need to capitalize all the three dimensions of each actor’s behaviors to fully grasp the complexity of terrorists’ operating profiles.

\begin{figure}[!hbt]
\centering
\includegraphics[scale=0.35]{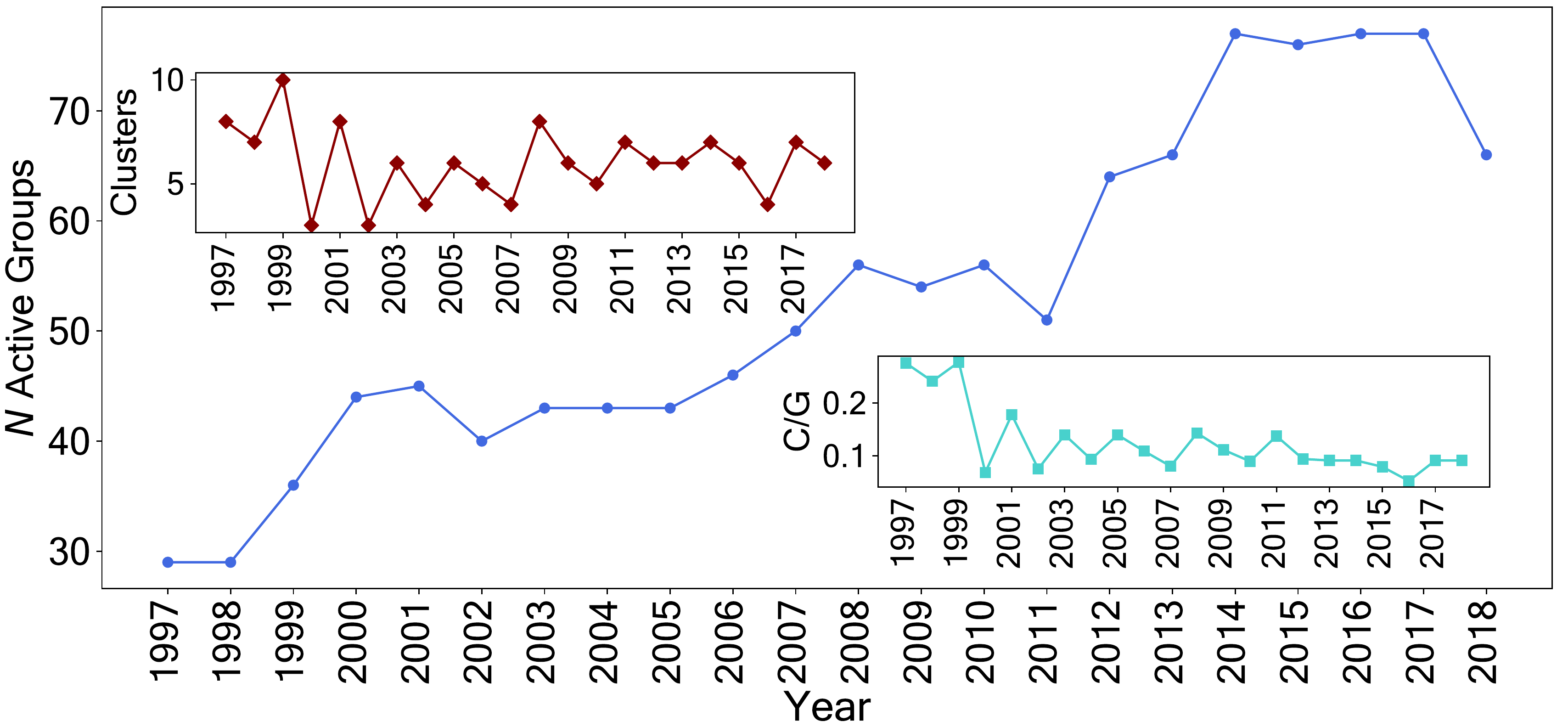}
\caption{\textit{Trend of detected clusters over time (top-left panel), number of active terrorist organizations in each year (main panel), and cluster to groups ratio (bottom-right panel). While the number of active groups has been significantly increasing from 1997 to 2018, the number of clusters remains almost stable, leading to a significantly decreasing trend in the cluster to groups ratio. This illustrates an overall reduction in the variety of operational behaviors and a tendency over homogeneity.}}
\label{fig:cluster_ot}
\end{figure}

The temporal trends of the yearly number of detected clusters, active groups, and the derived ratio of clusters to groups are instead displayed in Figure \ref{fig:cluster_ot}. By only paying attention to the trend in the number of detected clusters, oscillations emerge but no clear pattern can be disentangled. However, when this trend is compared with the one reporting the number of active terrorist organizations in each year, and particularly when the ratio of these two quantities is taken into account, a significant downward trend can be appreciated.

This pinpoints that over the course of the period 1997–2018, terrorist groups have reduced their heterogeneity in terms of deployed tactics, selected targets, and utilized weapons. Despite a clear increase in the number of organizations, the number of clusters remains very similar, thus pointing in the direction of the presence of a sizeable amount of terrorist organizations being very close to each other operationally. Additionally, we do not report any particular trend in the number of outliers, i.e., extremely prolific groups with unique operational profiles that are isolates as they are not clustered with any other group (see Supplementary Information Fig. \ref{fig:out}), ruling out the possibility that the detected homogeneity hides the presence of a tiny but increasing minority of anomalous organizations. In light of this, we conclude that, besides the constant presence of a varying amount of outliers, a critical mass of organizations, representing the vast majority of active groups in each year, became more and more cohesive and thus similar from an operational point of view.

\begin{figure*}[t]
\centering
\includegraphics[scale=0.4]{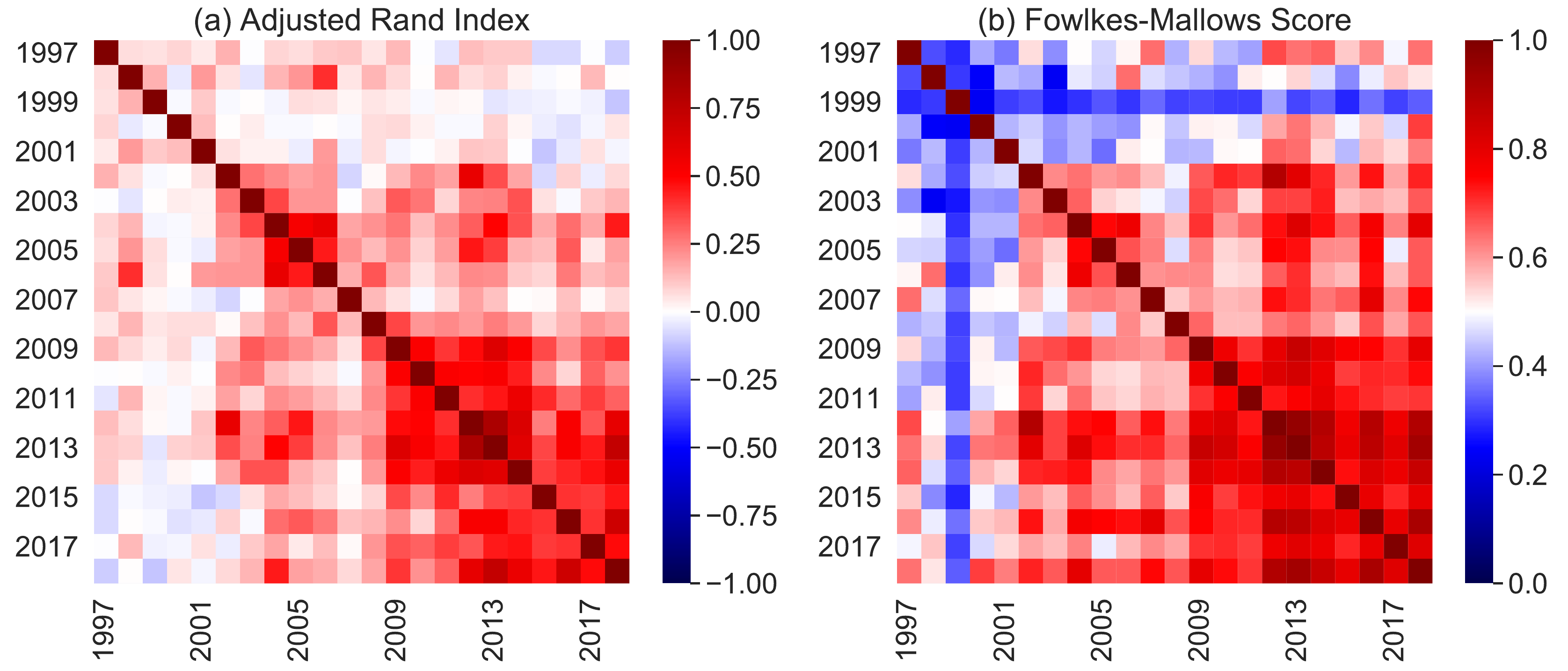}
\caption{\textit{ARI (left) and FMS (right) showing the degree of organizations' co-clustering stability over the years. Both metrics express a certain degree of stability from 2002 to 2018, with two separate sub-regions where stability is stronger, namely 2002-2006 and 2009-2018. Conversely, before 2002, terrorist organizations were more inclined to change their operations substantially from one year to another.}}
\label{fig:sim}
\end{figure*}

\subsection*{Stability of similarity}
Consistent outcomes emerge concerning co-clustering stability, despite the different ways in which the two metrics are computed (Figure \ref{fig:sim}). Across both measures, we can identify a temporal region of high stability from 2009 to 2018. Similarly, noticeable stability is identified also in the years 2002–2006. Terrorist organizations operating in these two temporal frames tended to be clustered with the same groups year after year, suggesting a limited degree of variations in behaviors and, consequently, a certain level of consistency in operations over time.

The picture is sensibly different in relation to co-clustering of terrorist organizations before 2002. From 1997 to 2001, the probability that two groups that are clustered together in a given year will also be co-clustered in the following years is low, pointing in the direction of a significant amount of groups varying their operational patterns and decision-making processes.

These outcomes should be read in conjunction with the ones presented in the previous subsection \ref{section: trends}, where we showcased a consistent downward trend in terrorists' operational diversity marking more recent years. Results are corroborated by the robustness check with the enlarged sample (see Supplementary Information Fig. \ref{fig:ot_rob}).

Additionally, it is worth noting that stability is not automatically higher between consecutive years. Intuitively, co-clustering should increase when considering temporally closer years. At the same time—and trivially—stability in co-clustering is expected to decrease as the distance between two years increases. Instead, there are cases in which clusterings are much more similar comparing years that are more distant in time. For instance, both in the ARI and FMS scenarios, taking as reference year 2018 we document higher clustering similarity with year 2016 (ARI = 0.688, FMS = 0.916), rather than year 2017 (ARI = 0.479, FMS = 0.809). The same occurs for years 2013 and 2018 (ARI = 0.732, FMS = 0.920), compared to years 2017 and 2018. This signifies that organizations not only change their operational dynamics over time: sometimes, some groups temporarily alter their profiles and even return to resemble behaviors that already characterized their operations in the past.

\begin{figure}[!hbt]
\centering
\includegraphics[scale=0.5]{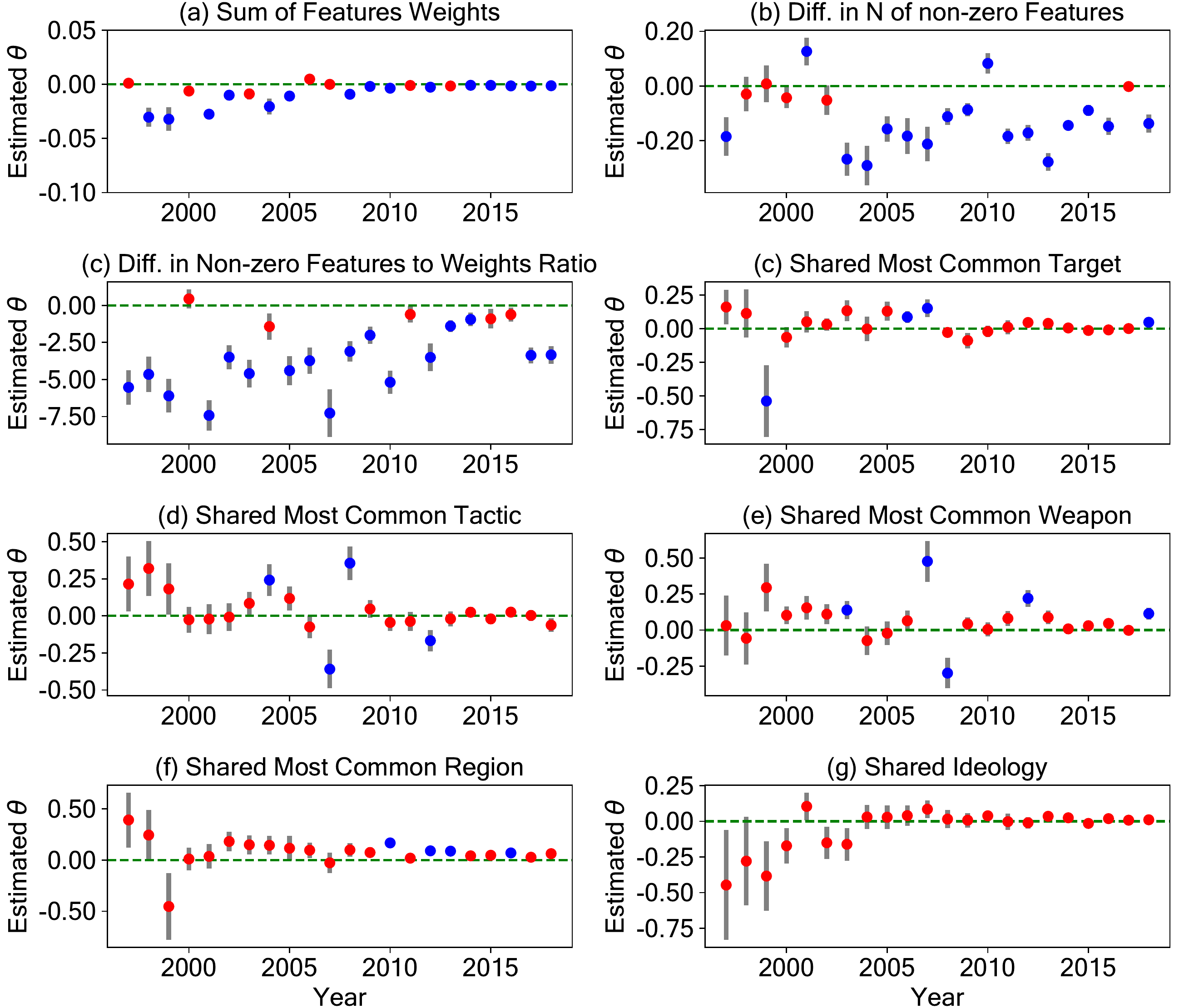}
\caption{\textit{Estimated coefficients for each covariate included in the analysis of drivers of co-clustering, with standard error bars. Blue dots are statistically significant at the 95 percent level (red are not), the dashed green horizontal lines at the 0 level aim at facilitating the interpretation of the coefficients in each subplot. (a) “Sum of Features Weights,” capturing two groups’ overall activity, is a consistently significant predictor in determining the probability of being clustered together: the negative estimated $\theta$ uggest the higher the sum of features weights between two groups, the lower the probability they will be clustered together. (b) The difference in non-zero features, defined as a proxy for repertoire diversity, and particularly the absolute difference between two organizations is the other consistent driver of operational similarity. Results indicate that the more two groups are different in terms of repertoire diversity, the lower the chances of being clustered together. The difference in non-zero features to weights ratio (c) also explains co-clustering: the lower the difference in this joint measure of activity and diversity, the higher the likelihood of co-clustering.}}
\label{fig:ergm}
\end{figure}

\subsection*{Drivers of similarity}
Three factors appear to be the main drivers of operational similarity in the form of co-clustering, as reported by Figure \ref{fig:ergm} and Table \ref{tab:tab-res} (in-depth quantitative description explanation of the ERGM results is provided in the Supplementary Information section \ref{explanation}). 

First, as shown by subfigure (a), ``Sum of Features Weights'' significantly impacts the probability of co-clustering (avg. $\theta_{1997-2018}=-0.007$, significant at the 95\% level in $68.18\%$ of the years). The lower the summed amount of activity and resources characterizing two groups, the higher the likelihood that they will be connected to the same cluster. Conversely, the higher the sum, the lower the probability of co-clustering. High-resource groups tend to be isolated as they possess unique profiles, and groups with low resources are not clustered together with high-resource organizations as their resource mismatch is too large to allow for proper similarity detection.

Second, subfigure (b) highlights that similarity in the yearly operational repertoire correlates with co-clustering (avg. $\theta_{1997-2018}=-0.116$, significant at the 95\% level in $77.27\%$ of the years). In fact, the lower the difference in terms of non-zero features between two organizations, the higher the likelihood of them being operationally similar. Hence, organizations with highly diverse yearly repertoires will be more likely to be in the same clusters with other terrorist groups characterized by the same behavioral heterogeneity. Trivially, the same applies to pairs of organizations with highly uniform yearly repertoires.

Third, also the difference in non-zero features to weights ratio strongly influences the odds of co-clustering for two terrorist organizations. The combination of both features weights—as a proxy of overall activity—and non-zero features—as a measure of heterogeneity—captures operational patterns of similarity among organizations (avg. $\theta_{1997-2018}=-3.332$, significant at the 95\% level in $77.27\%$ of the years).

On the other hand, the three other covariates mapping homophily in operational preferences (i.e., most common target, most common tactic, and most common weapon) do not exhibit correlations with co-clustering, with the exceptions of few years. Coefficients are mostly non-significant (in 18.18 percent of the total number of years for tactics and targets, and 22.72 percent for weapons), and their direction is not easily identifiable, especially for shared most common tactics (two significant positive coefficients and two negative ones). The lack of consistent findings for these three variables investigating operational homophily suggests that similarity between two groups is not only a matter of raw preference over a particular target, tactic, or weapon. Instead, it involves the complex spectrum of multiple decision-making layers, captured by our models through the interdependencies between overall activity and the heterogeneity in each organization’s operational portfolio.

\begin{center}
\begin{table}[t]
\setlength{\tabcolsep}{1.8pt} 
\renewcommand{\arraystretch}{1}
\centering
\footnotesize
\begin{tabular}{llccc}
\hline
\textbf{Covariate} & \textbf{Operational Definition} & \textbf{\begin{tabular}[c]{@{}c@{}}Avg  $\theta$\\ (1997-2018)\end{tabular}} & \textbf{\begin{tabular}[c]{@{}c@{}}St. Dev.   $\theta$\\(1997-2018)\end{tabular}} & \textbf{\begin{tabular}[c]{@{}c@{}}\% Sig. \\ 95\% level\\ \end{tabular}} \\ \hline
Sum of  Features Weights & \begin{tabular}[c]{@{}l@{}}Sum of two groups' overall\\ level of activity\end{tabular} & -0.007 & 0.0105 & 68.18 \\ \hline
Difference in  N of non-zero Features & \begin{tabular}[c]{@{}l@{}}Abs. difference in two groups' \\ repertoire  diversity\end{tabular} & -0.116 & 0.1103 & 77.27 \\ \hline
Difference in Non-zero Features to Weights Ratio & \begin{tabular}[c]{@{}l@{}}Abs. difference in two groups' \\ combined overall activity and\\ repertoire diversity\end{tabular} & -3.332 & 2.2101 & 77.27 \\ \hline
Shared Most  Common Tactic & \begin{tabular}[c]{@{}l@{}}Two groups share same \\ preference in terms of \\ deployed tactics\end{tabular} & 0.0351 & 0.1406 & 18.18 \\ \hline
Shared Most  Common Target & \begin{tabular}[c]{@{}l@{}}Two groups share same\\ preference in terms of\\ attacked targets\end{tabular} & 0.0117 & 0.1594 & 18.18 \\ \hline
Shared Most  Common Weapon & \begin{tabular}[c]{@{}l@{}}Two groups share same\\ preference in terms of\\ utilized weapons\end{tabular} & 0.0707 & 0.1468 & 22.72 \\ \hline
Shared Most Common Region & \begin{tabular}[c]{@{}l@{}}Two groups have homophily in \\terms of most attacked region\end{tabular} & 0.1127 & 0.2301 & 13.63 \\ \hline
Shared  Ideology & \begin{tabular}[c]{@{}l@{}}Two groups have homophily \\in terms of ideology\end{tabular} & -0.0535 & 0.1488 & 0.00 \\ \hline
\end{tabular}
\caption{\textit{Operational definition for each covariate included in each yearly ERGM. Average $\theta$, along with their standard deviation and the percentage of significant coefficients (at the 95 percent level) over the 1997–2018 period are also reported. “Difference in N of non-zero Features” and “Difference in Non-zero Features to Weights Ratio” are the coefficients with the highest percentages of significant $\theta$, followed by ``Sum of Featues Weights''. ollowed by “Sum of Featues Weights.” Ideological homophily is never detected as a predictor of co-clustering.}}
\label{tab:tab-res}
\end{table}
\end{center}
Similarly, being active in the same world region does not provide enduring evidence of increased (or decreased) likelihood of co-clustering for two given terrorist groups, as the covariate is found to be significant only in 13.63 percent of the twenty-two years of the analysis. Notably, this statistical outcome posits the absence of geographical homophily mechanisms: there is no statistically significant difference in the probability of co-clustering for two pairs of groups, one being characterized by organizations plotting attacks in two distinct regions and another being characterized by spatially closer organizations.

Finally, even clearer evidence demonstrates that operational similarity is not driven by or correlated with organizations’ ideologies, as the covariate is never found to be statistically significant at the conventional 95 percent level. While the literature has posited that ideology influences terrorists’ decision-making processes \cite{DrakeIdeology1998,AhmedTerroristIdeologiesTarget2018b}, our results argue that two groups acting with the same ideology have a probability of being clustered together which is not significantly different from the probability of two groups characterized by distinct ideologies being in the same cluster. Besides the absence of statistical significance associated with ideology, it is worth noting that the direction of the coefficients seems at least to follow clear stable patterns: while in the first part of the period under scrutiny the (non-significant) coefficients were negative, from 2003 on, ideology estimates became positive, although in the last years they got closer again to zero.

Diagnostic checks performed for each yearly model confirmed the good quality of the estimates (details are reported in the Supplementary Information Tables \ref{diag1997}-\ref{diag2018}).

Also in this case the results regarding the drivers of similarity are empirically corroborated by robustness models presented in the Supplementary Information subsection \ref{robus_simil}.

\section*{Discussion and Conclusions}
Little is known about patterns of similarity among organizations engaging in political violence. To address this research problem, we have here presented the results of a computational framework for detecting clusters of terrorist organizations that are similar in their operations relying on the representational power of multi-modal networks to characterize terrorist behaviors. We considered groups that have plotted at least 50 attacks worldwide in the period 1997–2018 and gathered yearly clusters obtained from multi-modal networks that map terrorist behaviors in terms of their selection of tactics, targets, and weapons.

We highlighted that when coupling the relatively little variability in the number of clusters over the period 1997–2018 with the yearly growing number of active groups, a clear downward trend in the cluster to groups ratio is found. Relatedly, we document an increase in the density of the RBG modal networks from which clusters are derived, along with an increase in clustering coefficient and a reduction in the number of components, indicating decreasing overall heterogeneity in terrorist operations globally and increasing cohesiveness.

Furthermore, we demonstrated that prior to 2002, year-to-year stability of co-clustering was very low, suggesting a relevant level of behavioral and operational variability for terrorist organizations involved. Conversely, stability increases after 2002, reaching high levels after 2009, thus corroborating the intuition that terrorist groups became more operationally consistent in recent years. We also showed how in certain cases there is a higher co-clustering similarity between years that are distant in time compared to consecutive years. This delineates the presence of groups that not only modify their behaviors, but do so temporarily, being able to return to previous operating profiles after years. Such finding helps to shed additional light on the revisited concept of terrorist innovation as the byproduct of a broad set of activity changes \cite{LubranoNavigatingTerroristInnovation2021}. Moreover, it reinforces the need for dynamic behavioral monitoring to track operational trajectories evolving over time.

Finally, we have disentangled drivers of co-clustering, which maps operational similarity, finding that similarity is mainly driven by groups’ yearly amount of activity, homophily in repertoire diversity, and a measure of the two combined. The other tested measures, namely homophily in operational preferences, geographical homophily, and ideological homophily revealed their non-significant role in driving co-clustering. With regards to geospatial dynamics, future research should explore whether finer-grained spatial resolutions provide different insights on geographical micro-patterns of terrorist behaviors. Concerning ideology, while extant work showed that sharing the same ideology drives inter-group alliances \cite{AsalFriendsTheseWhy2016, PhillipsTerroristGroupRivalries2019a}, the models demonstrated that when organizations are primarily represented by their mere operations, ideology fails to hold its driving assortative force, as it does not explain operational similarity. All our results are robust, as models re-estimated on an enlarged sample of terrorist organizations demonstrate.

Our work does not come without limitations. First, although the GTD is recognized as the most comprehensive and rigorous data source for information on terrorist events at the global level, there is a chance that some events may have escaped the data gathering radar at the source. Nonetheless, we believe this chance remains limited in its capacity to impact our results, especially given our explicit sampling strategy that focuses on major terrorist groups whose actions have been constantly under the spotlight in the last two decades, significantly increasing the likelihood of attacks’ documentation.

Second, the impossibility to systematically rely on data reporting events occurred prior to 1997 limits the historical variability of several variables, particularly the one mapping organizations’ ideology. Events from 1997 to 2018 led to a likely over-representation of certain ideologies, such as Islamist/Jihadist and Ethno-nationalist, compared to the larger picture that would have emerged considering attacks perpetrated from 1970 to 1996. Albeit our choice to maintain this specific time window is motivated by the higher reliability of fundamental pieces of information used to process our data, we recognize that a wider time frame would have provided richer results on the long-term historical trends observed in terrorism globally.

Third, our yearly-based apparatus may hide meso- or micro-temporal shifts in operational choices that could provide us with richer insights on the dynamic, evolving behavioral complexity of terrorist behaviors. Previous research has shown that the distribution of terrorist attacks is often clustered in short time frames as a consequence of proactive or reactive decisions \cite{LewisSelfexcitingpointprocess2012c, TenchSpatiotemporalpatternsIED2016a, ClarkModelingestimationselfexciting2018a, ChuangLocalalliancesrivalries2019c}. More recently, scholars  also showed that event features can be useful to learn patterns of behaviors at the micro-temporal level for forecasting purposes \cite{CampedelliLearningfutureterrorist2021a}. In light of these aspects, we plan to expand our work to assess whether more localized temporal trends are present, and how this may modify clustering.

Fourth, our framework cannot currently offer causal explanations behind patterns of operational similarity as we lack a sufficient information infrastructure to investigate etiological mechanisms that could explain what causes operational shifts or behavioral convergences in terrorist behaviors. Particularly, we only rely on endogenous variables to study operational similarity and operational patterns over time, without integrating exogenous events, such as regime changes or military campaigns, that can help us highlight when and why certain groups significantly modify their behaviors. Currently, we provide sets of statistical results that demonstrate the nature and characteristics of evolving operational patterns across terrorist organizations. Future endeavors will seek to frame these findings in a causal research design.

In spite of these limitations, however, our work advances the scientific understanding of the complex dynamics typifying terrorist organizations in two different ways.

We primarily provide insights on a largely overlooked research problem, namely the investigation of operational similarity patterns across terrorist organizations, offering possible new ways to think about terrorist decision-making processes and terrorist innovation and creativity.

Furthermore, our work contributes to research developing computationally-oriented systems for counter-terrorism, an area of inquiry that has gained momentum in the last years \cite{ChuangLocalalliancesrivalries2019c, YangQuantifyingfuturelethality2019b}. With this regard, our approach may be useful to develop dynamic assessment tools for intelligence monitoring. Generally, risk assessment tools aim at protecting locations or targets \cite{GuoRetoolAIforecast2018b} or preventing the radicalization of individuals \cite{Monahanindividualriskassessment2012a}. Yet, adopting a complementary approach to scan behavioral changes in terrorist groups can be highly effective in enriching data-driven support for terrorism prevention. The computational efficiency and the scalability of our framework allow for the adaptation of our approach to larger samples, finer-grained geographical units and temporal windows. This flexibility offers a versatile tool that can facilitate the analysis of groups’ behavioral trajectories, tailoring the context of monitoring based on the needs of practitioners and policy-makers.

The results of this study can inform counter-terrorism practitioners and analysts on both the macro-level trends characterizing terrorism at the global level and the micro-level hidden behavioral mechanics that describe organizations’ strategies and tactics. Insights offered by this study demonstrate that counter-terrorism measures specifically dedicated to organizations’ impairment or resource targeting should keep in mind the adaptive character of these actors. Static accounts fall short in capturing the complexity of inherently dynamic entities and behaviors. Additionally, the outcomes of our work fundamentally reappraise the role of ideology as a discriminant in specific operational patterns. This finding may help reconsider the often overlooked similarities existing among organizations engaging in political violence for very different reasons, motives, objectives.

As political violence evolves and new terrorist actors emerge in the global scenario, detecting operational affinity among different organizations can offer critical insights for the design and deployment of counter-terrorism policies, anticipating possible trends and providing insights on which organizations should be prioritized in the fight against terrorist violence.

\section*{Code Availability}
Source code and data are accessible at \href{https://github.com/ijcruic/Multi-Modal-Networks-Reveal-Patterns-of-Operational-Similarity-of-Terrorist-Organizations}{https://github.com/ijcruic/Multi-Modal-Networks-Reveal-Patterns-of-Operational-Similarity-of-Terrorist-Organizations}. Furthermore, the MVMC method is available as a tool within the ORA-Pro software \cite{CarleyORAToolkitDynamic2017}.
\section*{Acknowledgments}
We wish to thank Victor Asal, Bruce Desmarais, Maria Rita D'Orsogna, the participants to the Trento Center for Social Research Methods seminar series, and the participants to the 2021 APSA Political Networks Conference for their insightful comments on this manuscript. We are also grateful to the two anonymous reviewers for their comments.
This work was supported by the Department of Excellence initiative of the Italian Ministry of University and Research and in part by the Knight Foundation and the Office of Naval Research Grants N000141812106 and
N000141812108. The views and conclusions
contained in this document are those of the authors and should not be interpreted as representing the official policies,
either expressed or implied, of the Knight Foundation, Office of Naval Research or the U.S. government. 
\section*{Disclosure Statement}
No potential competing interest was reported by the authors.
\bibliographystyle{apalike}
\bibliography{PNAS_Ter.bib}

\begin{thebibliography}{}

\bibitem[Ahmed, 2018]{AhmedTerroristIdeologiesTarget2018b}
Ahmed, R. (2018).
\newblock Terrorist {Ideologies} and {Target} {Selection}.
\newblock {\em Journal of Applied Security Research}, 13(3):376--390.
\newblock Publisher: Routledge \_eprint:
  https://doi.org/10.1080/19361610.2018.1463140.

\bibitem[Asal and Rethemeyer, 2006]{AsalResearchingTerroristNetworks2006}
Asal, V. and Rethemeyer, R.~K. (2006).
\newblock Researching {Terrorist} {Networks}.
\newblock {\em Journal of Security Education}, 1(4):65--74.
\newblock Publisher: Routledge \_eprint:
  https://doi.org/10.1300/J460v01n04\_06.

\bibitem[Asal et~al., 2011]{AsalBigAlliedDangerous2011}
Asal, V., Rethemeyer, R.~K., and Anderson, I. (2011).
\newblock Big {Allied} and {Dangerous} ({BAAD}) {Database} 1 - {Lethality}
  {Data}, 1998-2005.
\newblock Publisher: Harvard Dataverse type: dataset.

\bibitem[Asal et~al., 2016]{AsalFriendsTheseWhy2016}
Asal, V.~H., Park, H.~H., Rethemeyer, R.~K., and Ackerman, G. (2016).
\newblock With {Friends} {Like} {These} … {Why} {Terrorist} {Organizations}
  {Ally}.
\newblock {\em International Public Management Journal}, 19(1):1--30.
\newblock Publisher: Routledge \_eprint:
  https://doi.org/10.1080/10967494.2015.1027431.

\bibitem[Asal et~al., 2009]{AsalSoftestTargetsStudy2009c}
Asal, V.~H., Rethemeyer, K.~R., Anderson, I., Stein, A., Rizzo, J., and Rozea,
  M. (2009).
\newblock The {Softest} of {Targets}: {A} {Study} on {Terrorist} {Target}
  {Selection}.
\newblock {\em Journal of Applied Security Research}, 4(3):258--278.
\newblock Publisher: Routledge \_eprint:
  https://doi.org/10.1080/19361610902929990.

\bibitem[Bouchard, 2017]{BouchardSocialNetworksTerrorism2017}
Bouchard, M., editor (2017).
\newblock {\em Social {Networks}, {Terrorism} and {Counter}-terrorism}.
\newblock Routledge, London, 1 edition edition.

\bibitem[Campedelli et~al., 2021]{CampedelliLearningfutureterrorist2021a}
Campedelli, G.~M., Bartulovic, M., and Carley, K.~M. (2021).
\newblock Learning future terrorist targets through temporal meta-graphs.
\newblock {\em Scientific Reports}, 11(1):8533.

\bibitem[Campedelli et~al., 2019]{Campedellicomplexnetworksapproach2019}
Campedelli, G.~M., Cruickshank, I., and Carley, K.~M. (2019).
\newblock A complex networks approach to find latent clusters of terrorist
  groups.
\newblock {\em Applied Network Science}, 4(1):1--22.

\bibitem[Carley, 2017]{CarleyORAToolkitDynamic2017}
Carley, K.~M. (2017).
\newblock {ORA}: {A} {Toolkit} for {Dynamic} {Network} {Analysis} and
  {Visualization}.
\newblock In Alhajj, R. and Rokne, J., editors, {\em Encyclopedia of {Social}
  {Network} {Analysis} and {Mining}}, pages 1--10. Springer New York, New York,
  NY.

\bibitem[Chuang et~al., 2019]{ChuangLocalalliancesrivalries2019c}
Chuang, Y.-L., Ben-Asher, N., and D’Orsogna, M.~R. (2019).
\newblock Local alliances and rivalries shape near-repeat terror activity of
  al-{Qaeda}, {ISIS}, and insurgents.
\newblock {\em Proceedings of the National Academy of Sciences},
  116(42):20898--20903.

\bibitem[Clark and Dixon, 2018]{ClarkModelingestimationselfexciting2018a}
Clark, N.~J. and Dixon, P.~M. (2018).
\newblock Modeling and estimation for self-exciting spatio-temporal models of
  terrorist activity.
\newblock {\em Annals of Applied Statistics}, 12(1):633--653.
\newblock Publisher: Institute of Mathematical Statistics.

\bibitem[Clauset and Gleditsch,
  2012]{ClausetDevelopmentalDynamicsTerrorist2012b}
Clauset, A. and Gleditsch, K.~S. (2012).
\newblock The {Developmental} {Dynamics} of {Terrorist} {Organizations}.
\newblock {\em PLoS ONE}, 7(11):e48633.

\bibitem[Clutterbuck, 1993]{ClutterbuckTrendsterroristweaponry1993}
Clutterbuck, R. (1993).
\newblock Trends in terrorist weaponry.
\newblock {\em Terrorism and Political Violence}, 5(2):130--139.
\newblock Publisher: Routledge \_eprint:
  https://doi.org/10.1080/09546559308427213.

\bibitem[Cranmer and Desmarais, 2011]{CranmerInferentialNetworkAnalysis2011b}
Cranmer, S.~J. and Desmarais, B.~A. (2011).
\newblock Inferential {Network} {Analysis} with {Exponential} {Random} {Graph}
  {Models}.
\newblock {\em Political Analysis}, 19(1):66--86.
\newblock Publisher: [Oxford University Press, Society for Political
  Methodology].

\bibitem[Cranmer et~al., 2012]{CranmerComplexDependenciesAlliance2012}
Cranmer, S.~J., Desmarais, B.~A., and Menninga, E.~J. (2012).
\newblock Complex {Dependencies} in the {Alliance} {Network}.
\newblock {\em Conflict Management and Peace Science}, 29(3):279--313.

\bibitem[Crenshaw, 2001]{CrenshawInnovationDecisionPoints2001}
Crenshaw, M. (2001).
\newblock Innovation: {Decision} {Points} in the {Trajectory} of {Terrorism}.
\newblock Harvard University.

\bibitem[Cruickshank, 2020]{CruickshankMultiviewClusteringSocialbased2020}
Cruickshank, I.~J. (2020).
\newblock {\em Multi-view {Clustering} of {Social}-based {Data}}.
\newblock Ph.{D}. {Dissertation}, Institute for Software Research, School of
  Computer Science - Carnegie Mellon University.

\bibitem[Cruickshank and Carley,
  2020]{cruickshank2020multiviewclusteringhashtags}
Cruickshank, I.~J. and Carley, K.~M. (2020).
\newblock Characterizing communities of hashtag usage on twitter during the
  2020 covid-19 pandemic by multi-view clustering.
\newblock {\em Applied Network Science}, 5(1):1--40.

\bibitem[Desmarais and Cranmer,
  2012]{DesmaraisStatisticalmechanicsnetworks2012}
Desmarais, B.~A. and Cranmer, S.~J. (2012).
\newblock Statistical mechanics of networks: {Estimation} and uncertainty.
\newblock {\em Physica A: Statistical Mechanics and its Applications},
  391(4):1865--1876.
\newblock Publisher: Elsevier.

\bibitem[Desmarais and Cranmer,
  2013]{DesmaraisForecastinglocationaldynamics2013}
Desmarais, B.~A. and Cranmer, S.~J. (2013).
\newblock Forecasting the locational dynamics of transnational terrorism: a
  network analytic approach.
\newblock {\em Security Informatics}, 2(1):8.

\bibitem[Dolnik, 2007]{DolnikUnderstandingTerroristInnovation2007b}
Dolnik, A. (2007).
\newblock {\em Understanding {Terrorist} {Innovation}: {Technology}, {Tactics}
  and {Global} {Trends}}.
\newblock Routledge.

\bibitem[Drake, 1998]{DrakeIdeology1998}
Drake, C. J.~M. (1998).
\newblock Ideology.
\newblock In Drake, C. J.~M., editor, {\em Terrorists’ {Target} {Selection}},
  pages 16--34. Palgrave Macmillan UK, London.

\bibitem[Duxbury and Haynie, 2018]{DuxburyNetworkStructureOpioid2018a}
Duxbury, S.~W. and Haynie, D.~L. (2018).
\newblock The {Network} {Structure} of {Opioid} {Distribution} on a {Darknet}
  {Cryptomarket}.
\newblock {\em Journal of Quantitative Criminology}, 34(4):921--941.

\bibitem[Fern and Brodley, 2004]{Fern2004BGPA}
Fern, X.~Z. and Brodley, C.~E. (2004).
\newblock Solving cluster ensemble problems by bipartite graph partitioning.
\newblock In {\em Proceedings of the Twenty-first International Conference on
  Machine Learning}, ICML '04, pages 36--, New York, NY, USA. ACM.

\bibitem[Fienberg and Wasserman, 1981]{FienbergCategoricalDataAnalysis1981a}
Fienberg, S.~E. and Wasserman, S.~S. (1981).
\newblock Categorical {Data} {Analysis} of {Single} {Sociometric} {Relations}.
\newblock {\em Sociological Methodology}, 12:156--192.
\newblock Publisher: [American Sociological Association, Wiley, Sage
  Publications, Inc.].

\bibitem[{Fortunato} and {Barthelemy}, 2007]{Fortunato2007resolution}
{Fortunato}, S. and {Barthelemy}, M. (2007).
\newblock {Resolution limit in community detection}.
\newblock {\em Proceedings of the National Academy of Science}, 104(1):36--41.

\bibitem[Guo et~al., 2018]{GuoRetoolAIforecast2018b}
Guo, W., Gleditsch, K., and Wilson, A. (2018).
\newblock Retool {AI} to forecast and limit wars.
\newblock {\em Nature}, 562(7727):331--333.
\newblock Number: 7727 Publisher: Nature Publishing Group.

\bibitem[Hanneke et~al., 2010]{HannekeDiscretetemporalmodels2010a}
Hanneke, S., Fu, W., and Xing, E.~P. (2010).
\newblock Discrete temporal models of social networks.
\newblock {\em Electronic Journal of Statistics}, 4(none).

\bibitem[Hoffman and McCormick, 2004]{HoffmanTerrorismSignalingSuicide2004}
Hoffman, B. and McCormick, G.~H. (2004).
\newblock Terrorism, {Signaling}, and {Suicide} {Attack}.
\newblock {\em Studies in Conflict \& Terrorism}, 27(4):243--281.
\newblock Publisher: Routledge \_eprint:
  https://doi.org/10.1080/10576100490466498.

\bibitem[Holland and Leinhardt, 1981]{HollandExponentialFamilyProbability1981}
Holland, P.~W. and Leinhardt, S. (1981).
\newblock An {Exponential} {Family} of {Probability} {Distributions} for
  {Directed} {Graphs}.
\newblock {\em Journal of the American Statistical Association},
  76(373):33--50.
\newblock Publisher: [American Statistical Association, Taylor \& Francis,
  Ltd.].

\bibitem[Hou et~al., 2020]{HouIntroducingExtendedData2020a}
Hou, D., Gaibulloev, K., and Sandler, T. (2020).
\newblock Introducing {Extended} {Data} on {Terrorist} {Groups} ({EDTG}), 1970
  to 2016.
\newblock {\em Journal of Conflict Resolution}, 64(1):199--225.

\bibitem[Hunter et~al., 2008]{ergmpackage}
Hunter, D.~R., Handcock, M.~S., Butts, C.~T., Goodreau, S.~M., and Morris, M.
  (2008).
\newblock ergm: A package to fit, simulate and diagnose exponential-family
  models for networks.
\newblock {\em Journal of Statistical Software}, 24(3):1--29.

\bibitem[Jackson, 2001]{JacksonTechnologyAcquisitionTerrorist2001a}
Jackson, B.~A. (2001).
\newblock Technology {Acquisition} by {Terrorist} {Groups}: {Threat}
  {Assessment} {Informed} by {Lessons} from {Private} {Sector} {Technology}
  {Adoption}.
\newblock {\em Studies in Conflict \& Terrorism}, 24(3):183--213.

\bibitem[Koehler-Derrick and Milton,
  2019]{Koehler-DerrickChooseYourWeapon2019a}
Koehler-Derrick, G. and Milton, D.~J. (2019).
\newblock Choose {Your} {Weapon}: {The} {Impact} of {Strategic}
  {Considerations} and {Resource} {Constraints} on {Terrorist} {Group} {Weapon}
  {Selection}.
\newblock {\em Terrorism and Political Violence}, 31(5):909--928.

\bibitem[Krebs, 2002]{KrebsUncloakingTerroristNetworks2002a}
Krebs, V. (2002).
\newblock Uncloaking {Terrorist} {Networks}.
\newblock {\em First Monday}.

\bibitem[Krivitsky and Handcock, 2014]{KrivitskySeparableModelDynamic2014a}
Krivitsky, P.~N. and Handcock, M.~S. (2014).
\newblock A {Separable} {Model} for {Dynamic} {Networks}.
\newblock {\em Journal of the Royal Statistical Society. Series B, Statistical
  Methodology}, 76(1):29--46.

\bibitem[LaFree and Dugan, 2007]{LaFreeIntroducingGlobalTerrorism2007}
LaFree, G. and Dugan, L. (2007).
\newblock Introducing the {Global} {Terrorism} {Database}.
\newblock {\em Terrorism and Political Violence}, 19(2):181--204.

\bibitem[Lewis et~al., 2012]{LewisSelfexcitingpointprocess2012c}
Lewis, E., Mohler, G., Brantingham, P.~J., and Bertozzi, A.~L. (2012).
\newblock Self-exciting point process models of civilian deaths in {Iraq}.
\newblock {\em Security Journal}, 25(3):244--264.

\bibitem[Lubrano, 2021]{LubranoNavigatingTerroristInnovation2021}
Lubrano, M. (2021).
\newblock Navigating {Terrorist} {Innovation}: {A} {Proposal} for a
  {Conceptual} {Framework} on {How} {Terrorists} {Innovate}.
\newblock {\em Terrorism and Political Violence}, 0(0):1--16.
\newblock Publisher: Routledge \_eprint:
  https://doi.org/10.1080/09546553.2021.1903440.

\bibitem[Maier et~al., 2009]{MaierOptimalconstructionknearestneighbor2009}
Maier, M., Hein, M., and von Luxburg, U. (2009).
\newblock Optimal construction of k-nearest-neighbor graphs for identifying
  noisy clusters.
\newblock {\em Theoretical Computer Science}, 410(19):1749--1764.

\bibitem[McCormick, 2003]{McCormickTerroristDecisionMaking2003}
McCormick, G.~H. (2003).
\newblock Terrorist {Decision} {Making}.
\newblock {\em Annual Review of Political Science}, 6(1):473--507.

\bibitem[Medina, 2014]{MedinaSocialNetworkAnalysis2014a}
Medina, R.~M. (2014).
\newblock Social {Network} {Analysis}: {A} case study of the {Islamist}
  terrorist network.
\newblock {\em Security Journal}, 27(1):97--121.

\bibitem[Merari, 1999]{MerariTerrorismstrategystruggle1999a}
Merari, A. (1999).
\newblock Terrorism as a strategy of struggle: {Past} and future.
\newblock {\em Terrorism and Political Violence}, 11(4):52--65.
\newblock Publisher: Routledge \_eprint:
  https://doi.org/10.1080/09546559908427531.

\bibitem[Monahan, 2012]{Monahanindividualriskassessment2012a}
Monahan, J. (2012).
\newblock The individual risk assessment of terrorism.
\newblock {\em Psychology, Public Policy, and Law}, 18(2):167--205.

\bibitem[Perliger and Pedahzur, 2011]{PerligerSocialNetworkAnalysis2011b}
Perliger, A. and Pedahzur, A. (2011).
\newblock Social {Network} {Analysis} in the {Study} of {Terrorism} and
  {Political} {Violence}.
\newblock {\em PS: Political Science and Politics}, 44(1):45--50.
\newblock Publisher: [American Political Science Association, Cambridge
  University Press].

\bibitem[Phillips, 2019]{PhillipsTerroristGroupRivalries2019a}
Phillips, B.~J. (2019).
\newblock Terrorist {Group} {Rivalries} and {Alliances}: {Testing} {Competing}
  {Explanations}.
\newblock {\em Studies in Conflict \& Terrorism}, 42(11):997--1019.
\newblock Publisher: Routledge \_eprint:
  https://doi.org/10.1080/1057610X.2018.1431365.

\bibitem[Polo and Gleditsch, 2016]{PoloTwistingarmssending2016e}
Polo, S.~M. and Gleditsch, K.~S. (2016).
\newblock Twisting arms and sending messages: {Terrorist} tactics in civil war.
\newblock {\em Journal of Peace Research}, 53(6):815--829.

\bibitem[Qiao et~al., 2018]{QiaoDatadrivengraphconstruction2018}
Qiao, L., Zhang, L., Chen, S., and Shen, D. (2018).
\newblock Data-driven graph construction and graph learning: {A} review.
\newblock {\em Neurocomputing}, 312:336--351.

\bibitem[Rahmah and Sitanggang, 2016]{RahmahDeterminationOptimalEpsilon2016}
Rahmah, N. and Sitanggang, I.~S. (2016).
\newblock Determination of {Optimal} {Epsilon} ({Eps}) {Value} on {DBSCAN}
  {Algorithm} to {Clustering} {Data} on {Peatland} {Hotspots} in {Sumatra}.
\newblock {\em IOP Conference Series: Earth and Environmental Science},
  31:012012.
\newblock Publisher: IOP Publishing.

\bibitem[Reichardt and Bornholdt, 2006]{Reichardt2006resolutionmodularity}
Reichardt, J. and Bornholdt, S. (2006).
\newblock Statistical mechanics of community detection.
\newblock {\em Phys. Rev. E}, 74:016110.

\bibitem[Santifort et~al., 2013]{SantifortTerroristattacktarget2013c}
Santifort, C., Sandler, T., and Brandt, P.~T. (2013).
\newblock Terrorist attack and target diversity: {Changepoints} and their
  drivers.
\newblock {\em Journal of Peace Research}, 50(1):75--90.

\bibitem[Shapiro, 2012]{ShapiroTerroristDecisionMakingInsights2012a}
Shapiro, J.~N. (2012).
\newblock Terrorist {Decision}-{Making}: {Insights} from {Economics} and
  {Political} {Science}.
\newblock {\em Perspectives on Terrorism}, 6(4/5):5--20.
\newblock Publisher: Terrorism Research Institute.

\bibitem[START, 2017]{STARTGTDCodebookInclusion2017}
START (2017).
\newblock {GTD} {Codebook}: {Inclusion} {Criteria} and {Variables}.
\newblock Technical report, University of Maryland.

\bibitem[Strauss and Ikeda, 1990]{StraussPseudolikelihoodEstimationSocial1990a}
Strauss, D. and Ikeda, M. (1990).
\newblock Pseudolikelihood {Estimation} for {Social} {Networks}.
\newblock {\em Journal of the American Statistical Association},
  85(409):204--212.
\newblock Publisher: Taylor \& Francis \_eprint:
  https://www.tandfonline.com/doi/pdf/10.1080/01621459.1990.10475327.

\bibitem[Strehl and Ghosh, 2003a]{StrehlClusterensemblesknowledge2003a}
Strehl, A. and Ghosh, J. (2003a).
\newblock Cluster ensembles --- a knowledge reuse framework for combining
  multiple partitions.
\newblock {\em The Journal of Machine Learning Research}, 3(null):583--617.

\bibitem[Strehl and Ghosh, 2003b]{Strehl2003clusterensembling}
Strehl, A. and Ghosh, J. (2003b).
\newblock Cluster ensembles --- a knowledge reuse framework for combining
  multiple partitions.
\newblock {\em J. Mach. Learn. Res.}, 3:583--617.

\bibitem[Tench et~al., 2016]{TenchSpatiotemporalpatternsIED2016a}
Tench, S., Fry, H., and Gill, P. (2016).
\newblock Spatio-temporal patterns of {IED} usage by the {Provisional} {Irish}
  {Republican} {Army}.
\newblock {\em European Journal of Applied Mathematics}, 27(3):377--402.

\bibitem[{The Beacham Group}, 2021]{TheBeachamGroupTRACTerrorismResearch2021}
{The Beacham Group} (2021).
\newblock {TRAC} - {Terrorism} {Research} \& {Analysis} {Consortium}
  {Platform}.

\bibitem[Wasserman and Pattison, 1996]{WassermanLogitmodelslogistic1996}
Wasserman, S. and Pattison, P. (1996).
\newblock Logit models and logistic regressions for social networks: {I}. {An}
  introduction to {Markov} graphs and p*.
\newblock {\em Psychometrika}, 61(3):401--425.
\newblock Place: Germany Publisher: Springer.

\bibitem[Yang et~al., 2019]{YangQuantifyingfuturelethality2019b}
Yang, Y., Pah, A.~R., and Uzzi, B. (2019).
\newblock Quantifying the future lethality of terror organizations.
\newblock {\em Proceedings of the National Academy of Sciences},
  116(43):21463--21468.

\end{thebibliography}

\pagebreak

\appendix
\begin{center}
\textbf{\Large Supplementary Information: \\
Multi-Modal Networks Reveal Patterns of \\ Operational Similarity of Terrorist Organizations}
\end{center}
\setcounter{equation}{0}
\setcounter{section}{0}
\setcounter{figure}{0}
\setcounter{table}{0}

\makeatletter
\renewcommand{\theequation}{S\arabic{equation}}
\renewcommand{\thefigure}{S\arabic{figure}}
\renewcommand{\thetable}{S\arabic{table}}

\section{Supplementary Results}

\subsection{RBG Network Dynamics}
\begin{figure}[!hbt]
    \centering
    \includegraphics[scale=0.55]{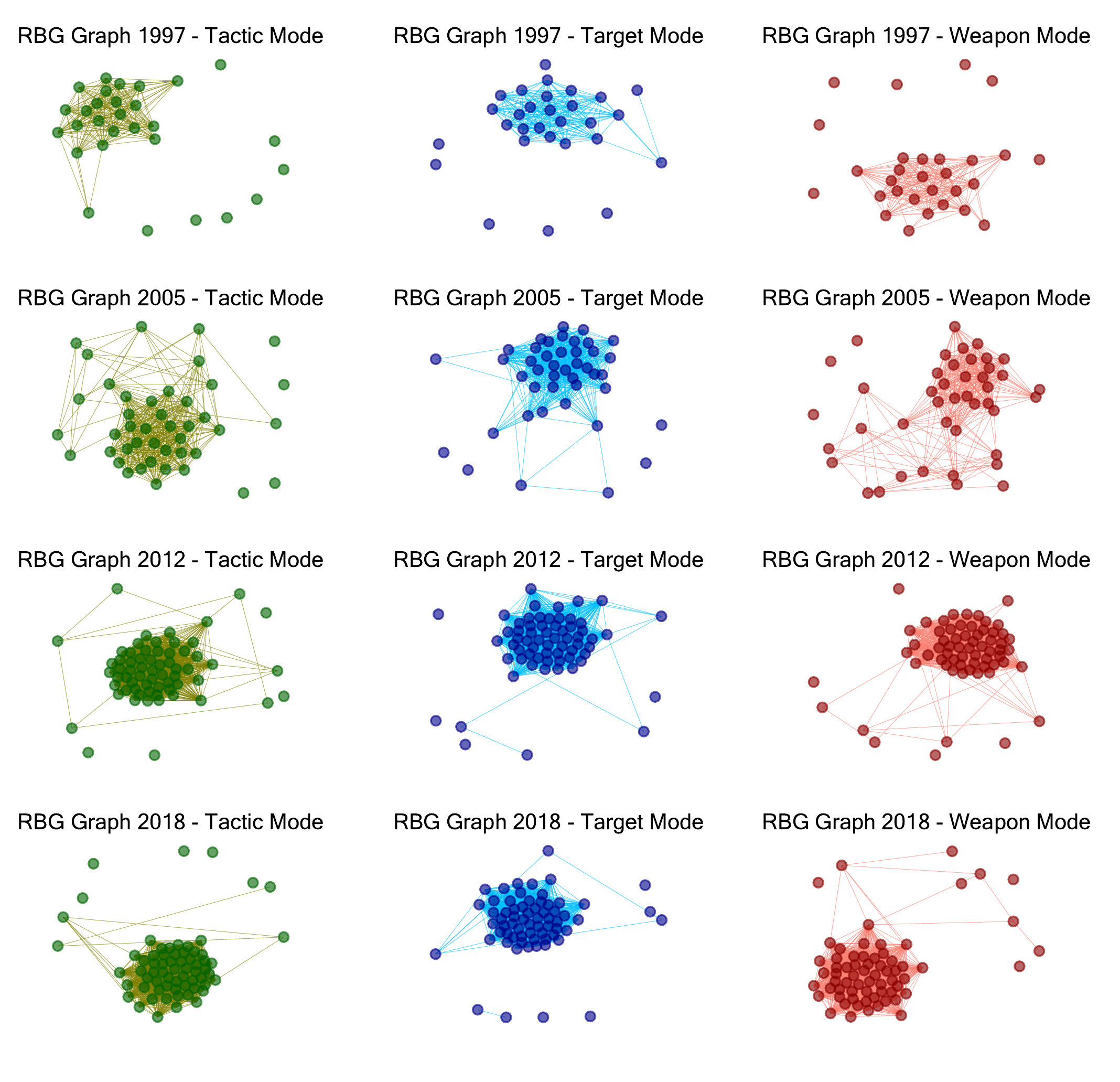}
    \caption{Visualization of three modal RBG networks in four sample years (1997, 2005, 2012, 2018)}
    \label{fig:rbg_net_dynamics}
\end{figure}

\begin{figure}[!hbt]
    \centering
    \includegraphics[scale=0.6]{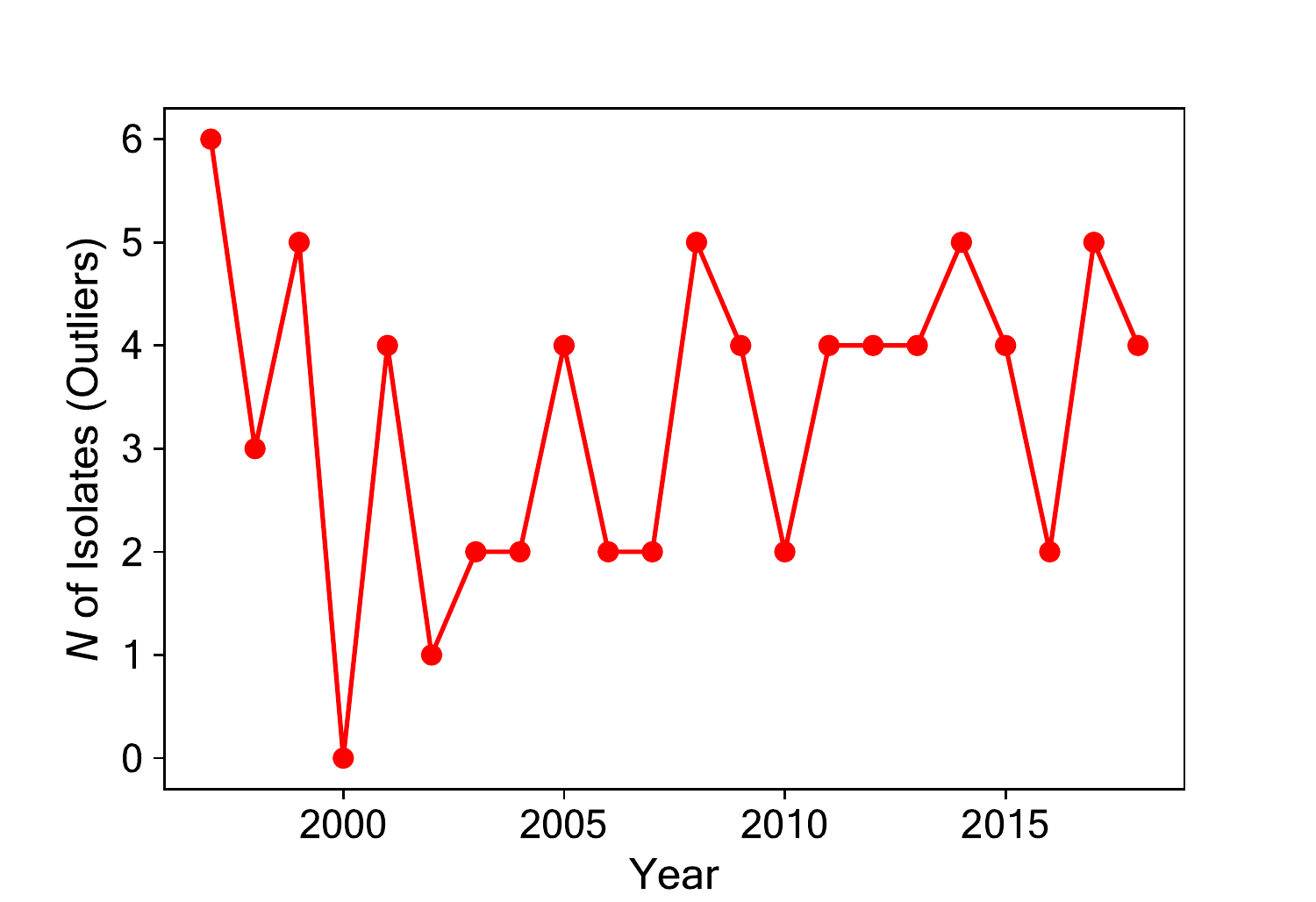}
    \caption{Number of outliers (i.e., isolates, namely terrorist organizations that are not clustered with any other group) in each year}
    \label{fig:out}
\end{figure}

A set of sample visualizations for RBG networks per each of the data modes is provided in Figure \ref{fig:rbg_net_dynamics}. It is straightforward to note how, in each mode and in each year, outliers are present. Outliers are groups that due to their peculiar operational patterns do not share commonalities with other organizations. It might be that, in a given year, one group is an outlier in one specific mode but not in another. The number of outliers, however, does not exhibit a particular increasing or decreasing trend over the years, as displayed by Figure \ref{fig:out}. Years 2000 is the only year in which we do not find isolates, with the highest amount of outliers being the first year of our analysis.  Besides outliers and beyond case-wise differences, all networks share a common pattern: most of the nodes are part of a critical mass of organizations sharing similar operational patterns.

\subsection{Cluster Performance}

We also looked at patterns in the performance of the MVMC clusters to get a sense of how well the method was performing and aspects of the data. In particular, we recorded the average view resolutions, $\gamma^v$, and average view weights, $w^v$, for each year and view of the data. In general, larger denser graphs typically require a higher resolution parameter, or more weight placed on the null model in the modularity function, in order to discover clusters in the data \cite{Fortunato2007resolution}. And, the weight parameter correlates to those views which have more impact on the clustering results; having a higher relative view weight means that that view is more important to the found clusters. The following figure, Figure \ref{fig:performance}, displays these values for the whole data set.

\begin{figure}[!hbt]
    \centering
    \includegraphics[scale=0.5]{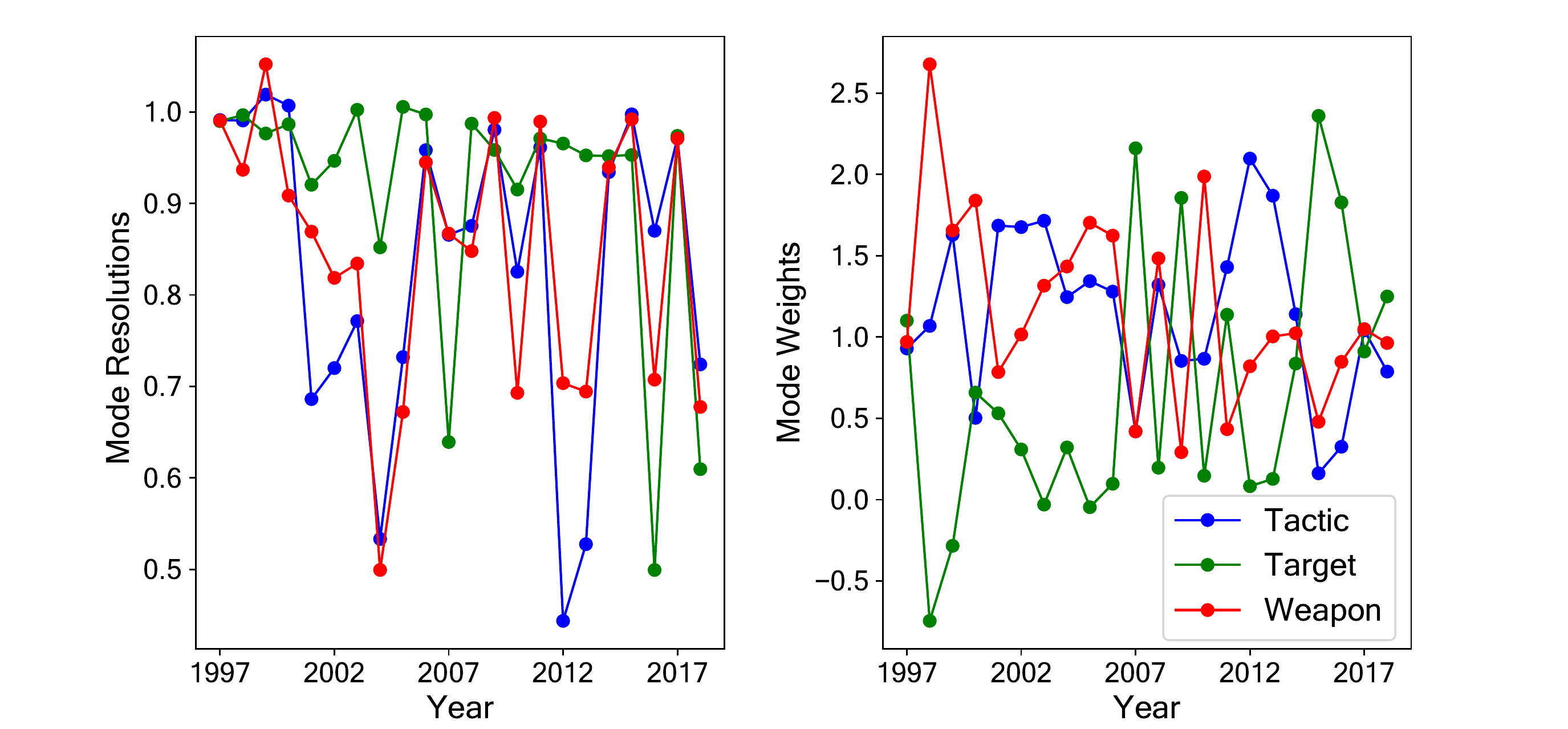}
    \caption{Clustering procedure resolution and weights for each mode in each year}
    \label{fig:performance}
\end{figure}

The view resolutions generally remain around 1, which is the standard resolution parameter from base modularity (i.e. a balance between the actual edges and the configuration null model in the modularity function). We do observe, however, the resolution parameter can often be less than one in our data set, which can occur when there are many clusters relative to the size of the graphs. For the view weights, we generally observe that Tactic and Weapon have higher view weights and more similar view weights than Target, however, especially from 2007 onward, Target was occasionally the highest weighted view. We expected tactic and weapon to have a similar impact on the clustering structure, since these two modes can correlate (i.e. certain tactics, like a bombing, require certain, limited types of weapons). So, from the mere point of view of the data contained in the modes,  clustering in any given year is influenced by shifts between years and supports the idea that terrorist violence is a dynamic phenomenon.

\subsection{Robustness: Temporal Trends}
\begin{figure}[!hbt]
    \centering
    \includegraphics[scale=0.3]{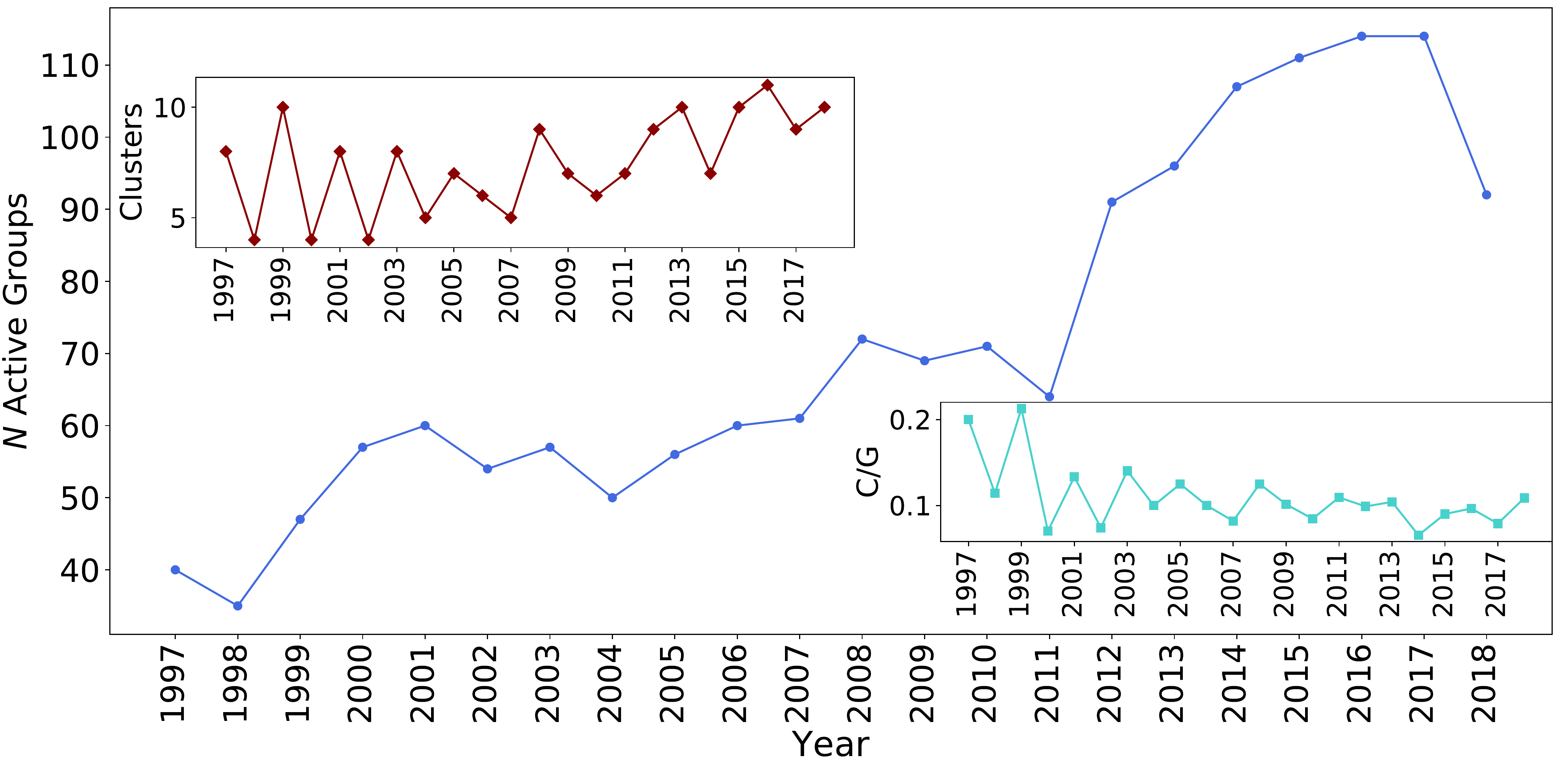}
    \caption{Trend of detected clusters over time (top-left panel), number of active terroris torganizations in each year (main panel) and cluster to groups ratio (bottom-right panel). Robustness results with enlarged sample.}
    \label{fig:ot_rob}
\end{figure}

To verify that our results were not influenced by the selection of our sample, we have computed again our clusters using an enlarged set of terrorist organizations, by focusing on those groups that have plotted at least 30 attacks over the period 1997-2018. The enlarged sample resulted in a total of 164 organizations, accounting for a $\sim$57\% increase in the number of analyzed organizations. Results are visualized in Figure \ref{fig:ot_rob}. 

The trends found in the main analysis are mostly detected also with the enlarged sample. The only slight difference is given by the increasing trend in the number of found clusters from 2009 onwards. Nonetheless, the clusters-to-groups ratio still exhibits a decreasing shape, due to the steeper increase in terms of active groups after 2011. 
This finding points again in the direction of an overall reduction in variability and heterogeneity of terrorist operational patterns over the years. 

\subsection{Robustness: Co-Clustering Stability}
\begin{figure}[!hbt]
    \centering
    \includegraphics[scale=0.373]{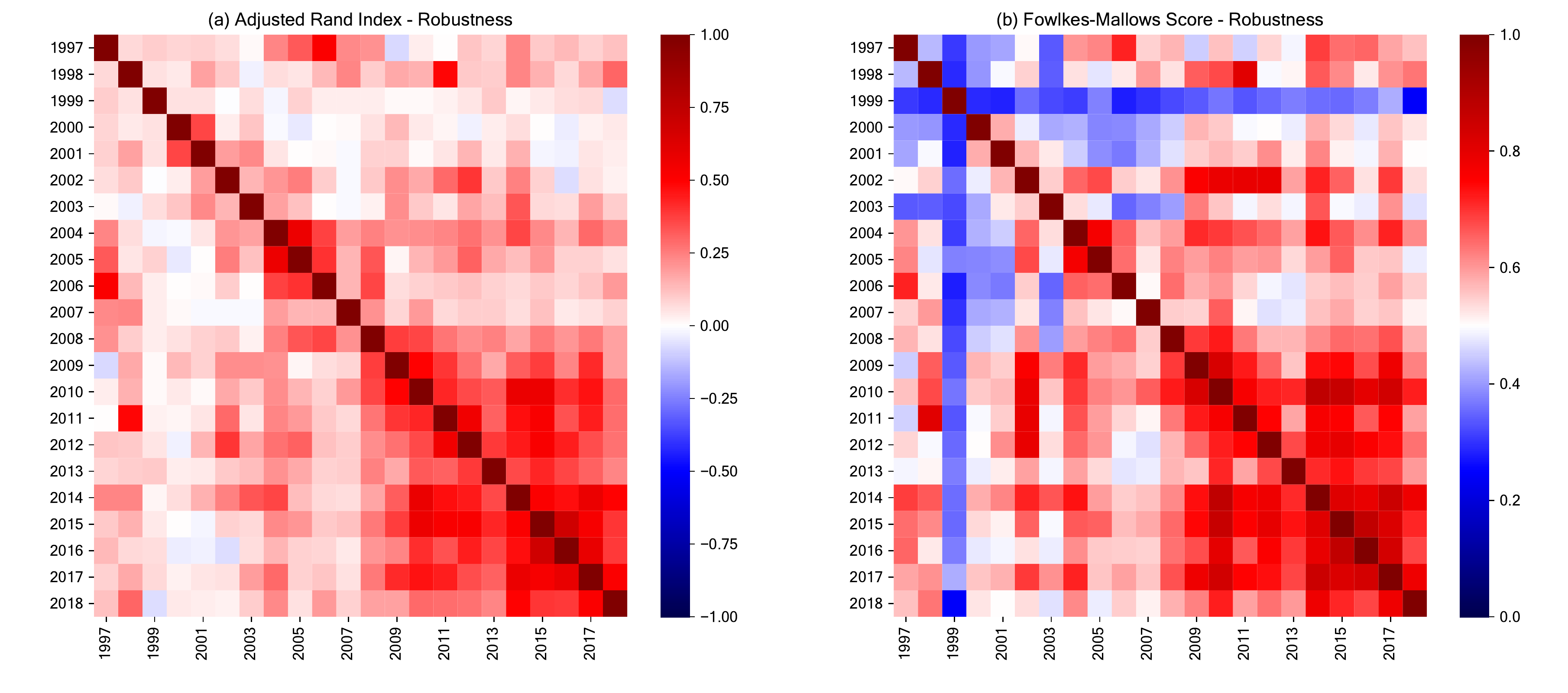}
    \caption{ARI and FMS robustness results with the enlarged sample containing a total of 164 organizations.}
    \label{fig:stab_robustness}
\end{figure}

The co-clustering analysis also was repeated using the enlarged sample for robustness. Results are visualized in Figure \ref{fig:stab_robustness}. As commented in the main analysis, it is clear how stability in year-to-year co-clustering has been higher in the 2009-2018 period. Before 2002, however, both ARI and FMS indicate a low level of stability, with groups changing their operational behaviors from one year to the other. 

\newpage

\subsection{Full Outcomes: Drivers of Similarity}\label{explanation}

\subsubsection{Explanation: Intro}
We used the package the \texttt{ergm} package in the \texttt{Statnet} suite \cite{ergmpackage} available in \texttt{R} for estimating our bipartite ERGM, with networks having two node-sets: groups $O$ active in year $t$ and the clusters found in the same year $C$. An edge exists if group $o$ is associated with cluster $c$.
In terms of syntax and, most importantly, specifications of the covariates, "Sum of Features Weights" is a \texttt{b1cov} term, "Diff. in N of non-zero Features" and "Diff. in Non-Zero Features to Weights Ratio" are coded as two distinct \texttt{absdiff} terms and the other four covariates are coded as \texttt{b1nodematch} factors, specifically intended to capture homophily. All covariates refer to the $O$ nodeset of groups. Below three examples of our bipartite networks are provided, for the years 1999, 2008 and 2017 respectively.
\begin{figure}[!hbt]
    \centering
    \includegraphics[scale=0.25]{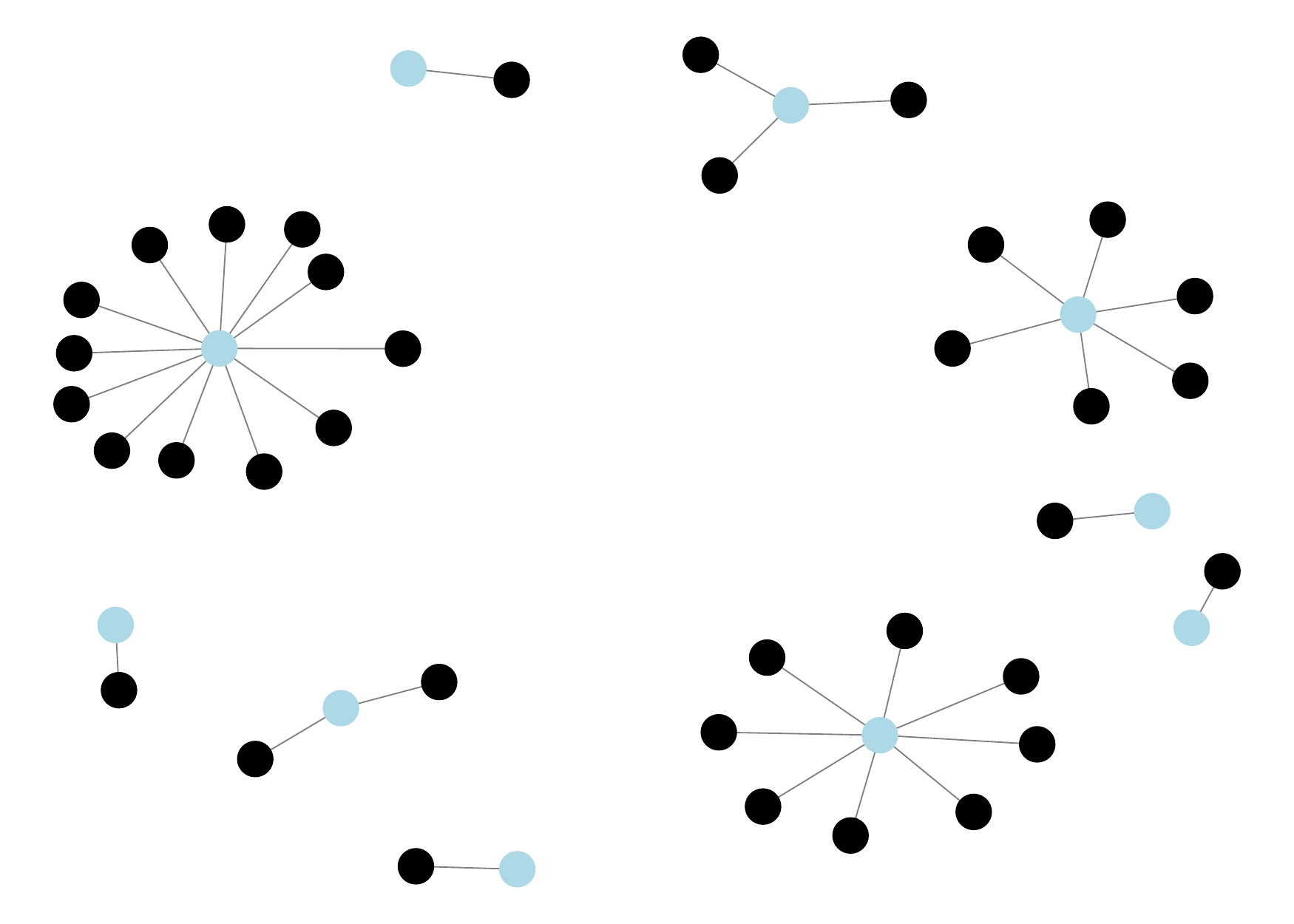}
    \caption{$O\times C$ Bipartite Network for the year 1999. Black nodes represent groups in the $O$ set, light blue nodes are clusters in the $C$ set.}
    \label{fig:my_label}
\end{figure}

\begin{figure}[!hbt]
    \centering
    \includegraphics[scale=0.25]{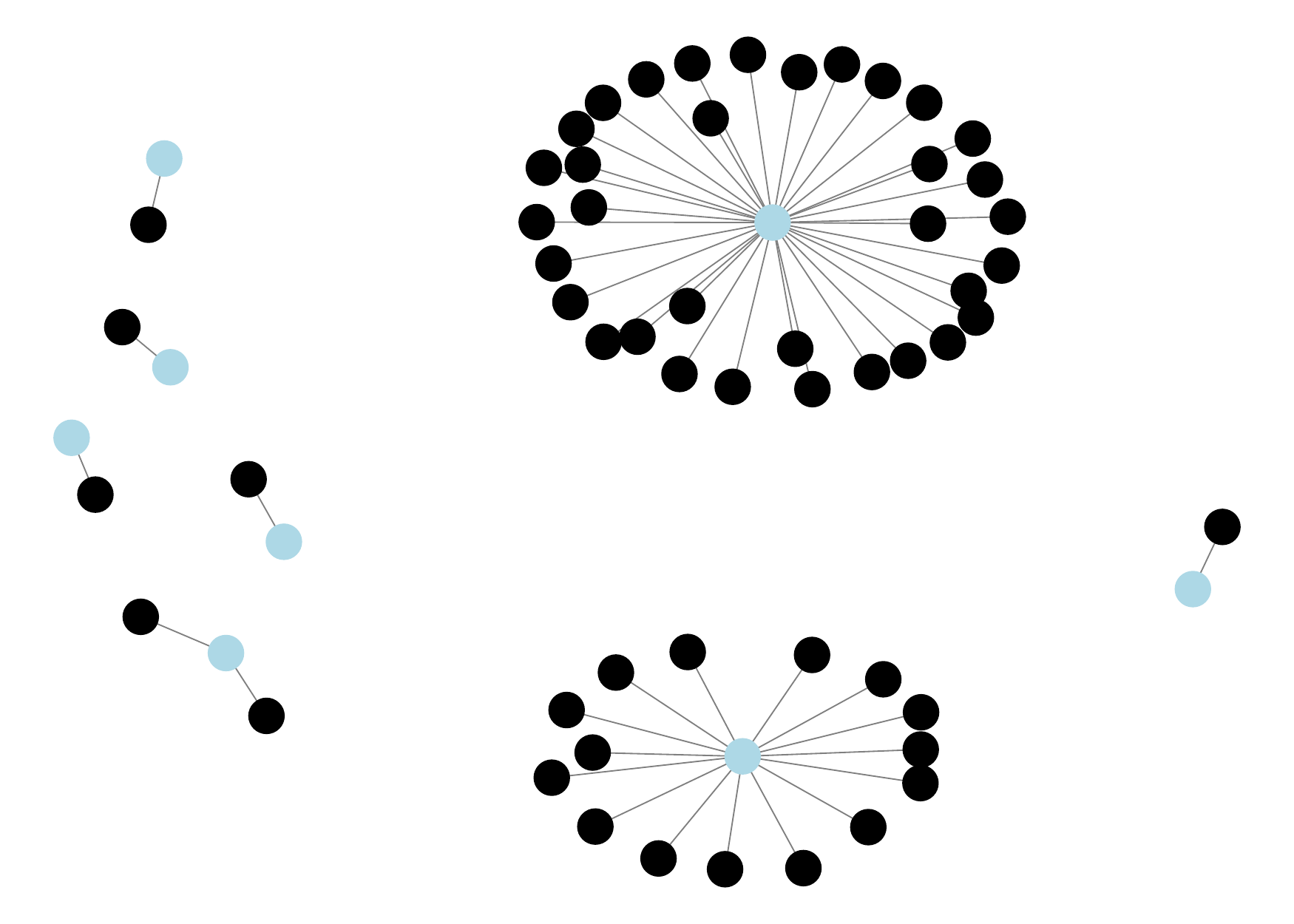}
    \caption{$O\times C$ Bipartite Network for the year 2008. Black nodes represent groups in the $O$ set, light blue nodes are clusters in the $C$ set.}
    \label{fig:my_label}
\end{figure}

\begin{figure}[!hbt]
    \centering
    \includegraphics[scale=0.25]{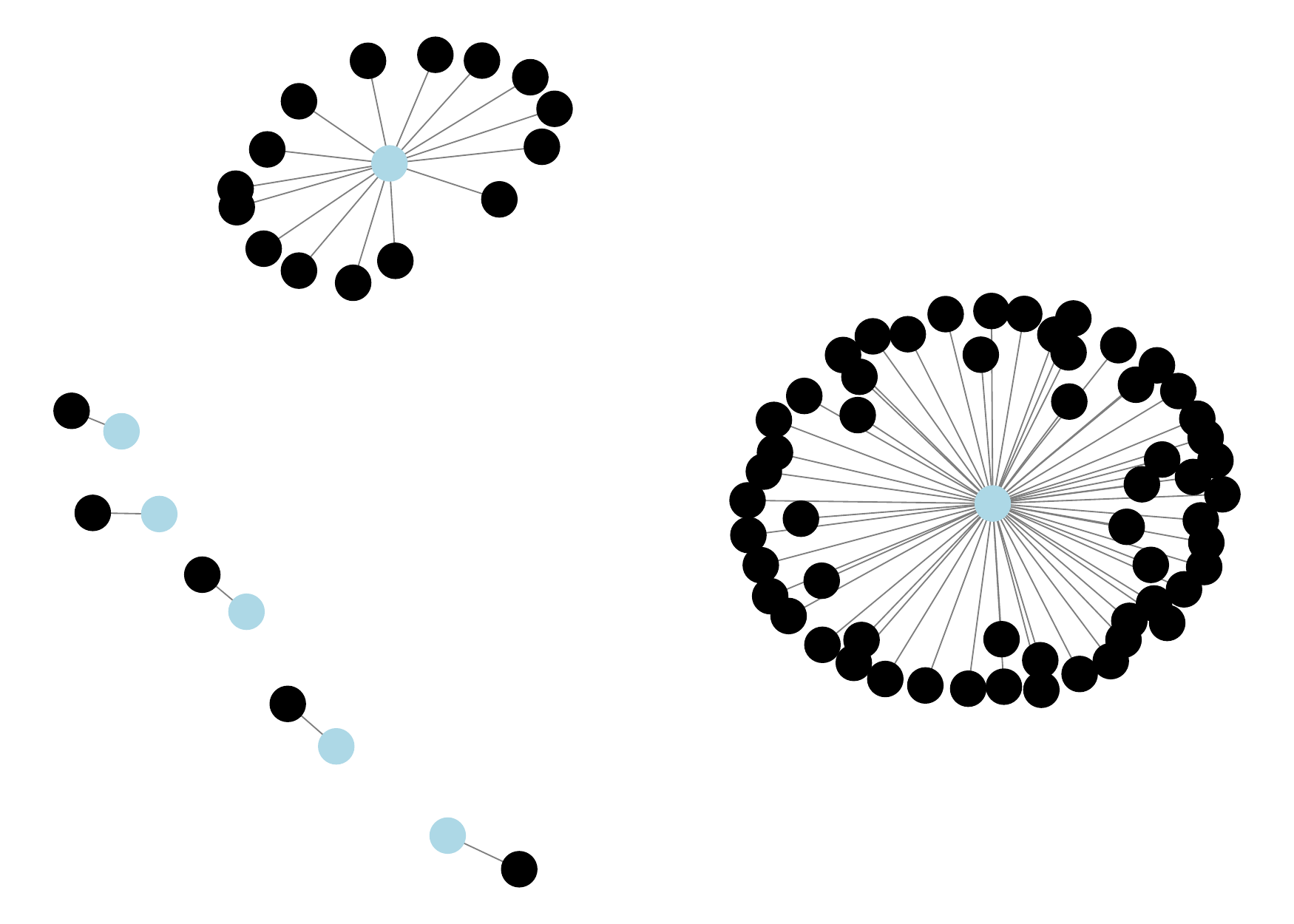}
    \caption{$O\times C$ Bipartite Network for the year 2017. Black nodes represent groups in the $O$ set, light blue nodes are clusters in the $C$ set.}
    \label{fig:my_label}
\end{figure}

The interpretation of the estimated $\theta$ varies based on the type of variable. We provide some examples to facilitate the understanding of the coefficients. 
It is worth noting that an ERGM can be also expressed in terms of the conditional log odds that two groups $o_i$ and $o_j$ are connected to the same cluster $c_k$ as:

\begin{equation}
    logit(y_{o_{i}\rightarrow c_k}=1; y_{o_{j}\rightarrow c_k}=1)= \theta^{\mathrm{T}}\Delta [g(y,X)]_{o_{i},o_{j}}
\end{equation}

where $\Delta [g(y,X)]_{o_{i},o_{j}}$ is the change in $g(y, X)$ when the values of $y_{o_{i}\rightarrow c_k}; y_{o_{j}\rightarrow c_k}$ are both 0. 

\subsubsection{Sum of Features Weights}
\begin{figure}[!hbt]
    \centering
    \includegraphics[scale=0.3]{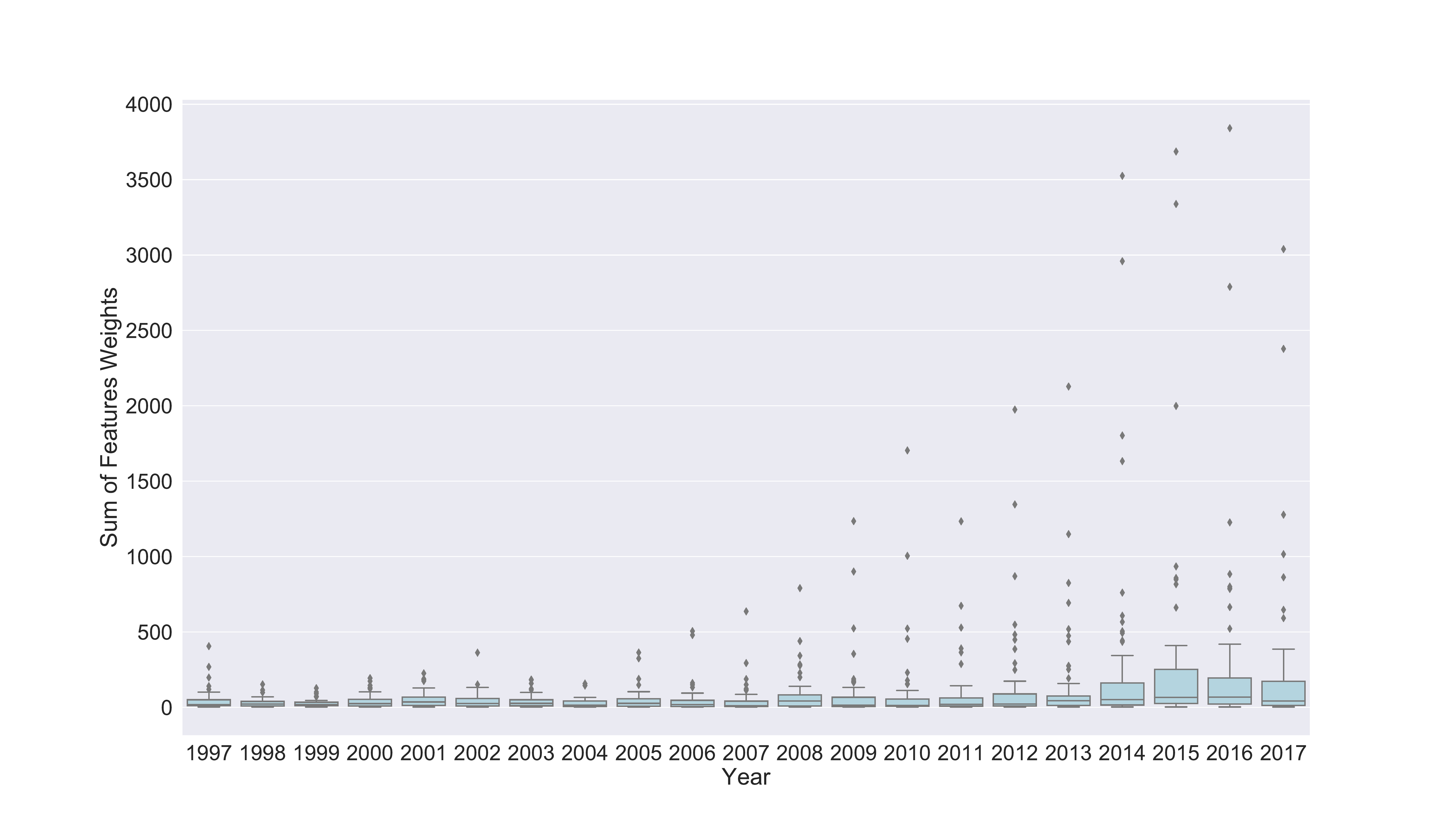}
    \caption{Yearly Distribution of Sum of Features Weights}
    \label{fig:sup_sum_dist}
\end{figure}

For what concerns the "Sum of Features Weights" variable, the conditional log-odds that two groups $o_i$ and $o_j$ are in the same cluster $c_k$ - assuming that all the other variables are held constant - is computed as follows:

\begin{equation}
    logit(p(y_{o_{i}\rightarrow c_k}=1; y_{o_{j}\rightarrow c_k}=1))=\theta_{\mathrm{Sum\: of\: Feature\: Weights}}\times (\mathrm{Sum\: of\: Feature\: Weights}_{o_i}+\mathrm{Sum\: of\: Feature\: Weights}_{o_j})
\end{equation}
The corresponding probability is then obtained by taking the expit of $logit(p(y_{o_{i}\rightarrow c_k}=1; y_{o_{j}\rightarrow c_k}=1))$:
\begin{equation}
    p(y_{o_{i}\rightarrow c_k}=1; y_{o_{j}\rightarrow c_k}=1))=\frac{\mathrm{exp}(logit(p(y))}{1+\mathrm{exp}(logit(p(y))}
    \label{expit}
\end{equation}
Using as an example the year 2005, where the estimated $\theta$ is significant at the 99\% level and is qual to $-0.0108$, we consider four groups with very different Sum of Features Weights features: \textit{National Liberation Front of Tripura (NLFT)} ("Sum of Features Weights"=3), \textit{Shining Path} ("Sum of Features Weights"=5), \textit{Al-Qaida in Iraq} ("Sum of Features Weights"=189, and \textit{Liberation Tigers of Tamil Eelam (LTTE)} ("Sum of Features Weights"=325). The conditional log-odds that the \textit{National Liberation Front of Tripura (NLFT)} and \textit{Shining Path} are in the same are both connected to the same cluster, is given by $-0.0108\times(3+5)=-0.0864$, which in terms of probability is transformed in $\mathrm{exp}(-0.0864)/(1+\mathrm{exp}(-0.0864))=0.47$. Instead, the conditional log-odds that \textit{Al-Qaida in Iraq} and \textit{Liberation Tigers of Tamil Eelam (LTTE)} are in the same cluster are computed as $-0.0108\times(189+325)=-5.5512$. Therefore, the related probability is $0.003$. Finally, the conditional log-odds that \textit{Shining Path} is clustered together with \textit{Al-Qaida in Iraq} is equal to $-2.0952$, with a probability of $0.1095$. 
These results indicate that it is much more likely that two groups with low levels of activity end up being clustered together, compared to the other two scenarios. Particularly, the higher the sum, as in the case of the two most active groups, the lower the likelihood that they can be considered as operationally similar. This further indicates how the outliers in our sample of groups are significantly different from one another. Figure \ref{fig:sup_sum_dist} displays the distribution of the "Sum of Features Weights" variable at the yearly level in order to provide reference on the magnitude of the possible $\mathrm{min-max}$ possible values. It is worth noting how, in recent years, there has been a considerable increase in outlier groups that are characterized by extremely high amount of activity.

\subsubsection{Difference in Number of Non-zero Features}
The conditional log-odds that two groups $o_i$ and $o_j$ are connected to the same cluster $c$ is calculated as:

\begin{equation}
    logit(p(y_{o_{i}\rightarrow c_k}=1; y_{o_{j}\rightarrow c_k}=1))=\theta_{\mathrm{Non-zero \:Features}}\times (|\mathrm{Non-zero\: Features}_{o_i}-\mathrm{Non-zero\: Features}_{o_j}|)
\end{equation}
with the related probability calculated using the same formula seen in Equation \ref{expit}. To exemplify, we take again into consideration year 2005,  and particularly four groups: \textit{Lashkar-e-Jhangvi} ("Non-zero Features"=3), \textit{Revolutionary Armed Forces of Colombia (FARC)} ("Non-zero Features"=18), \textit{Animal Liberation Front (ALF)} ("Non-zero Features"=4), and the \textit{Taliban} ("Non-zero Features"=23). The conditional log-odds (assuming that all the other variables are held constant) that \textit{Lashkar-e-Jhangvi} is clustered together with the \textit{Animal Liberation Front (ALF)} is equal to $-0.1574 \times (|4-3|)=-0.1574$, which corresponds to a probability of $0.4607$. The conditional log-odds that the \textit{Taliban} are clustered with the \textit{Animal Liberation Front (ALF)} are given by $-0.1574 \times (|23-4|)=-2.9906$, corresponding to a probability of $0.0478$. Finally, the probability that \textit{Revolutionary Armed Forces of Colombia (FARC)} are in the same cluster with the \textit{Taliban} is equal to $0.3128$, given the conditional log-odds computed as   $-0.1574 \times (|18-23|)=-0.787$. 

\begin{figure}[!hbt]
    \centering
    \includegraphics[scale=0.3]{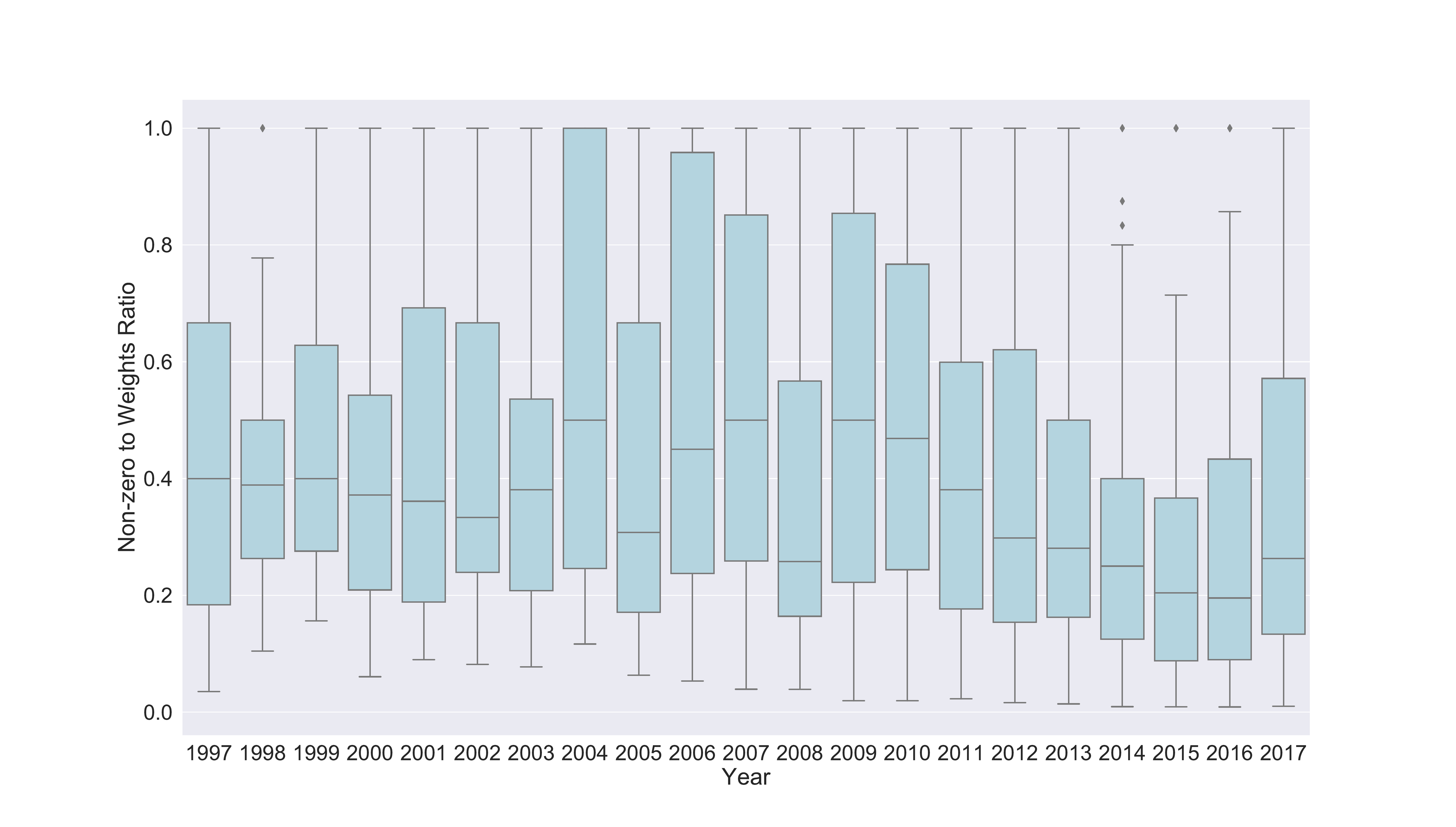}
    \caption{Yearly Distribution of Non-zero Features}
    \label{fig:sup_nonzero_dist}
\end{figure}

These calculations show that two terrorist groups have higher odds of being found operationally similar if the number of non-zero features for a given pair of groups is similar, regardless of the magnitude of the actual number of non-zero features. Two pairs of groups, with one having  $5$ and $7$ non-zero features and the other having $18$ and $16$ have the same probability of being clustered together. A visualization showing the distribution of the variable is provided in Figure \ref{fig:sup_nonzero_dist}.

\subsubsection{Difference in Non-zero Features to Weights Ratio}
The same calculation steps seen in the previous paragraph applies also the the "Difference in Number of Non-zero Features to Weights Ratio" variable. This covariate captures dynamics from both the previous two: it maps both the information layer available when considering the overall amount of activity of a group, along with the layer that derives from its repertoire characterization.

\begin{figure}[!hbt]
    \centering
    \includegraphics[scale=0.3]{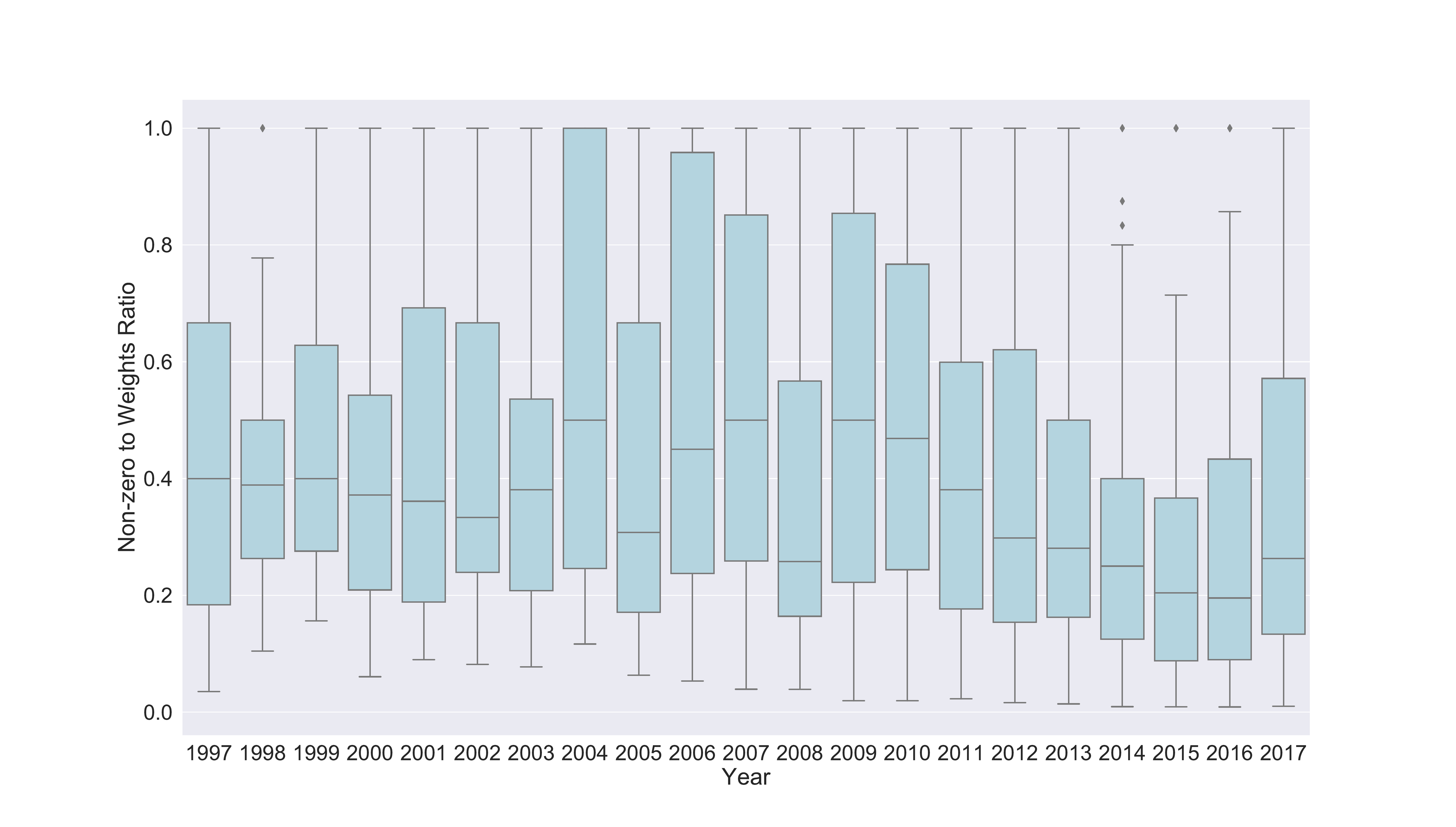}
    \caption{Yearly Distribution of Non-zero to Sum Ratio}
    \label{fig:sup_nzerotosum_dist}
\end{figure}

To exemplify, we take year 2005 as reference again, and we use the following four groups as reference: \textit{Free Aceh Movement} ("Non-zero Features to Weights Ratio"=1), \textit{Taliban} ("Non-zero Features to Weights Ratio"=0.06), \textit{Revolutionary Armed Forces of Colombia (FARC)} ("Non-zero Features to Weights Ratio"=0.17), \textit{Palestinian Islamic Jihad (PIJ)}("Non-zero Features to Weights Ratio"=0.16). The conditional log-odds for a co-clustering between the \textit{Free Aceh Movement} and \textit{Al Qaeda in Iraq}, which are the groups with the maximum and minimum "Non-zero Features to Weight Ratio" is given by $-4.438 \times (|1-0.06|)=-4.1717$, which equals a probability of $0.015$. As it can be seen, the likelihood that these two groups are clustered together is extremely low. The likelihood is higher when the considered pair includes the \textit{Taliban} and 
the \textit{Palestinian Islamic Jihad (PIJ)}: conditional log-odds equal to $-0.44$ and probability of $0.39$. Finally, if when investigating the likelihood of a co-clustering between \textit{Palestinian Islamic Jihad (PIJ)} and the \textit{Revolutionary Armed Forces of Colombia (FARC)}, the conditional log-odds is $-0.04$ and the related probability is $0.49$. It is worth noting that while these two groups are the ones, among the examples here presented, to have the higher likelihood of being clustered together, they are quite different in terms of "Sum of Features Weights" and "Number of non-zero Features". In fact, while \textit{Palestinian Islamic Jihad (PIJ)} has "Sum of Features Weights"$=49$ and $8$ as "Number of non-zero Features", the \textit{Revolutionary Armed Forces of Colombia (FARC)} has "Sum of Features Weights"$=104$ and "Number of non-zero Features"=$8$. This testifies that this variable is critical in capturing complex repertoire dynamics that are emerging from the normalized comparative accounts of organizations with distinct resource portfolios. Figure \ref{fig:sup_nzerotosum_dist} shows the distribution of the variable across the years of our analysis. 

\subsubsection{Homophily terms: Shared Most Common Target, Shared Most Common Tactic, Shared Most Common Weapon, Shared Most Common Region, Shared Ideology}

All the remaining variables are considered as categorical homophily terms. For a given homophily variable $\mathrm{h}$, the computation of conditional log-odds is straightforward, and is expressed as: 

\begin{equation}
    logit(p(y))=\left\{\begin{matrix} \theta_{\mathrm{h}} & \mathrm{if} & \mathrm{h}_{o_{1}}=\mathrm{h}_{o_{2}} \\ 0 & & \mathrm{otherwise} \end{matrix}\right.
\end{equation}

To provide a further example, we take into consideration the year 2007, in which a homophily pattern in terms of operational preferences of weapons selection has been detected. Using as examples two groups, namely the \textit{Abu Sayyaf Group (ASG)} and the \textit{Al-Aqsa Martyrs Brigade}, we compute the conditional log-odds that these two groups are clustered together derived from the estimated coefficient of "Shared Most Common Weapon", assuming all other variables are held constant, as $0.4766$, with a computed probability of $0.61$.

\subsubsection{A Complete Example}
So far, we have only considered conditional log-odds and probabilities that two groups $o_i$ and $o_j$ are clustered together, i.e., are both connected to cluster $c_k$, considering one covariate of interest at a time, and assuming all other variables were held constant. Here below, we provide a full example to express the probability that three specific groups are clustered together in a given year. 
The example year is 2005. We specifically use three groups: \textit{National Liberation Army of Colombia (ELN)}, \textit{Jemaah Islamiya (JI)} and \textit{Liberation Tigers of Tamil Eelam (LTTE)}. For the year 2005, the estimated $\theta$ for each variable is reported in Table \ref{tab: theta}. It is worth noting that the example below should be treated as an example only, as we will treat the reported coefficient as if they were all significant for illustrating a complete equation.

\begin{table}[!hbt]
\caption{Estimated $\theta$ for each variable in the model - year 2005}
\centering
\begin{tabular}{lc}
\hline
\textbf{Variable} & \textbf{$\theta$ (2005)} \\ \hline
Sum of Features Weights & -0.0108 \\
Diff. in N of Non-zero Features & -0.1574 \\
Diff. in Non-zero Features to Weights Ratio & -4.4038 \\
Shared Most Common Target & 0.1309 \\
Shared Most Common Tactic & 0.1175 \\
Shared Most Common Weapon & -0.0216 \\
Shared Most Common Region & 0.0394 \\
Shared Ideology & 0.1155 \\ \hline
\end{tabular}
\label{tab: theta}
\end{table}

Table \ref{tab: ref2005}
instead provides the covariate statistics for each group.

\begin{table}[!hbt]
\footnotesize
\caption{Variable values associated with ELN, JI and LTTE - Year 2005 }
\centering
\begin{tabular}{lccc}
\hline
\textbf{Group} & \textbf{ELN} & \textbf{JI} & \textbf{LTTE} \\ \hline
Sum of Features Weights & 21 & 27 & 325 \\
N of non-zero Features & 10 & 7 & 23 \\
Non-zero Features to Weights Ratio & 0.471 & 0.259 & 0.07 \\
Most Common Target & Private Citizens \& Property & Business & Private Citizens \& Property \\
Most Common Tactic & Bombing/Explosion & Bombing/Explosion & Armed Assault \\
Most Common Weapon & Explosives & Explosives & Firearms \\
Most Common Region & South America & Southeast Asia & South Asia \\
Ideology & Left \& Ethno & Isl/Jihadism & Left \& Ethno \\ \hline
\end{tabular}
\label{tab: ref2005}
\end{table}

The conditional log odds for the three resulting pairs are then calculated as: 

\begin{multline}
    logit(y_{ELN\rightarrow c_k}=1; y_{JI\rightarrow c_k}=1)=\\ [-0.0108*(21+27)]+[-0.1574*(|10-7|)]+[-4.4034*(|0.471-0.259|)]+0.011-0.0216=-1.93
\end{multline}

\begin{multline}
    logit(y_{ELN\rightarrow c_k}=1; y_{LTTE\rightarrow c_k}=1)=\\ [-0.0108*(21+325)]+[-0.1574*(|10-23|)]+[-4.4034*(|0.471-0.07|)]+0.1309+0.1155=-7.30
\end{multline}

\begin{equation}
logit(y_{JI\rightarrow c_k}=1; y_{LTTE\rightarrow c_k}=1)= [-0.0108*(27+325)]+[-0.1574*(|7-23|)]+[-4.4034*(|0.259-0.07|)]=-7.15
\end{equation}

These conditional log-odds can be expressed in terms of probability, leading to: 

\begin{equation}
    p(y_{ELN\rightarrow c_k}=1; y_{JI\rightarrow c_k}=1)= \mathrm{exp(-1.93)}/(1+\mathrm{exp}(-1.93))=0.126
\end{equation}

\begin{equation}
    p(y_{ELN\rightarrow c_k}=1; y_{LTTE\rightarrow c_k}=1)= \mathrm{exp(-7.30)}/(1+\mathrm{exp}(-7.30))=0.0006
\end{equation}

\begin{equation}
    p(y_{JI\rightarrow c_k}=1; y_{LTTE\rightarrow c_k}=1)= \mathrm{exp(-7.15)}/(1+\mathrm{exp}(-7.15))=0.0007
\end{equation}

These equations show that the pair that has the higher chances to end up in the same cluster, and is therefore more similar, is the ELN-JI one. Conversely, LTTE demonstrates to be extremely peculiar, given its outlier amount of activity, leading therefore to a very low likelihood of being clustered with any of the other two groups.

\subsubsection{Model Outcomes}
Below are presented the full results for the yearly ERGM estimated on the bipartite networks. All the models have been fit using Markov chain Monte Carlo (MCMC) and coefficients are estimated using Monte Carlo Maximum Likelihood Estimation (MCMLE).

\begin{table}[!hbt]
\setlength{\tabcolsep}{3pt}
\footnotesize
\begin{center}
\caption{ERGM Full Results - Years 1997, 1998, 1999, 2000, 2001, 2002}
\begin{tabular}{l c c c c c c}
\hline
 & Year 1997 & Year 1998 & Year 1999 & Year 2000 & Year 2001 & Year 2002 \\
\hline
Sum of Features Weights                      & $0.0011$        & $-0.0303^{***}$ & $-0.0322^{**}$  & $-0.0062$  & $-0.0276^{***}$ & $-0.0100^{*}$   \\
                                            & $(0.0034)$      & $(0.0087)$      & $(0.0109)$      & $(0.0040)$ & $(0.0045)$      & $(0.0046)$      \\
Diff in N of non-zero Features              & $-0.1853^{**}$  & $-0.0299$       & $0.0078$        & $-0.0433$  & $0.1266^{*}$    & $-0.0521$       \\
                                            & $(0.0708)$      & $(0.0633)$      & $(0.0667)$      & $(0.0381)$ & $(0.0505)$      & $(0.0538)$      \\
Diff. in Non-zero Features to Weights Ratio & $-5.5308^{***}$ & $-4.6537^{***}$ & $-6.1039^{***}$ & $0.4376$   & $-7.4228^{***}$ & $-3.4924^{***}$ \\
                                            & $(1.1583)$      & $(1.1966)$      & $(1.1339)$      & $(0.6498)$ & $(1.0381)$      & $(0.8172)$      \\
Shared Most Common Target                   & $0.1619$        & $0.1144$        & $-0.5383^{*}$   & $-0.0640$  & $0.0518$        & $0.0326$        \\
                                            & $(0.1289)$      & $(0.1800)$      & $(0.2660)$      & $(0.0761)$ & $(0.0782)$      & $(0.0456)$      \\
Shared Most Common Tactic                   & $0.2144$        & $0.3196$        & $0.1814$        & $-0.0258$  & $-0.0213$       & $-0.0081$       \\
                                            & $(0.1845)$      & $(0.1860)$      & $(0.1743)$      & $(0.0867)$ & $(0.1014)$      & $(0.0944)$      \\
Shared Most Common Weapon                   & $0.0311$        & $-0.0570$       & $0.2942$        & $0.1027$   & $0.1546$        & $0.1102$        \\
                                            & $(0.2070)$      & $(0.1806)$      & $(0.1653)$      & $(0.0617)$ & $(0.0810)$      & $(0.0696)$      \\
Shared Ideology                  & $-0.4467$       & $-0.2793$       & $-0.3840$       & $-0.1720$  & $0.1040$        & $-0.1507$       \\
                                            & $(0.3850)$      & $(0.3107)$      & $(0.2429)$      & $(0.1244)$ & $(0.0977)$      & $(0.1125)$      \\
Shared Most Common Region                            & $0.3915$        & $0.2440$        & $-0.4533$       & $0.0105$   & $0.0359$        & $0.1815$        \\
                                            & $(0.2686)$      & $(0.2435)$      & $(0.3276)$      & $(0.1123)$ & $(0.1194)$      & $(0.0967)$      \\
\hline
AIC                                         & $188.1583$      & $190.4504$      & $257.2862$      & $184.1833$ & $282.9729$      & $151.4950$      \\
BIC                                         & $215.7322$      & $216.9560$      & $288.3750$      & $207.2457$ & $314.0617$      & $173.7949$      \\
Log Likelihood                              & $-86.0791$      & $-87.2252$      & $-120.6431$     & $-84.0916$ & $-133.4864$     & $-67.7475$      \\
\hline
\multicolumn{7}{l}{\scriptsize{$^{***}p<0.001$; $^{**}p<0.01$; $^{*}p<0.05$}}
\end{tabular}

\label{table:coefficients}
\end{center}
\end{table}

\begin{table}[!hbt]
\centering
\setlength{\tabcolsep}{3pt}
\footnotesize
\begin{center}
\caption{ERGM Full Results - Years 2003, 2004, 2005, 2006, 2007, 2008}
\begin{tabular}{l c c c c c c}
\hline
 & Year 2003 & Year 2004 & Year 2005 & Year 2006 & Year 2007 & Year 2008 \\
\hline
Sum of Features Weights                      & $-0.0088$       & $-0.0205^{**}$  & $-0.0108^{**}$  & $0.0048$        & $0.0000$        & $-0.0092^{***}$ \\
                                            & $(0.0050)$      & $(0.0075)$      & $(0.0040)$      & $(0.0028)$      & $(0.0023)$      & $(0.0028)$      \\
Diff in N of non-zero Features              & $-0.2681^{***}$ & $-0.2913^{***}$ & $-0.1574^{***}$ & $-0.1835^{**}$  & $-0.2127^{***}$ & $-0.1124^{***}$ \\
                                            & $(0.0601)$      & $(0.0721)$      & $(0.0466)$      & $(0.0663)$      & $(0.0637)$      & $(0.0306)$      \\
Diff. in Non-zero Features to Weights Ratio & $-4.6008^{***}$ & $-1.4293$       & $-4.4038^{***}$ & $-3.7335^{***}$ & $-7.2693^{***}$ & $-3.1076^{***}$ \\
                                            & $(0.9362)$      & $(0.8946)$      & $(0.9669)$      & $(0.8804)$      & $(1.5961)$      & $(0.6915)$      \\
Shared Most Common Target                   & $0.1343$        & $-0.0018$       & $0.1309$        & $0.0870^{*}$    & $0.1527^{*}$    & $-0.0274$       \\
                                            & $(0.0767)$      & $(0.0913)$      & $(0.0698)$      & $(0.0425)$      & $(0.0646)$      & $(0.0382)$      \\
Shared Most Common Tactic                   & $0.0841$        & $0.2414^{*}$    & $0.1175$        & $-0.0743$       & $-0.3588^{**}$  & $0.3553^{**}$   \\
                                            & $(0.0764)$      & $(0.1085)$      & $(0.0811)$      & $(0.0757)$      & $(0.1301)$      & $(0.1136)$      \\
Shared Most Common Weapon                   & $0.1389^{*}$    & $-0.0729$       & $-0.0216$       & $0.0646$        & $0.4766^{***}$  & $-0.2980^{**}$  \\
                                            & $(0.0633)$      & $(0.0997)$      & $(0.0825)$      & $(0.0698)$      & $(0.1421)$      & $(0.1063)$      \\
Shared Ideology                   & $-0.1612$       & $0.0298$        & $0.0285$        & $0.0394$        & $0.0844$        & $0.0152$        \\
                                            & $(0.1148)$      & $(0.0832)$      & $(0.0818)$      & $(0.0721)$      & $(0.0614)$      & $(0.0633)$      \\
Shared Most Common Region                           & $0.1492$        & $0.1439$        & $0.1155$        & $0.0962$        & $-0.0272$       & $0.0989$        \\
                                            & $(0.0931)$      & $(0.0904)$      & $(0.1197)$      & $(0.0749)$      & $(0.1007)$      & $(0.0668)$      \\
\hline
AIC                                         & $207.8771$      & $172.4371$      & $223.2313$      & $232.0810$      & $145.1315$      & $381.7142$      \\
BIC                                         & $236.3008$      & $197.6170$      & $251.6549$      & $259.5856$      & $171.5181$      & $414.5525$      \\
Log Likelihood                              & $-95.9386$      & $-78.2185$      & $-103.6156$     & $-108.0405$     & $-64.5658$      & $-182.8571$     \\
\hline
\multicolumn{7}{l}{\scriptsize{$^{***}p<0.001$; $^{**}p<0.01$; $^{*}p<0.05$}}
\end{tabular}

\label{table:coefficients}
\end{center}
\end{table}

\begin{table}[!hbt]
\setlength{\tabcolsep}{3pt}
\footnotesize
\begin{center}
\caption{ERGM Full Results - Years 2010, 2011, 2012, 2013, 2014, 2015}
\begin{tabular}{l c c c c c c}
\hline
 & Year 2009 & Year 2010 & Year 2011 & Year 2012 & Year 2013 & Year 2014 \\
\hline
Sum of Features Weights                      & $-0.0019^{*}$   & $-0.0036^{**}$  & $-0.0010$       & $-0.0026^{**}$  & $-0.0015$       & $-0.0009^{**}$  \\
                                            & $(0.0010)$      & $(0.0012)$      & $(0.0010)$      & $(0.0008)$      & $(0.0008)$      & $(0.0003)$      \\
Diff in N of non-zero Features              & $-0.0873^{***}$ & $0.0827^{*}$    & $-0.1844^{***}$ & $-0.1719^{***}$ & $-0.2780^{***}$ & $-0.1445^{***}$ \\
                                            & $(0.0238)$      & $(0.0382)$      & $(0.0287)$      & $(0.0292)$      & $(0.0325)$      & $(0.0181)$      \\
Diff. in Non-zero Features to Weights Ratio & $-2.0154^{***}$ & $-5.1853^{***}$ & $-0.6121$       & $-3.5098^{***}$ & $-1.3977^{***}$ & $-0.9507^{*}$   \\
                                            & $(0.5819)$      & $(0.7880)$      & $(0.5565)$      & $(0.9371)$      & $(0.3964)$      & $(0.4668)$      \\
Shared Most Common Target                   & $-0.0880$       & $-0.0205$       & $0.0115$        & $0.0468$        & $0.0404$        & $0.0060$        \\
                                            & $(0.0570)$      & $(0.0425)$      & $(0.0533)$      & $(0.0340)$      & $(0.0358)$      & $(0.0175)$      \\
Shared Most Common Tactic                   & $0.0466$        & $-0.0446$       & $-0.0373$       & $-0.1672^{*}$   & $-0.0200$       & $0.0244$        \\
                                            & $(0.0586)$      & $(0.0569)$      & $(0.0638)$      & $(0.0712)$      & $(0.0497)$      & $(0.0358)$      \\
Shared Most Common Weapon                   & $0.0431$        & $0.0042$        & $0.0810$        & $0.2200^{***}$  & $0.0877$        & $0.0085$        \\
                                            & $(0.0442)$      & $(0.0486)$      & $(0.0509)$      & $(0.0582)$      & $(0.0473)$      & $(0.0362)$      \\
Shared Ideology                   & $0.0075$        & $0.0383$        & $-0.0026$       & $-0.0105$       & $0.0347$        & $0.0244$        \\
                                            & $(0.0510)$      & $(0.0336)$      & $(0.0576)$      & $(0.0404)$      & $(0.0350)$      & $(0.0255)$      \\
Shared Most Common Region                          & $0.0736$        & $0.1673^{***}$  & $0.0186$        & $0.0899^{*}$    & $0.0871^{**}$   & $0.0413$        \\
                                            & $(0.0516)$      & $(0.0409)$      & $(0.0472)$      & $(0.0385)$      & $(0.0307)$      & $(0.0301)$      \\
\hline
AIC                                         & $346.2004$      & $303.3060$      & $331.9040$      & $272.9121$      & $315.2215$      & $491.5163$      \\
BIC                                         & $376.4463$      & $332.3843$      & $362.9259$      & $304.5172$      & $347.0728$      & $525.8340$      \\
Log Likelihood                              & $-165.1002$     & $-143.6530$     & $-157.9520$     & $-128.4560$     & $-149.6107$     & $-237.7581$     \\
\hline
\multicolumn{7}{l}{\scriptsize{$^{***}p<0.001$; $^{**}p<0.01$; $^{*}p<0.05$}}
\end{tabular}

\label{table:coefficients}
\end{center}
\end{table}
\newpage
\begin{table}[!hbt]
\setlength{\tabcolsep}{3pt}
\footnotesize
\begin{center}
\caption{ERGM Full Results - Years 2016, 2017, 2018}
\begin{tabular}{l c c c c}
\hline
 & Year 2015 & Year 2016 & Year 2017 & Year 2018 \\
\hline
Sum of Features Weights                      & $-0.0010^{*}$   & $-0.0014^{*}$   & $-0.0016^{**}$  & $-0.0013^{*}$   \\
                                            & $(0.0004)$      & $(0.0006)$      & $(0.0006)$      & $(0.0006)$      \\
Diff in N of non-zero Features              & $-0.0896^{***}$ & $-0.1479^{***}$ & $-0.0021$       & $-0.1374^{***}$ \\
                                            & $(0.0205)$      & $(0.0313)$      & $(0.0158)$      & $(0.0329)$      \\
Diff. in Non-zero Features to Weights Ratio & $-0.9059$       & $-0.6203$       & $-3.3708^{***}$ & $-3.3394^{***}$ \\
                                            & $(0.6521)$      & $(0.4719)$      & $(0.5319)$      & $(0.5955)$      \\
Shared Most Common Target                   & $-0.0125$       & $-0.0083$       & $0.0015$        & $0.0484^{*}$    \\
                                            & $(0.0202)$      & $(0.0171)$      & $(0.0159)$      & $(0.0202)$      \\
Shared Most Common Tactic                   & $-0.0204$       & $0.0252$        & $0.0028$        & $-0.0620$       \\
                                            & $(0.0351)$      & $(0.0357)$      & $(0.0367)$      & $(0.0458)$      \\
Shared Most Common Weapon                   & $0.0304$        & $0.0463$        & $-0.0016$       & $0.1144^{**}$   \\
                                            & $(0.0350)$      & $(0.0330)$      & $(0.0340)$      & $(0.0406)$      \\
Shared Ideology                  & $-0.0151$       & $0.0184$        & $0.0075$        & $0.0103$        \\
                                            & $(0.0316)$      & $(0.0219)$      & $(0.0263)$      & $(0.0252)$      \\
Shared Most Common Region                             & $0.0468$        & $0.0696^{*}$    & $0.0276$        & $0.0625$        \\
                                            & $(0.0245)$      & $(0.0293)$      & $(0.0308)$      & $(0.0360)$      \\
\hline
AIC                                         & $509.5102$      & $347.2990$      & $588.2078$      & $354.6986$      \\
BIC                                         & $542.4902$      & $377.1398$      & $622.5256$      & $386.5499$      \\
Log Likelihood                              & $-246.7551$     & $-165.6495$     & $-286.1039$     & $-169.3493$     \\
\hline
\multicolumn{5}{l}{\scriptsize{$^{***}p<0.001$; $^{**}p<0.01$; $^{*}p<0.05$}}
\end{tabular}

\label{table:coefficients}
\end{center}
\end{table}

\clearpage
\subsubsection{ERGM Diagnostics}
Tables S7-S28 report the results of the diagnostic checks made on each yearly ERGM to verify that the MCMC simulations are reliable. We specifically report the sample statistic auto-correlation of each covariate in each model which indicates the correlation between sample statistics at different lags across the MCMC chain. A good MCMC simulation would lead to low auto-correlation values (the closer to 0 the better) at each lag (lag 0 excluded). The diagnostic check on each model signaled very low correlation in all models, with the only exception in year 2007 which displayed relatively high correlation at across the sampled lags. 

To complement our diagnostic check and avoid relying on a single diagnostic indicator, we also used the sample statistic burn-in diagnostic (in the tables labeled as "Geweke Diagnostic"). The Geweke Diagnostic offers a measure of convergence comparing means of the sample statistic at distinct points in the Markov chain. A good MCMC simulation should lead to equal means at distinct points, and therefore higher p-values suggest MCMC quality. Across all models, very few variables exhibit p-values lower or equal to 0.05: Diff in N of non-zero Features in 2009 (p=0.05), Diff in Non-zero Features to Weights Ratio in 2015 (p=0.05), Shared Most Common Target in 2015 (p=0.01), and Shared Most Common Weapon in 2015 (p=0.02), accounting for the 2.27\% of all the variables by year observations here analyzed. 

Overall, our diagnostic checks indicate that the MCMC simulations that generated the ERGM estimates for each year are reliable. Although the sample statistic auto-correlation check for year 2007 seemed to suggest poor fit, the analysis of the Geweke Diagnostic did not display any relevant problem, significantly limiting the concern for the sample statistic auto-correlation anomalous outcomes. 

\begin{table}[!h]
\caption{MCMC Diagnostics - ERGM Year 1997}
\setlength{\tabcolsep}{3pt}
\centering
\footnotesize
\begin{tabular}{lllllll|cc}
\hline
 & \multicolumn{6}{c}{\begin{tabular}[c]{@{}c@{}}Sample statistic \\ auto-correlation\end{tabular}} & \multicolumn{2}{c}{\begin{tabular}[c]{@{}c@{}}Geweke\\ Diagnostic\end{tabular}} \\ \hline
Covariate & \multicolumn{1}{c}{\begin{tabular}[c]{@{}c@{}}Lag \\ 0\end{tabular}} & \multicolumn{1}{c}{\begin{tabular}[c]{@{}c@{}}Lag \\ 1024\end{tabular}} & \multicolumn{1}{c}{\begin{tabular}[c]{@{}c@{}}Lag \\ 2048\end{tabular}} & \multicolumn{1}{c}{\begin{tabular}[c]{@{}c@{}}Lag \\ 3072\end{tabular}} & \multicolumn{1}{c}{\begin{tabular}[c]{@{}c@{}}Lag\\ 4096\end{tabular}} & \multicolumn{1}{c|}{\begin{tabular}[c]{@{}c@{}}Lag\\ 5120\end{tabular}} & \multicolumn{1}{l}{Statistic} & \multicolumn{1}{l}{p-value} \\ \hline
Sum of Features Weights & 1.00 & 0.006 & -0.021 & -0.042 & 0.023 & 0.004 & 1.256 & 0.20 \\
Diff in N of non-zero Features & 1.00 & 0.015 & -0.016 & -0.012 & 0.002 & 0.004 & 0.538 & 0.59 \\
Diff in Non-zero Features to Weights Ratio & 1.00 & 0.017 & -0.010 & 0.000 & 0.000 & -0.008 & 0.698 & 0.48 \\
Shared Most Common Target & 1.00 & 0.022 & -0.007 & -0.013 & -0.000 & -0.038 & 0.811 & 0.41 \\
Shared Most Common Tactic & 1.00 & 0.091 & -0.007 & 0.000 & 0.005 & 0.011 & 1.033 & 0.30 \\
Shared Most Common Weapon & 1.00 & 0.066 & -0.010 & -0.019 & 0.002 & 0.007 & 0.526 & 0.59 \\
Shared Ideology & 1.00 & 0.059 & 0.004 & 0.001 & 0.012 & 0.009 & 0.686 & 0.49 \\
Shared Most Common Region & 1.00 & 0.070 & 0.000 & 0.000 & 0.009 & 0.002 & 1.845 & 0.06 \\ \hline
\end{tabular}
\label{diag1997}
\end{table}

\begin{table}[!h]
\caption{MCMC Diagnostics - ERGM Year 1998}
\setlength{\tabcolsep}{3pt}
\centering
\footnotesize
\begin{tabular}{lllllll|cc}
\hline
 & \multicolumn{6}{c}{\begin{tabular}[c]{@{}c@{}}Sample statistic \\ auto-correlation\end{tabular}} & \multicolumn{2}{c}{\begin{tabular}[c]{@{}c@{}}Geweke\\ Diagnostic\end{tabular}} \\ \hline
Covariate & \multicolumn{1}{c}{\begin{tabular}[c]{@{}c@{}}Lag \\ 0\end{tabular}} & \multicolumn{1}{c}{\begin{tabular}[c]{@{}c@{}}Lag \\ 1024\end{tabular}} & \multicolumn{1}{c}{\begin{tabular}[c]{@{}c@{}}Lag \\ 2048\end{tabular}} & \multicolumn{1}{c}{\begin{tabular}[c]{@{}c@{}}Lag \\ 3072\end{tabular}} & \multicolumn{1}{c}{\begin{tabular}[c]{@{}c@{}}Lag\\ 4096\end{tabular}} & \multicolumn{1}{c|}{\begin{tabular}[c]{@{}c@{}}Lag\\ 5120\end{tabular}} & \multicolumn{1}{l}{Statistic} & \multicolumn{1}{l}{p-value} \\ \hline
Sum of Features Weights & 1.00 & 0.039 & 0.013 & 0.003 & -0.005 & 0.005 & 0.013 & 0.98 \\
Diff in N of non-zero Features & 1.00 & 0.028 & -0.010 & 0.004 & -0.007 & 0.005 & -0.323 & 0.74 \\
Diff in Non-zero Features to Weights Ratio & 1.00 & 0.0254 & 0.005 & 0.011 & -0.022 & 0.002 & 0.357 & 0.72 \\
Shared Most Common Target & 1.00 & 0.036 & 0.015 & 0.003 & 0.005 & -0.000 & 0.157 & 0.87 \\
Shared Most Common Tactic & 1.00 & 0.071 & 0.012 & 0.005 & -0.001 & 0.001 & 0.110 & 0.91 \\
Shared Most Common Weapon & 1.00 & 0.063 & 0.015 & 0.010 & 0.001 & 0.004 & 0.250 & 0.80 \\
Shared Ideology & 1.00 & 0.037 & 0.011 & 0.000 & 0.008 & 0.003 & 0.646 & 0.51 \\
Shared Most Common Region & 1.00 & 0.028 & -0.003 & 0.007 & -0.009 & 0.006 & -0.140 & 0.88 \\ \hline
\end{tabular}
\end{table}

\begin{table}[!h]
\caption{MCMC Diagnostics - ERGM Year 1999}
\setlength{\tabcolsep}{3pt}
\centering
\footnotesize
\begin{tabular}{lllllll|cc}
\hline
 & \multicolumn{6}{c}{\begin{tabular}[c]{@{}c@{}}Sample statistic \\ auto-correlation\end{tabular}} & \multicolumn{2}{c}{\begin{tabular}[c]{@{}c@{}}Geweke\\ Diagnostic\end{tabular}} \\ \hline
Covariate & \multicolumn{1}{c}{\begin{tabular}[c]{@{}c@{}}Lag \\ 0\end{tabular}} & \multicolumn{1}{c}{\begin{tabular}[c]{@{}c@{}}Lag \\ 1024\end{tabular}} & \multicolumn{1}{c}{\begin{tabular}[c]{@{}c@{}}Lag \\ 2048\end{tabular}} & \multicolumn{1}{c}{\begin{tabular}[c]{@{}c@{}}Lag \\ 3072\end{tabular}} & \multicolumn{1}{c}{\begin{tabular}[c]{@{}c@{}}Lag\\ 4096\end{tabular}} & \multicolumn{1}{c|}{\begin{tabular}[c]{@{}c@{}}Lag\\ 5120\end{tabular}} & \multicolumn{1}{l}{Statistic} & \multicolumn{1}{l}{p-value} \\ \hline
Sum of Features Weights & 1.00 & 0.087 & 0.041 & 0.006 & 0.015 & 0.010 & 0.531 & 0.59 \\
Diff in N of non-zero Features & 1.00 & 0.140 & 0.062 & 0.033 & 0.000 & 0.003 & -0.198 & 0.84 \\
Diff in Non-zero Features to Weights Ratio & 1.00 & 0.248 & 0.095 & 0.043 & 0.014 & 0.001 & -0.014 & 0.98 \\
Shared Most Common Target & 1.00 & 0.249 & 0.132 & 0.072 & 0.031 & 0.018 & 0.362 & 0.71 \\
Shared Most Common Tactic & 1.00 & 0.392 & 0.177 & 0.078 & 0.033 & 0.018 & 0.0433 & 0.66 \\
Shared Most Common Weapon & 1.00 & 0.398 & 0.179 & 0.072 & 0.031 & 0.018 & 0.590 & 0.55 \\
Shared Ideology & 1.00 & 0.247 & 0.115 & 0.045 & 0.040 & 0.006 & 0.010 & 0.99 \\
Shared Most Common Region & 1.00 & 0.268 & 0.124 & 0.047 & 0.013 & 0.001 & 0.548 & 0.58 \\ \hline
\end{tabular}
\end{table}

\begin{table}[!h]
\caption{MCMC Diagnostics - ERGM Year 2000}
\setlength{\tabcolsep}{3pt}
\centering
\footnotesize
\begin{tabular}{lllllll|cc}
\hline
 & \multicolumn{6}{c}{\begin{tabular}[c]{@{}c@{}}Sample statistic \\ auto-correlation\end{tabular}} & \multicolumn{2}{c}{\begin{tabular}[c]{@{}c@{}}Geweke\\ Diagnostic\end{tabular}} \\ \hline
Covariate & \multicolumn{1}{c}{\begin{tabular}[c]{@{}c@{}}Lag \\ 0\end{tabular}} & \multicolumn{1}{c}{\begin{tabular}[c]{@{}c@{}}Lag \\ 1024\end{tabular}} & \multicolumn{1}{c}{\begin{tabular}[c]{@{}c@{}}Lag \\ 2048\end{tabular}} & \multicolumn{1}{c}{\begin{tabular}[c]{@{}c@{}}Lag \\ 3072\end{tabular}} & \multicolumn{1}{c}{\begin{tabular}[c]{@{}c@{}}Lag\\ 4096\end{tabular}} & \multicolumn{1}{c|}{\begin{tabular}[c]{@{}c@{}}Lag\\ 5120\end{tabular}} & \multicolumn{1}{l}{Statistic} & \multicolumn{1}{l}{p-value} \\ \hline
Sum of Features Weights & 1.00 & 0.012 & -0.010 & -0.012 & -0.003 & 0.001 & -0.030 & 0.76 \\
Diff in N of non-zero Features & 1.00 & 0.021 & -0.012 & 0.009 & 0.003 & -0.015 & -0.159 & 0.87 \\
Diff in Non-zero Features to Weights Ratio & 1.00 & 0.010 & -0.012 & 0.010 & -0.009 & -0.024 & -0.266 & 0.78 \\
Shared Most Common Target & 1.00 & 0.033 & -0.038 & 0.005 & -0.001 & -0.015 & 0.322 & 0.74 \\
Shared Most Common Tactic & 1.00 & 0.024 & -0.033 & -0.024 & 0.008 & 0.003 & -0.214 & 0.83 \\
Shared Most Common Weapon & 1.00 & 0.016 & -0.013 & -0.017 & -0.001 & 0.008 & -0.492 & 0.62 \\
Shared Ideology & 1.00 & 0.017 & -0.011 & -0.025 & 0.023 & -0.029 & -0.040 & 0.96 \\
Shared Most Common Region & 1.00 & 0.034 & -0.023 & -0.017 & 0.003 & -0.011 & -0.076 & 0.93 \\ \hline
\end{tabular}
\end{table}

\begin{table}[!h]
\caption{MCMC Diagnostics - ERGM Year 2001}
\setlength{\tabcolsep}{3pt}
\centering
\footnotesize
\begin{tabular}{lllllll|cc}
\hline
 & \multicolumn{6}{c}{\begin{tabular}[c]{@{}c@{}}Sample statistic \\ auto-correlation\end{tabular}} & \multicolumn{2}{c}{\begin{tabular}[c]{@{}c@{}}Geweke\\ Diagnostic\end{tabular}} \\ \hline
Covariate & \multicolumn{1}{c}{\begin{tabular}[c]{@{}c@{}}Lag \\ 0\end{tabular}} & \multicolumn{1}{c}{\begin{tabular}[c]{@{}c@{}}Lag \\ 1024\end{tabular}} & \multicolumn{1}{c}{\begin{tabular}[c]{@{}c@{}}Lag \\ 2048\end{tabular}} & \multicolumn{1}{c}{\begin{tabular}[c]{@{}c@{}}Lag \\ 3072\end{tabular}} & \multicolumn{1}{c}{\begin{tabular}[c]{@{}c@{}}Lag\\ 4096\end{tabular}} & \multicolumn{1}{c|}{\begin{tabular}[c]{@{}c@{}}Lag\\ 5120\end{tabular}} & \multicolumn{1}{l}{Statistic} & \multicolumn{1}{l}{p-value} \\ \hline
Sum of Features Weights & 1.00 & 0.157 & 0.043 & 0.021 & 0.023 & 0.017 & -0.596 & 0.55 \\
Diff in N of non-zero Features & 1.00 & 0.218 & 0.079 & 0.038 & 0.021 & 0.022 & 0.064 & 0.51 \\
Diff in Non-zero Features to Weights Ratio & 1.00 & 0.189 & 0.073 & 0.026 & 0.028 & 0.011 & 1.067 & 0.28 \\
Shared Most Common Target & 1.00 & 0.263 & 0.101 & 0.046 & 0.018 & 0.012 & -0.021 & 0.98 \\
Shared Most Common Tactic & 1.00 & 0.305 & 0.128 & 0.055 & 0.033 & 0.028 & 0.307 & 0.75 \\
Shared Most Common Weapon & 1.00 & 0.304 & 0.123 & 0.050 & 0.036 & 0.026 & 0.185 & 0.85 \\
Shared Ideology & 1.00 & 0.275 & 0.131 & 0.047 & 0.020 & 0.007 & 0.158 & 0.87 \\
Shared Most Common Region & 1.00 & 0.304 & 0.126 & 0.043 & 0.012 & -0.002 & -0.003 & 0.99 \\ \hline
\end{tabular}
\end{table}

\begin{table}[!h]
\caption{MCMC Diagnostics - ERGM Year 2002}
\setlength{\tabcolsep}{3pt}
\centering
\footnotesize
\begin{tabular}{lllllll|cc}
\hline
 & \multicolumn{6}{c}{\begin{tabular}[c]{@{}c@{}}Sample statistic \\ auto-correlation\end{tabular}} & \multicolumn{2}{c}{\begin{tabular}[c]{@{}c@{}}Geweke\\ Diagnostic\end{tabular}} \\ \hline
Covariate & \multicolumn{1}{c}{\begin{tabular}[c]{@{}c@{}}Lag \\ 0\end{tabular}} & \multicolumn{1}{c}{\begin{tabular}[c]{@{}c@{}}Lag \\ 1024\end{tabular}} & \multicolumn{1}{c}{\begin{tabular}[c]{@{}c@{}}Lag \\ 2048\end{tabular}} & \multicolumn{1}{c}{\begin{tabular}[c]{@{}c@{}}Lag \\ 3072\end{tabular}} & \multicolumn{1}{c}{\begin{tabular}[c]{@{}c@{}}Lag\\ 4096\end{tabular}} & \multicolumn{1}{c|}{\begin{tabular}[c]{@{}c@{}}Lag\\ 5120\end{tabular}} & \multicolumn{1}{l}{Statistic} & \multicolumn{1}{l}{p-value} \\ \hline
Sum of Features Weights & 1.000 & 0.033 & 0.005 & -0.003 & 0.029 & -0.002 & -1.744 & 0.08 \\
Diff in N of non-zero Features & 1.000 & 0.048 & -0.002 & 0.008 & 0.004 & -0.010 & -1.398 & 0.16 \\
Diff in Non-zero Features to Weights Ratio & 1.000 & 0.012 & -0.005 & -0.019 & 0.009 & 0.016 & -0.256 & 0.79 \\
Shared Most Common Target & 1.000 & 0.046 & -0.011 & 0.009 & 0.020 & 0.012 & -1.691 & 0.09 \\
Shared Most Common Tactic & 1.000 & 0.050 & -0.011 & 0.005 & 0.005 & -0.002 & -1.220 & 0.22 \\
Shared Most Common Weapon & 1.000 & 0.051 & -0.010 & 0.003 & 0.006 & -0.003 & -1.048 & 0.29 \\
Shared Ideology & 1.000 & 0.052 & -0.004 & 0.004 & 0.015 & 0.015 & -1.093 & 0.27 \\
Shared Most Common Region & 1.000 & 0.059 & -0.013 & 0.009 & 0.008 & 0.004 & 0.824 & 0.40 \\ \hline
\end{tabular}
\end{table}

\begin{table}[!h]
\caption{MCMC Diagnostics - ERGM Year 2003}
\setlength{\tabcolsep}{3pt}
\centering
\footnotesize
\begin{tabular}{lllllll|cc}
\hline
 & \multicolumn{6}{c}{\begin{tabular}[c]{@{}c@{}}Sample statistic \\ auto-correlation\end{tabular}} & \multicolumn{2}{c}{\begin{tabular}[c]{@{}c@{}}Geweke\\ Diagnostic\end{tabular}} \\ \hline
Covariate & \multicolumn{1}{c}{\begin{tabular}[c]{@{}c@{}}Lag \\ 0\end{tabular}} & \multicolumn{1}{c}{\begin{tabular}[c]{@{}c@{}}Lag \\ 1024\end{tabular}} & \multicolumn{1}{c}{\begin{tabular}[c]{@{}c@{}}Lag \\ 2048\end{tabular}} & \multicolumn{1}{c}{\begin{tabular}[c]{@{}c@{}}Lag \\ 3072\end{tabular}} & \multicolumn{1}{c}{\begin{tabular}[c]{@{}c@{}}Lag\\ 4096\end{tabular}} & \multicolumn{1}{c|}{\begin{tabular}[c]{@{}c@{}}Lag\\ 5120\end{tabular}} & \multicolumn{1}{l}{Statistic} & \multicolumn{1}{l}{p-value} \\ \hline
Sum of Features Weights & 1.000 & 0.361 & 0.151 & 0.085 & 0.046 & 0.016 & 0.073 & 0.94 \\
Diff in N of non-zero Features & 1.000 & 0.380 & 0.172 & 0.098 & 0.062 & 0.036 & -0.211 & 0.83 \\
Diff in Non-zero Features to Weights Ratio & 1.000 & 0.348 & 0.172 & 0.107 & 0.067 & 0.023 & 0.243 & 0.80 \\
Shared Most Common Target & 1.000 & 0.432 & 0.221 & 0.138 & 0.082 & 0.043 & -0.060 & 0.95 \\
Shared Most Common Tactic & 1.000 & 0.460 & 0.230 & 0.126 & 0.071 & 0.038 & 0.098 & 0.92 \\
Shared Most Common Weapon & 1.000 & 0.472 & 0.231 & 0.128 & 0.075 & 0.039 & 0.181 & 0.85 \\
Shared Ideology & 1.000 & 0.423 & 0.198 & 0.101 & 0.071 & 0.037 & -0.031 & 0.97 \\
Shared Most Common Region & 1.000 & 0.439 & 0.204 & 0.101 & 0.062 & 0.036 & -0.37 & 0.70 \\ \hline
\end{tabular}
\end{table}

\begin{table}[!h]
\caption{MCMC Diagnostics - ERGM Year 2004}
\setlength{\tabcolsep}{3pt}
\centering
\footnotesize
\begin{tabular}{lllllll|cc}
\hline
 & \multicolumn{6}{c}{\begin{tabular}[c]{@{}c@{}}Sample statistic \\ auto-correlation\end{tabular}} & \multicolumn{2}{c}{\begin{tabular}[c]{@{}c@{}}Geweke\\ Diagnostic\end{tabular}} \\ \hline
Covariate & \multicolumn{1}{c}{\begin{tabular}[c]{@{}c@{}}Lag \\ 0\end{tabular}} & \multicolumn{1}{c}{\begin{tabular}[c]{@{}c@{}}Lag \\ 1024\end{tabular}} & \multicolumn{1}{c}{\begin{tabular}[c]{@{}c@{}}Lag \\ 2048\end{tabular}} & \multicolumn{1}{c}{\begin{tabular}[c]{@{}c@{}}Lag \\ 3072\end{tabular}} & \multicolumn{1}{c}{\begin{tabular}[c]{@{}c@{}}Lag\\ 4096\end{tabular}} & \multicolumn{1}{c|}{\begin{tabular}[c]{@{}c@{}}Lag\\ 5120\end{tabular}} & \multicolumn{1}{l}{Statistic} & \multicolumn{1}{l}{p-value} \\ \hline
Sum of Features Weights & 1.000 & 0.170 & 0.059 & 0.011 & 0.020 & 0.016 & 0.277 & 0.78 \\
Diff in N of non-zero Features & 1.000 & 0.138 & 0.020 & -0.002 & 0.006 & -0.001 & 0.596 & 0.55 \\
Diff in Non-zero Features to Weights Ratio & 1.000 & 0.143 & 0.040 & 0.007 & 0.011 & 0.001 & 0.522 & 0.60 \\
Shared Most Common Target & 1.000 & 0.199 & 0.072 & 0.035 & 0.015 & 0.018 & 0.255 & 0.79 \\
Shared Most Common Tactic & 1.000 & 0.238 & 0.071 & 0.031 & 0.022 & 0.013 & 0.805 & 0.42 \\
Shared Most Common Weapon & 1.000 & 0.244 & 0.077 & 0.033 & 0.024 & 0.014 & 0.848 & 0.39 \\
Shared Ideology & 1.000 & 0.210 & 0.070 & 0.022 & 0.010 & 0.004 & 0.847 & 0.39 \\
Shared Most Common Region & 1.000 & 0.224 & 0.070 & 0.016 & 0.019 & 0.008 & 0.293 & 0.76 \\ \hline
\end{tabular}
\end{table}

\begin{table}[!h]
\caption{MCMC Diagnostics - ERGM Year 2005}
\setlength{\tabcolsep}{3pt}
\centering
\footnotesize
\begin{tabular}{lllllll|cc}
\hline
 & \multicolumn{6}{c}{\begin{tabular}[c]{@{}c@{}}Sample statistic \\ auto-correlation\end{tabular}} & \multicolumn{2}{c}{\begin{tabular}[c]{@{}c@{}}Geweke\\ Diagnostic\end{tabular}} \\ \hline
Covariate & \multicolumn{1}{c}{\begin{tabular}[c]{@{}c@{}}Lag \\ 0\end{tabular}} & \multicolumn{1}{c}{\begin{tabular}[c]{@{}c@{}}Lag \\ 1024\end{tabular}} & \multicolumn{1}{c}{\begin{tabular}[c]{@{}c@{}}Lag \\ 2048\end{tabular}} & \multicolumn{1}{c}{\begin{tabular}[c]{@{}c@{}}Lag \\ 3072\end{tabular}} & \multicolumn{1}{c}{\begin{tabular}[c]{@{}c@{}}Lag\\ 4096\end{tabular}} & \multicolumn{1}{c|}{\begin{tabular}[c]{@{}c@{}}Lag\\ 5120\end{tabular}} & \multicolumn{1}{l}{Statistic} & \multicolumn{1}{l}{p-value} \\ \hline
Sum of Features Weights & 1.000 & 0.201 & 0.095 & 0.023 & 0.016 & -0.012 & 0.449 & 0.65 \\
Diff in N of non-zero Features & 1.000 & 0.236 & 0.076 & 0.024 & 0.006 & -0.028 & 0.182 & 0.85 \\
Diff in Non-zero Features to Weights Ratio & 1.000 & 0.245 & 0.086 & 0.039 & -0.001 & -0.039 & 0.313 & 0.75 \\
Shared Most Common Target & 1.000 & 0.343 & 0.105 & 0.023 & -0.006 & -0.025 & 0.210 & 0.83 \\
Shared Most Common Tactic & 1.000 & 0.353 & 0.132 & 0.039 & 0.003 & -0.029 & 0.377 & 0.70 \\
Shared Most Common Weapon & 1.000 & 0.337 & 0.126 & 0.039 & 0.003 & -0.031 & 0.072 & 0.94 \\
Shared Ideology & 1.000 & 0.336 & 0.130 & 0.036 & 0.007 & -0.024 & 0.781 & 0.43 \\
Shared Most Common Region & 1.000 & 0.338 & 0.127 & 0.033 & 0.001 & -0.029 & 0.017 & 0.98 \\ \hline
\end{tabular}
\end{table}

\begin{table}[!h]
\caption{MCMC Diagnostics - ERGM Year 2006}
\setlength{\tabcolsep}{3pt}
\centering
\footnotesize
\begin{tabular}{lllllll|cc}
\hline
 & \multicolumn{6}{c}{\begin{tabular}[c]{@{}c@{}}Sample statistic \\ auto-correlation\end{tabular}} & \multicolumn{2}{c}{\begin{tabular}[c]{@{}c@{}}Geweke\\ Diagnostic\end{tabular}} \\ \hline
Covariate & \multicolumn{1}{c}{\begin{tabular}[c]{@{}c@{}}Lag \\ 0\end{tabular}} & \multicolumn{1}{c}{\begin{tabular}[c]{@{}c@{}}Lag \\ 1024\end{tabular}} & \multicolumn{1}{c}{\begin{tabular}[c]{@{}c@{}}Lag \\ 2048\end{tabular}} & \multicolumn{1}{c}{\begin{tabular}[c]{@{}c@{}}Lag \\ 3072\end{tabular}} & \multicolumn{1}{c}{\begin{tabular}[c]{@{}c@{}}Lag\\ 4096\end{tabular}} & \multicolumn{1}{c|}{\begin{tabular}[c]{@{}c@{}}Lag\\ 5120\end{tabular}} & \multicolumn{1}{l}{Statistic} & \multicolumn{1}{l}{p-value} \\ \hline
Sum of Features Weights & 1.000 & 0.025 & 0.004 & -0.005 & -0.008 & 0.021 & 0.385 & 0.69 \\
Diff in N of non-zero Features & 1.000 & 0.051 & -0.002 & -0.014 & -0.009 & 0.026 & -0.316 & 0.75 \\
Diff in Non-zero Features to Weights Ratio & 1.000 & 0.050 & -0.015 & 0.004 & -0.002 & 0.015 & -0.756 & 0.44 \\
Shared Most Common Target & 1.000 & 0.077 & -0.019 & -0.017 & -0.016 & -0.003 & 0.527 & 0.59 \\
Shared Most Common Tactic & 1.000 & 0.077 & 0.002 & -0.012 & 0.007 & 0.017 & 0.190 & 0.84 \\
Shared Most Common Weapon & 1.000 & 0.082 & -0.001 & -0.008 & 0.006 & 0.018 & 0.388 & 0.69 \\
Shared Ideology & 1.000 & 0.075 & -0.001 & -0.013 & 0.002 & 0.019 & 0.677 & 0.49 \\
Shared Most Common Region & 1.000 & 0.066 & -0.003 & -0.017 & 0.015 & 0.020 & 0.129 & 0.89 \\ \hline
\end{tabular}
\end{table}

\begin{table}[!h]
\caption{MCMC Diagnostics - ERGM Year 2007}
\setlength{\tabcolsep}{3pt}
\centering
\footnotesize
\begin{tabular}{lllllll|cc}
\hline
 & \multicolumn{6}{c}{\begin{tabular}[c]{@{}c@{}}Sample statistic \\ auto-correlation\end{tabular}} & \multicolumn{2}{c}{\begin{tabular}[c]{@{}c@{}}Geweke\\ Diagnostic\end{tabular}} \\ \hline
Covariate & \multicolumn{1}{c}{\begin{tabular}[c]{@{}c@{}}Lag \\ 0\end{tabular}} & \multicolumn{1}{c}{\begin{tabular}[c]{@{}c@{}}Lag \\ 1024\end{tabular}} & \multicolumn{1}{c}{\begin{tabular}[c]{@{}c@{}}Lag \\ 2048\end{tabular}} & \multicolumn{1}{c}{\begin{tabular}[c]{@{}c@{}}Lag \\ 3072\end{tabular}} & \multicolumn{1}{c}{\begin{tabular}[c]{@{}c@{}}Lag\\ 4096\end{tabular}} & \multicolumn{1}{c|}{\begin{tabular}[c]{@{}c@{}}Lag\\ 5120\end{tabular}} & \multicolumn{1}{l}{Statistic} & \multicolumn{1}{l}{p-value} \\ \hline
Sum of Features Weights & 1.000 & 0.291 & 0.214 & 0.164 & 0.117 & 0.087 & 0.878 & 0.37 \\
Diff in N of non-zero Features & 1.000 & 0.591 & 0.427 & 0.319 & 0.228 & 0.166 & 0.794 & 0.42 \\
Diff in Non-zero Features to Weights Ratio & 1.000 & 0.661 & 0.476 & 0.344 & 0.253 & 0.184 & 0.750 & 0.45 \\
Shared Most Common Target & 1.000 & 0.661 & 0.475 & 0.350 & 0.259 & 0.192 & 0.773 & 0.43 \\
Shared Most Common Tactic & 1.000 & 0.659 & 0.473 & 0.347 & 0.252 & 0.185 & 0.656 & 0.51 \\
Shared Most Common Weapon & 1.000 & 0.678 & 0.489 & 0.359 & 0.261 & 0.192 & 0.660 & 0.50 \\
Shared Ideology & 1.000 & 0.683 & 0.497 & 0.367 & 0.264 & 0.192 & 0.679 & 0.49 \\
Shared Most Common Region & 1.000 & 0.674 & 0.485 & 0.352 & 0.258 & 0.190 & 0.697 & 0.48 \\ \hline
\end{tabular}
\end{table}

\begin{table}[!h]
\caption{MCMC Diagnostics - ERGM Year 2008}
\setlength{\tabcolsep}{3pt}
\centering
\footnotesize
\begin{tabular}{lllllll|cc}
\hline
 & \multicolumn{6}{c}{\begin{tabular}[c]{@{}c@{}}Sample statistic \\ auto-correlation\end{tabular}} & \multicolumn{2}{c}{\begin{tabular}[c]{@{}c@{}}Geweke\\ Diagnostic\end{tabular}} \\ \hline
Covariate & \multicolumn{1}{c}{\begin{tabular}[c]{@{}c@{}}Lag \\ 0\end{tabular}} & \multicolumn{1}{c}{\begin{tabular}[c]{@{}c@{}}Lag \\ 1024\end{tabular}} & \multicolumn{1}{c}{\begin{tabular}[c]{@{}c@{}}Lag \\ 2048\end{tabular}} & \multicolumn{1}{c}{\begin{tabular}[c]{@{}c@{}}Lag \\ 3072\end{tabular}} & \multicolumn{1}{c}{\begin{tabular}[c]{@{}c@{}}Lag\\ 4096\end{tabular}} & \multicolumn{1}{c|}{\begin{tabular}[c]{@{}c@{}}Lag\\ 5120\end{tabular}} & \multicolumn{1}{l}{Statistic} & \multicolumn{1}{l}{p-value} \\ \hline
Sum of Features Weights & 1.000 & 0.203 & 0.060 & 0.030 & 0.027 & -0.027 & -0.289 & 0.77 \\
Diff in N of non-zero Features & 1.000 & 0.161 & 0.060 & 0.015 & 0.021 & -0.018 & -0.438 & 0.66 \\
Diff in Non-zero Features to Weights Ratio & 1.000 & 0.183 & 0.078 & 0.020 & 0.035 & 0.015 & -0.053 & 0.95 \\
Shared Most Common Target & 1.000 & 0.241 & 0.107 & 0.029 & 0.025 & -0.007 & 0.045 & 0.96 \\
Shared Most Common Tactic & 1.000 & 0.302 & 0.114 & 0.037 & 0.038 & -0.015 & 0.029 & 0.97 \\
Shared Most Common Weapon & 1.000 & 0.284 & 0.102 & 0.031 & 0.033 & -0.022 & -0.075 & 0.93 \\
Shared Ideology & 1.000 & 0.288 & 0.111 & 0.011 & 0.031 & -0.004 & 0.137 & 0.89 \\
Shared Most Common Region & 1.000 & 0.312 & 0.135 & 0.041 & 0.038 & 0.003 & -0.046 & 0.96 \\ \hline
\end{tabular}
\end{table}

\begin{table}[!h]
\caption{MCMC Diagnostics - ERGM Year 2009}
\setlength{\tabcolsep}{3pt}
\centering
\footnotesize
\begin{tabular}{lllllll|cc}
\hline
 & \multicolumn{6}{c}{\begin{tabular}[c]{@{}c@{}}Sample statistic \\ auto-correlation\end{tabular}} & \multicolumn{2}{c}{\begin{tabular}[c]{@{}c@{}}Geweke\\ Diagnostic\end{tabular}} \\ \hline
Covariate & \multicolumn{1}{c}{\begin{tabular}[c]{@{}c@{}}Lag \\ 0\end{tabular}} & \multicolumn{1}{c}{\begin{tabular}[c]{@{}c@{}}Lag \\ 1024\end{tabular}} & \multicolumn{1}{c}{\begin{tabular}[c]{@{}c@{}}Lag \\ 2048\end{tabular}} & \multicolumn{1}{c}{\begin{tabular}[c]{@{}c@{}}Lag \\ 3072\end{tabular}} & \multicolumn{1}{c}{\begin{tabular}[c]{@{}c@{}}Lag\\ 4096\end{tabular}} & \multicolumn{1}{c|}{\begin{tabular}[c]{@{}c@{}}Lag\\ 5120\end{tabular}} & \multicolumn{1}{l}{Statistic} & \multicolumn{1}{l}{p-value} \\ \hline
Sum of Features Weights & 1.000 & 0.021 & -0.023 & -0.012 & -0.004 & -0.019 & -1.068 & 0.28 \\
Diff in N of non-zero Features & 1.000 & 0.052 & 0.027 & -0.006 & 0.020 & -0.008 & -1.942 & 0.05 \\
Diff in Non-zero Features to Weights Ratio & 1.000 & 0.094 & 0.040 & 0.001 & 0.032 & 0.008 & -1.659 & 0.09 \\
Shared Most Common Target & 1.000 & 0.081 & 0.023 & -0.008 & 0.015 & 0.023 & -0.396 & 0.69 \\
Shared Most Common Tactic & 1.000 & 0.148 & 0.049 & 0.004 & 0.038 & 0.024 & -1.189 & 0.23 \\
Shared Most Common Weapon & 1.000 & 0.154 & 0.047 & 0.015 & 0.049 & 0.022 & -0.907 & 0.36 \\
Shared Ideology & 1.000 & 0.149 & 0.048 & 0.014 & 0.037 & 0.018 & 1.355 & 0.17 \\
Shared Most Common Region & 1.000 & 0.145 & 0.045 & 0.012 & 0.035 & 0.007 & -0.373 & 0.70 \\ \hline
\end{tabular}
\end{table}

\begin{table}[!h]
\caption{MCMC Diagnostics - ERGM Year 2010}
\setlength{\tabcolsep}{3pt}
\centering
\footnotesize
\begin{tabular}{lllllll|cc}
\hline
 & \multicolumn{6}{c}{\begin{tabular}[c]{@{}c@{}}Sample statistic \\ auto-correlation\end{tabular}} & \multicolumn{2}{c}{\begin{tabular}[c]{@{}c@{}}Geweke\\ Diagnostic\end{tabular}} \\ \hline
Covariate & \multicolumn{1}{c}{\begin{tabular}[c]{@{}c@{}}Lag \\ 0\end{tabular}} & \multicolumn{1}{c}{\begin{tabular}[c]{@{}c@{}}Lag \\ 1024\end{tabular}} & \multicolumn{1}{c}{\begin{tabular}[c]{@{}c@{}}Lag \\ 2048\end{tabular}} & \multicolumn{1}{c}{\begin{tabular}[c]{@{}c@{}}Lag \\ 3072\end{tabular}} & \multicolumn{1}{c}{\begin{tabular}[c]{@{}c@{}}Lag\\ 4096\end{tabular}} & \multicolumn{1}{c|}{\begin{tabular}[c]{@{}c@{}}Lag\\ 5120\end{tabular}} & \multicolumn{1}{l}{Statistic} & \multicolumn{1}{l}{p-value} \\ \hline
Sum of Features Weights & 1.000 & 0.047 & -0.017 & 0.032 & -0.012 & 0.001 & 0.328 & 0.74 \\
Diff in N of non-zero Features & 1.000 & 0.103 & 0.050 & 0.024 & -0.010 & -0.025 & -0.882 & 0.37 \\
Diff in Non-zero Features to Weights Ratio & 1.000 & 0.092 & 0.051 & 0.007 & -0.017 & -0.026 & -0.933 & 0.34 \\
Shared Most Common Target & 1.000 & 0.090 & 0.034 & 0.002 & 0.002 & 0.007 & 0.030 & 0.97 \\
Shared Most Common Tactic & 1.000 & 0.132 & 0.033 & 0.012 & -0.010 & -0.008 & -0.818 & 0.41 \\
Shared Most Common Weapon & 1.000 & 0.149 & 0.046 & 0.009 & 0.002 & -0.007 & -0.834 & 0.40 \\
Shared Ideology & 1.000 & 0.147 & 0.034 & 0.000 & 0.002 & -0.002 & -0.466 & 0.64 \\
Shared Most Common Region & 1.000 & 0.208 & 0.066 & 0.003 & -0.007 & 0.006 & -1.376 & 0.16 \\ \hline
\end{tabular}
\end{table}

\begin{table}[!h]
\caption{MCMC Diagnostics - ERGM Year 2011}
\setlength{\tabcolsep}{3pt}
\centering
\footnotesize
\begin{tabular}{lllllll|cc}
\hline
 & \multicolumn{6}{c}{\begin{tabular}[c]{@{}c@{}}Sample statistic \\ auto-correlation\end{tabular}} & \multicolumn{2}{c}{\begin{tabular}[c]{@{}c@{}}Geweke\\ Diagnostic\end{tabular}} \\ \hline
Covariate & \multicolumn{1}{c}{\begin{tabular}[c]{@{}c@{}}Lag \\ 0\end{tabular}} & \multicolumn{1}{c}{\begin{tabular}[c]{@{}c@{}}Lag \\ 1024\end{tabular}} & \multicolumn{1}{c}{\begin{tabular}[c]{@{}c@{}}Lag \\ 2048\end{tabular}} & \multicolumn{1}{c}{\begin{tabular}[c]{@{}c@{}}Lag \\ 3072\end{tabular}} & \multicolumn{1}{c}{\begin{tabular}[c]{@{}c@{}}Lag\\ 4096\end{tabular}} & \multicolumn{1}{c|}{\begin{tabular}[c]{@{}c@{}}Lag\\ 5120\end{tabular}} & \multicolumn{1}{l}{Statistic} & \multicolumn{1}{l}{p-value} \\ \hline
Sum of Features Weights & 1.000 & 0.038 & 0.001 & 0.017 & 0.002 & 0.047 & -0.729 & 0.46 \\
Diff in N of non-zero Features & 1.000 & 0.040 & 0.018 & 0.007 & -0.007 & 0.024 & -0.753 & 0.45 \\
Diff in Non-zero Features to Weights Ratio & 1.000 & 0.062 & 0.007 & 0.027 & 0.000 & -0.019 & -1.259 & 0.20 \\
Shared Most Common Target & 1.000 & 0.084 & 0.008 & 0.028 & 0.009 & 0.012 & -0.975 & 0.32 \\
Shared Most Common Tactic & 1.000 & 0.133 & 0.060 & 0.010 & 0.018 & -0.001 & -0.778 & 0.43 \\
Shared Most Common Weapon & 1.000 & 0.143 & 0.062 & 0.022 & 0.021 & -0.007 & -0.979 & 0.32 \\
Shared Ideology & 1.000 & 0.101 & 0.034 & 0.027 & 0.009 & 0.001 & -1.002 & 0.31 \\
Shared Most Common Region & 1.000 & 0.092 & 0.010 & 0.032 & 0.008 & -0.017 & 0.226 & 0.82 \\ \hline
\end{tabular}
\end{table}

\begin{table}[!h]
\caption{MCMC Diagnostics - ERGM Year 2012}
\setlength{\tabcolsep}{3pt}
\centering
\footnotesize
\begin{tabular}{lllllll|cc}
\hline
 & \multicolumn{6}{c}{\begin{tabular}[c]{@{}c@{}}Sample statistic \\ auto-correlation\end{tabular}} & \multicolumn{2}{c}{\begin{tabular}[c]{@{}c@{}}Geweke\\ Diagnostic\end{tabular}} \\ \hline
Covariate & \multicolumn{1}{c}{\begin{tabular}[c]{@{}c@{}}Lag \\ 0\end{tabular}} & \multicolumn{1}{c}{\begin{tabular}[c]{@{}c@{}}Lag \\ 1024\end{tabular}} & \multicolumn{1}{c}{\begin{tabular}[c]{@{}c@{}}Lag \\ 2048\end{tabular}} & \multicolumn{1}{c}{\begin{tabular}[c]{@{}c@{}}Lag \\ 3072\end{tabular}} & \multicolumn{1}{c}{\begin{tabular}[c]{@{}c@{}}Lag\\ 4096\end{tabular}} & \multicolumn{1}{c|}{\begin{tabular}[c]{@{}c@{}}Lag\\ 5120\end{tabular}} & \multicolumn{1}{l}{Statistic} & \multicolumn{1}{l}{p-value} \\ \hline
Sum of Features Weights & 1.000 & 0.185 & 0.121 & 0.073 & 0.083 & 0.052 & -1.220 & 0.22 \\
Diff in N of non-zero Features & 1.000 & 0.414 & 0.236 & 0.165 & 0.123 & 0.106 & -0.526 & 0.59 \\
Diff in Non-zero Features to Weights Ratio & 1.000 & 0.384 & 0.223 & 0.136 & 0.111 & 0.096 & -0.784 & 0.43 \\
Shared Most Common Target & 1.000 & 0.479 & 0.270 & 0.160 & 0.127 & 0.104 & -0.641 & 0.52 \\
Shared Most Common Tactic & 1.000 & 0.534 & 0.311 & 0.191 & 0.128 & 0.105 & -0.775 & 0.43 \\
Shared Most Common Weapon & 1.000 & 0.550 & 0.323 & 0.200 & 0.135 & 0.111 & -0.791 & 0.42 \\
Shared Ideology & 1.000 & 0.502 & 0.292 & 0.188 & 0.128 & 0.111 & -0.884 & 0.37 \\
Shared Most Common Region & 1.000 & 0.514 & 0.298 & 0.184 & 0.128 & 0.113 & -0.638 & 0.52 \\ \hline
\end{tabular}
\end{table}

\begin{table}[!h]
\caption{MCMC Diagnostics - ERGM Year 2013}
\setlength{\tabcolsep}{3pt}
\centering
\footnotesize
\begin{tabular}{lllllll|cc}
\hline
 & \multicolumn{6}{c}{\begin{tabular}[c]{@{}c@{}}Sample statistic \\ auto-correlation\end{tabular}} & \multicolumn{2}{c}{\begin{tabular}[c]{@{}c@{}}Geweke\\ Diagnostic\end{tabular}} \\ \hline
Covariate & \multicolumn{1}{c}{\begin{tabular}[c]{@{}c@{}}Lag \\ 0\end{tabular}} & \multicolumn{1}{c}{\begin{tabular}[c]{@{}c@{}}Lag \\ 1024\end{tabular}} & \multicolumn{1}{c}{\begin{tabular}[c]{@{}c@{}}Lag \\ 2048\end{tabular}} & \multicolumn{1}{c}{\begin{tabular}[c]{@{}c@{}}Lag \\ 3072\end{tabular}} & \multicolumn{1}{c}{\begin{tabular}[c]{@{}c@{}}Lag\\ 4096\end{tabular}} & \multicolumn{1}{c|}{\begin{tabular}[c]{@{}c@{}}Lag\\ 5120\end{tabular}} & \multicolumn{1}{l}{Statistic} & \multicolumn{1}{l}{p-value} \\ \hline
Sum of Features Weights & 1.000 & 0.189 & 0.080 & 0.048 & 0.036 & 0.024 & 0.360 & 0.71 \\
Diff in N of non-zero Features & 1.000 & 0.313 & 0.151 & 0.073 & 0.056 & 0.024 & -0.090 & 0.92 \\
Diff in Non-zero Features to Weights Ratio & 1.000 & 0.278 & 0.123 & 0.030 & 0.031 & 0.002 & 0.231 & 0.81 \\
Shared Most Common Target & 1.000 & 0.410 & 0.201 & 0.110 & 0.056 & 0.031 & -0.594 & 0.55 \\
Shared Most Common Tactic & 1.000 & 0.440 & 0.250 & 0.138 & 0.078 & 0.047 & 0.239 & 0.81 \\
Shared Most Common Weapon & 1.000 & 0.449 & 0.249 & 0.136 & 0.079 & 0.046 & 0.196 & 0.84 \\
Shared Ideology & 1.000 & 0.401 & 0.205 & 0.122 & 0.072 & 0.028 & -0.130 & 0.89 \\
Shared Most Common Region & 1.000 & 0.442 & 0.216 & 0.105 & 0.062 & 0.033 & -0.075 & 0.93 \\ \hline
\end{tabular}
\end{table}

\begin{table}[!h]
\caption{MCMC Diagnostics - ERGM Year 2014}
\setlength{\tabcolsep}{3pt}
\centering
\footnotesize
\begin{tabular}{lllllll|cc}
\hline
 & \multicolumn{6}{c}{\begin{tabular}[c]{@{}c@{}}Sample statistic \\ auto-correlation\end{tabular}} & \multicolumn{2}{c}{\begin{tabular}[c]{@{}c@{}}Geweke\\ Diagnostic\end{tabular}} \\ \hline
Covariate & \multicolumn{1}{c}{\begin{tabular}[c]{@{}c@{}}Lag \\ 0\end{tabular}} & \multicolumn{1}{c}{\begin{tabular}[c]{@{}c@{}}Lag \\ 1024\end{tabular}} & \multicolumn{1}{c}{\begin{tabular}[c]{@{}c@{}}Lag \\ 2048\end{tabular}} & \multicolumn{1}{c}{\begin{tabular}[c]{@{}c@{}}Lag \\ 3072\end{tabular}} & \multicolumn{1}{c}{\begin{tabular}[c]{@{}c@{}}Lag\\ 4096\end{tabular}} & \multicolumn{1}{c|}{\begin{tabular}[c]{@{}c@{}}Lag\\ 5120\end{tabular}} & \multicolumn{1}{l}{Statistic} & \multicolumn{1}{l}{p-value} \\ \hline
Sum of Features Weights & 1.000 & 0.051 & -0.004 & -0.004 & 0.009 & -0.001 & 0.193 & 0.84 \\
Diff in N of non-zero Features & 1.000 & 0.120 & 0.063 & 0.017 & 0.021 & 0.026 & 0.031 & 0.97 \\
Diff in Non-zero Features to Weights Ratio & 1.000 & 0.133 & 0.057 & 0.023 & 0.001 & 0.007 & 0.774 & 0.43 \\
Shared Most Common Target & 1.000 & 0.288 & 0.115 & 0.045 & 0.014 & 0.007 & 1.687 & 0.09 \\
Shared Most Common Tactic & 1.000 & 0.334 & 0.147 & 0.065 & 0.040 & 0.032 & 1.841 & 0.06 \\
Shared Most Common Weapon & 1.000 & 0.329 & 0.148 & 0.066 & 0.042 & 0.029 & 1.757 & 0.07 \\
Shared Ideology & 1.000 & 0.318 & 0.127 & 0.055 & 0.023 & 0.021 & 1.676 & 0.09 \\
Shared Most Common Region & 1.000 & 0.319 & 0.129 & 0.053 & 0.018 & 0.004 & 1.902 & 0.06 \\ \hline
\end{tabular}
\end{table}

\begin{table}[!h]
\caption{MCMC Diagnostics - ERGM Year 2015}
\setlength{\tabcolsep}{3pt}
\centering
\footnotesize
\begin{tabular}{lllllll|cc}
\hline
 & \multicolumn{6}{c}{\begin{tabular}[c]{@{}c@{}}Sample statistic \\ auto-correlation\end{tabular}} & \multicolumn{2}{c}{\begin{tabular}[c]{@{}c@{}}Geweke\\ Diagnostic\end{tabular}} \\ \hline
Covariate & \multicolumn{1}{c}{\begin{tabular}[c]{@{}c@{}}Lag \\ 0\end{tabular}} & \multicolumn{1}{c}{\begin{tabular}[c]{@{}c@{}}Lag \\ 1024\end{tabular}} & \multicolumn{1}{c}{\begin{tabular}[c]{@{}c@{}}Lag \\ 2048\end{tabular}} & \multicolumn{1}{c}{\begin{tabular}[c]{@{}c@{}}Lag \\ 3072\end{tabular}} & \multicolumn{1}{c}{\begin{tabular}[c]{@{}c@{}}Lag\\ 4096\end{tabular}} & \multicolumn{1}{c|}{\begin{tabular}[c]{@{}c@{}}Lag\\ 5120\end{tabular}} & \multicolumn{1}{l}{Statistic} & \multicolumn{1}{l}{p-value} \\ \hline
Sum of Features Weights & 1.000 & 0.032 & -0.007 & -0.001 & 0.012 & 0.010 & -1.208 & 0.22 \\
Diff in N of non-zero Features & 1.000 & 0.009 & -0.002 & 0.022 & -0.019 & 0.037 & -1.756 & 0.08 \\
Diff in Non-zero Features to Weights Ratio & 1.000 & 0.011 & 0.004 & 0.007 & -0.011 & 0.026 & -1.957 & 0.05 \\
Shared Most Common Target & 1.000 & 0.044 & 0.023 & 0.016 & 0.020 & 0.026 & -3.267 & 0.01 \\
Shared Most Common Tactic & 1.000 & 0.074 & 0.003 & 0.005 & 0.006 & 0.028 & -1.741 & 0.08 \\
Shared Most Common Weapon & 1.000 & 0.084 & 0.013 & 0.008 & 0.017 & 0.036 & -2.413 & 0.02 \\
Shared Ideology & 1.000 & 0.078 & 0.022 & 0.006 & 0.023 & 0.031 & -2.137 & 0.32 \\
Shared Most Common Region & 1.000 & 0.146 & 0.019 & -0.014 & -0.011 & 0.033 & -1.357 & 0.17 \\ \hline
\end{tabular}
\end{table}

\begin{table}[!h]
\caption{MCMC Diagnostics - ERGM Year 2016}
\setlength{\tabcolsep}{3pt}
\centering
\footnotesize
\begin{tabular}{lllllll|cc}
\hline
 & \multicolumn{6}{c}{\begin{tabular}[c]{@{}c@{}}Sample statistic \\ auto-correlation\end{tabular}} & \multicolumn{2}{c}{\begin{tabular}[c]{@{}c@{}}Geweke\\ Diagnostic\end{tabular}} \\ \hline
Covariate & \multicolumn{1}{c}{\begin{tabular}[c]{@{}c@{}}Lag \\ 0\end{tabular}} & \multicolumn{1}{c}{\begin{tabular}[c]{@{}c@{}}Lag \\ 1024\end{tabular}} & \multicolumn{1}{c}{\begin{tabular}[c]{@{}c@{}}Lag \\ 2048\end{tabular}} & \multicolumn{1}{c}{\begin{tabular}[c]{@{}c@{}}Lag \\ 3072\end{tabular}} & \multicolumn{1}{c}{\begin{tabular}[c]{@{}c@{}}Lag\\ 4096\end{tabular}} & \multicolumn{1}{c|}{\begin{tabular}[c]{@{}c@{}}Lag\\ 5120\end{tabular}} & \multicolumn{1}{l}{Statistic} & \multicolumn{1}{l}{p-value} \\ \hline
Sum of Features Weights & 1.000 & 0.200 & 0.054 & 0.017 & 0.025 & 0.006 & -0.205 & 0.83 \\
Diff in N of non-zero Features & 1.000 & 0.243 & 0.069 & 0.020 & 0.004 & 0.000 & 1.413 & 0.15 \\
Diff in Non-zero Features to Weights Ratio & 1.000 & 0.138 & 0.022 & 0.014 & 0.007 & 0.001 & 1.846 & 0.06 \\
Shared Most Common Target & 1.000 & 0.252 & 0.088 & 0.040 & 0.009 & 0.002 & 0.356 & 0.72 \\
Shared Most Common Tactic & 1.000 & 0.325 & 0.106 & 0.018 & 0.001 & 0.007 & 0.512 & 0.60 \\
Shared Most Common Weapon & 1.000 & 0.321 & 0.105 & 0.015 & -0.001 & 0.006 & 0.497 & 0.62 \\
Shared Ideology & 1.000 & 0.284 & 0.081 & 0.014 & 0.008 & 0.008 & 0.663 & 0.50 \\
Shared Most Common Region & 1.000 & 0.314 & 0.085 & 0.028 & 0.021 & 0.011 & 0.846 & 0.39 \\ \hline
\end{tabular}
\end{table}

\begin{table}[!h]
\caption{MCMC Diagnostics - ERGM Year 2017}
\setlength{\tabcolsep}{3pt}
\centering
\footnotesize
\begin{tabular}{lllllll|cc}
\hline
 & \multicolumn{6}{c}{\begin{tabular}[c]{@{}c@{}}Sample statistic \\ auto-correlation\end{tabular}} & \multicolumn{2}{c}{\begin{tabular}[c]{@{}c@{}}Geweke\\ Diagnostic\end{tabular}} \\ \hline
Covariate & \multicolumn{1}{c}{\begin{tabular}[c]{@{}c@{}}Lag \\ 0\end{tabular}} & \multicolumn{1}{c}{\begin{tabular}[c]{@{}c@{}}Lag \\ 1024\end{tabular}} & \multicolumn{1}{c}{\begin{tabular}[c]{@{}c@{}}Lag \\ 2048\end{tabular}} & \multicolumn{1}{c}{\begin{tabular}[c]{@{}c@{}}Lag \\ 3072\end{tabular}} & \multicolumn{1}{c}{\begin{tabular}[c]{@{}c@{}}Lag\\ 4096\end{tabular}} & \multicolumn{1}{c|}{\begin{tabular}[c]{@{}c@{}}Lag\\ 5120\end{tabular}} & \multicolumn{1}{l}{Statistic} & \multicolumn{1}{l}{p-value} \\ \hline
Sum of Features Weights & 1.000 & 0.042 & -0.002 & -0.001 & -0.002 & -0.019 & -0.232 & 0.81 \\
Diff in N of non-zero Features & 1.000 & 0.072 & -0.002 & -0.022 & 0.007 & 0.031 & -0.717 & 0.47 \\
Diff in Non-zero Features to Weights Ratio & 1.000 & 0.055 & 0.008 & 0.001 & 0.017 & 0.014 & -0.811 & 0.41 \\
Shared Most Common Target & 1.000 & 0.091 & 0.023 & 0.018 & 0.016 & 0.007 & -0.469 & 0.63 \\
Shared Most Common Tactic & 1.000 & 0.091 & 0.004 & 0.008 & 0.013 & -0.001 & -0.410 & 0.68 \\
Shared Most Common Weapon & 1.000 & 0.095 & 0.018 & 0.008 & 0.018 & 0.010 & -0.088 & 0.92 \\
Shared Ideology & 1.000 & 0.087 & 0.005 & 0.007 & 0.028 & 0.014 & -0.525 & 0.59 \\
Shared Most Common Region & 1.000 & 0.114 & 0.004 & 0.015 & 0.003 & 0.004 & 1.135 & 0.25 \\ \hline
\end{tabular}
\end{table}

\begin{table}[!h]
\caption{MCMC Diagnostics - ERGM Year 2018}
\setlength{\tabcolsep}{3pt}
\centering
\footnotesize
\begin{tabular}{lllllll|cc}
\hline
 & \multicolumn{6}{c}{\begin{tabular}[c]{@{}c@{}}Sample statistic \\ auto-correlation\end{tabular}} & \multicolumn{2}{c}{\begin{tabular}[c]{@{}c@{}}Geweke\\ Diagnostic\end{tabular}} \\ \hline
Covariate & \multicolumn{1}{c}{\begin{tabular}[c]{@{}c@{}}Lag \\ 0\end{tabular}} & \multicolumn{1}{c}{\begin{tabular}[c]{@{}c@{}}Lag \\ 1024\end{tabular}} & \multicolumn{1}{c}{\begin{tabular}[c]{@{}c@{}}Lag \\ 2048\end{tabular}} & \multicolumn{1}{c}{\begin{tabular}[c]{@{}c@{}}Lag \\ 3072\end{tabular}} & \multicolumn{1}{c}{\begin{tabular}[c]{@{}c@{}}Lag\\ 4096\end{tabular}} & \multicolumn{1}{c|}{\begin{tabular}[c]{@{}c@{}}Lag\\ 5120\end{tabular}} & \multicolumn{1}{l}{Statistic} & \multicolumn{1}{l}{p-value} \\ \hline
Sum of Features Weights & 1.000 & 0.168 & 0.094 & 0.035 & 0.015 & 0.016 & 0.633 & 0.52 \\
Diff in N of non-zero Features & 1.000 & 0.286 & 0.136 & 0.038 & 0.005 & 0.021 & 1.600 & 0.10 \\
Diff in Non-zero Features to Weights Ratio & 1.000 & 0.231 & 0.077 & 0.037 & -0.018 & 0.002 & 1.214 & 0.22 \\
Shared Most Common Target & 1.000 & 0.373 & 0.180 & 0.067 & 0.021 & 0.019 & 1.058 & 0.29 \\
Shared Most Common Tactic & 1.000 & 0.413 & 0.182 & 0.071 & 0.022 & 0.012 & 1.735 & 0.08 \\
Shared Most Common Weapon & 1.000 & 0.410 & 0.179 & 0.077 & 0.019 & 0.014 & 1.672 & 0.09 \\
Shared Ideology & 1.000 & 0.380 & 0.172 & 0.064 & 0.022 & 0.029 & 1.171 & 0.24 \\
Shared Most Common Region & 1.000 & 0.411 & 0.166 & 0.075 & 0.044 & 0.027 & 1.497 & 0.13 \\ \hline
\end{tabular}
\label{diag2018}
\end{table}

\clearpage
\subsection{Robustness: Drivers of Similarity}\label{robus_simil}
\begin{figure}[!hbt]
    \centering
    \includegraphics[scale=0.5]{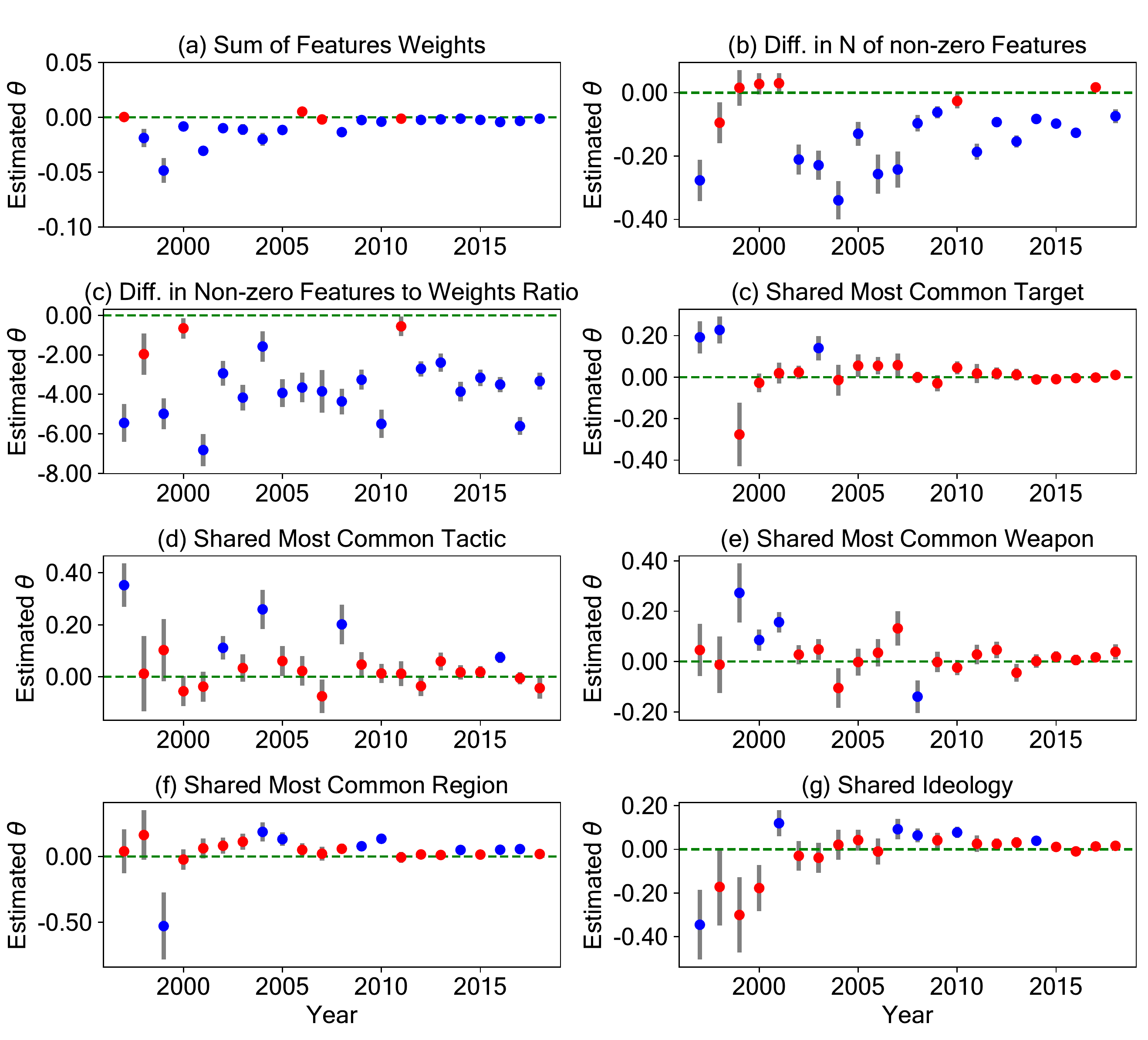}
    \caption{ERGM estimated coefficients for each covariate included in the model, using the enlarged sample for robustness check ($N$=164).}
    \label{ergm_robus}
\end{figure}

As for the co-clustering stability analysis, we have re-estimated ERGM using the enlarged sample for robustness purposes. Results of the models are reported in Figure \ref{ergm_robus} and showcase that the outcomes commented in the main text hold even when including those groups that have plotted at least 30 attacks over the period under consideration. 

Sum of Features Weights, Difference in Number of non-zero Features and Difference in Non-zero Features to Weights Ratio largely and stably remain the three main predictors explaining co-clustering, and therefore operational similarity. Interestingly, it should be noted that the estimated $\theta$ coefficients for Difference in Non-zero Features to Weights Ratio are considerably lower in magnitude compared to the ones seen in the main analysis. Nonetheless, their significance and direction remain clear. In terms of other differences, ideology appears to have a role in partially explaining co-clustering in certain years, although the wide majority of non-significant coefficients support the conclusion that ideological homophily cannot be considered a consistent driver of co-clustering. 

\newpage
\section{Complete List of Terrorist Groups in the Sample}
\color{black}
The terrorist organizations that are included in the sample are displayed below. 

\begin{table}[!h]
\color{black}
\caption{\color{black}Terrorist Organizations in the Original Sample (n=105), in Alphabetical Order}
\footnotesize
\centering
\begin{tabular}{l|l}
\hline
Abu Sayyaf Group   (ASG) & Islamic State of Iraq and the Levant (ISIL) \\
Al-Aqsa Martyrs Brigade & Islamist extremists \\
Algerian Islamic Extremists & Jaish-e-Mohammad (JeM) \\
Allied Democratic Forces (ADF) & Jamaat Nusrat al-Islam wal Muslimin (JNIM) \\
Al-Nusrah Front & Jamaat-E-Islami (Bangladesh) \\
Al-Qaida & Jama'atul Mujahideen Bangladesh (JMB) \\
Al-Qaida in Iraq & Janatantrik Terai Mukti Morcha- Jwala Singh (JTMM-J) \\
Al-Qaida in the Arabian Peninsula (AQAP) & Janjaweed \\
Al-Qaida in the Islamic Maghreb (AQIM) & Jemaah Islamiya (JI) \\
Al-Shabaab & Jihadi-inspired extremists \\
Anarchists & Khorasan Chapter of the Islamic State \\
Animal Liberation Front (ALF) & Kurdistan Workers' Party (PKK) \\
Ansar al-Sharia (Libya) & Lashkar-e-Islam (Pakistan) \\
Ansar Bayt al-Maqdis (Ansar Jerusalem) & Lashkar-e-Jhangvi \\
Anti-Abortion extremists & Lashkar-e-Taiba (LeT) \\
Anti-Balaka Militia & Liberation Tigers of Tamil Eelam (LTTE) \\
Anti-Muslim extremists & Lord's Resistance Army (LRA) \\
Armed Islamic Group (GIA) & Luhansk People's Republic \\
Asa'ib Ahl al-Haqq & Maoists \\
Badr Brigades & Mayi Mayi \\
Baloch Liberation Army (BLA) & Moro Islamic Liberation Front (MILF) \\
Baloch Liberation Front (BLF) & Movement for the Emancipation of the Niger Delta (MEND) \\
Baloch Republican Army (BRA) & Mozambique National Resistance Movement (MNR) \\
Bangladesh Nationalist Party (BNP) & Muslim extremists \\
Bangsamoro Islamic Freedom Movement   (BIFM) & National Democratic Front of Bodoland (NDFB) \\
Barisan Revolusi Nasional (BRN) & National Liberation Army of Colombia (ELN) \\
Barqa Province of the Islamic State & National Liberation Front of Tripura (NLFT) \\
Basque Fatherland and Freedom (ETA) & National Socialist Council of Nagaland-Isak-Muivah (NSCN-IM) \\
Boko Haram & National Union for the Total Independence of Angola (UNITA) \\
Chechen Rebels & New People's Army (NPA) \\
Communist Party of India - Maoist   (CPI-Maoist) & Niger Delta Avengers (NDA) \\
Communist Party of Nepal - Maoist   (CPN-Maoist-Chand) & Palestinian Extremists \\
Conspiracy of Cells of Fire & Palestinian Islamic Jihad (PIJ) \\
Corsican National Liberation Front   (FLNC) & Paraguayan People's Army (EPP) \\
Corsican National Liberation Front-   Historic Channel & People's Liberation Front of India \\
Democratic Front for the Liberation of   Rwanda (FDLR) & People's War Group (PWG) \\
Dissident Republicans & Popular Front for the Liberation of Palestine (PFLP) \\
Donetsk People's Republic & Popular Resistance Committees \\
Earth Liberation Front (ELF) & Pro Hartal Activists \\
Free Aceh Movement (GAM) & Revolutionary Armed Forces of Colombia (FARC) \\
Free Syrian Army & Revolutionary Armed Forces of Colombia (FARC) dissidents \\
Fulani extremists & Runda Kumpulan Kecil (RKK) \\
Garo National Liberation Army & Salafist Group for Preaching and Fighting (GSPC) \\
Gorkha Janmukti Morcha (GJM) & Shining Path (SL) \\
Hamas (Islamic Resistance Movement) & Sinai Province of the Islamic State \\
Haqqani Network & Sindhu Desh Liberation Army (SDLA) \\
Hezbollah & Sudan People's Liberation Movement in Opposition (SPLM-IO) \\
Hizbul Mujahideen (HM) & Taliban \\
Houthi extremists (Ansar Allah) & Tehrik-i-Taliban Pakistan (TTP) \\
Hutu extremists & Tripoli Province of the Islamic State \\
Indian Mujahideen & United Baloch Army (UBA) \\
Islamic State of Iraq (ISI) & United Liberation Front of Assam (ULFA) \\
 & United Self Defense Units of Colombia (AUC)\\\hline
\end{tabular}
\label{groups}
\end{table}

It should be noted that, from the pool of actors that have been responsible of at least 50 attacks from 1997 to 2018, we removed several of them. First, we filtered out ``Unknown'' perpetrators as it would have been meaningless to treat this category as an actual organization. Second, we also removed the following actors: ``Gunmen'', ``Separatists'', ``Militants'', ``Tribesmen'', ``Anti-Muslim Extremists''. This choice has been motivated by the fact that these categories are overly broad in the dataset, and do not refer to specific groups. For instance, ``Separatists'' refers to perpetrators of attacks in different parts of the world where independence instances are at play. This extremely heterogeneity leaving open too many possible interpretations prevented us to treat such actors are unique formations. Contrarily, other broad categories such as ``Jihadi-inspired extremists'' have been left in the sample because, in spite of geographical variability'', they can be more precisely linked to a motive and ideology.

\newpage


\end{document}